\documentclass[superscriptaddress,showpacs,pre,twocolumn,nofootinbib]{revtex4-1}
\usepackage{epsfig}
\usepackage{bm}
\usepackage{amsmath}
\usepackage{subfigure}
\usepackage{graphicx}
\usepackage{epstopdf}
\usepackage[pdftex]{color}
\usepackage{hyperref}
\hypersetup{
    colorlinks=true,       
    linkcolor=cyan,          
    citecolor=magenta,        
    filecolor=magenta,      
    urlcolor=cyan,           
    runcolor=cyan
}

\newcommand{\beq}{\begin{eqnarray}}
\newcommand{\eeq}{\end{eqnarray}}
\newcommand{\bes} {\begin{subequations}}
\newcommand{\ees} {\end{subequations}}

\newcommand{\byDL}[1]{{\textcolor{blue}{#1}}}

\newcommand{\HI}{H_{\textrm{Ising}}}

\newcommand{\TTS}{\textrm{TTS}}
\newcommand{\ignore}[1]{}
\newcommand{\vs}{\textit{vs} }

\begin{document}

\title{Probing for quantum speedup in spin glass problems with planted solutions}

\author{Itay Hen}
\affiliation{Information Sciences Institute, University of Southern California, Marina del Rey, CA 90292}

\author{Joshua Job}
\affiliation{Department of Physics and Astronomy, University of Southern California, Los Angeles, California 90089, USA}
\affiliation{Center for Quantum Information Science \& Technology, University of Southern California, Los Angeles, California 90089, USA}

\author{Tameem Albash}
\affiliation{Department of Physics and Astronomy, University of Southern California, Los Angeles, California 90089, USA}
\affiliation{Center for Quantum Information Science \& Technology, University of Southern California, Los Angeles, California 90089, USA}

\author{Troels F. R{\o}nnow}
\affiliation{Theoretische Physik, ETH Zurich, 8093 Zurich, Switzerland}
\author{Matthias Troyer}
\affiliation{Theoretische Physik, ETH Zurich, 8093 Zurich, Switzerland}

\author{Daniel Lidar$^\ast$}
\affiliation{Department of Physics and Astronomy, University of Southern California, Los Angeles, California 90089, USA}
\affiliation{Center for Quantum Information Science \& Technology, University of Southern California, Los Angeles, California 90089, USA}
\affiliation{Department of Electrical Engineering, University of Southern California, Los Angeles, California 90089, USA}
\affiliation{Department of Chemistry, University of Southern California, Los Angeles, California 90089, USA}
\email{lidar@usc.edu}

\date{\today}

\begin{abstract}
The availability of quantum annealing devices with hundreds of qubits has made the experimental demonstration of a quantum speedup for optimization problems a coveted, albeit elusive goal. 
Going beyond earlier studies of random Ising problems, here we introduce a method to construct a set of frustrated Ising-model optimization problems with tunable hardness. We study the performance of a D-Wave Two device (DW2) with up to $503$ qubits on these problems and compare it to a suite of classical algorithms, including a highly optimized algorithm designed to compete directly with the DW2. 
The problems are generated around predetermined ground-state configurations, called planted solutions, which makes them particularly suitable for benchmarking purposes. The problem set exhibits properties familiar from constraint satisfaction (SAT) problems, such as  
a peak in the typical hardness of the problems, determined by a tunable clause density parameter. 
We bound the hardness regime where the DW2 device either does not or might exhibit a quantum speedup for our problem set. While we do not find evidence for a speedup for the hardest and most frustrated problems in our problem set, we cannot rule out that a speedup might exist for some of the easier, less frustrated problems. Our empirical findings pertain to the specific D-Wave processor and problem set we studied and leave open the possibility that future processors might exhibit a quantum speedup on the same problem set.
\end{abstract}

\maketitle

\section{Introduction}
\label{sec:intro}

Interest in quantum computing is motivated by the potential for quantum speedup -- the ability of quantum computers, aided by uniquely quantum features such as entanglement and tunneling, to solve certain computational problems in a manner that scales better with problem size than is possible classically. 
Despite tremendous progress in building small-scale quantum information processors \cite{Ladd:10}, there is as of yet no conclusive experimental evidence of a quantum speedup. For this reason, there has been much recent interest  in building more specialized quantum information processing devices that can achieve relatively large scales, such as quantum simulators \cite{Cirac:2012pi} and quantum annealers \cite{DWave,EPJ-ST:2015}. The latter are designed to solve classically hard optimization problems by exploiting the phenomenon of quantum tunneling \cite{Apolloni:1989fj,PhysRevB.39.11828,Amara:1993rt,finnila_quantum_1994,Kadowaki_quantum_1998,sqa1,Santoro,Brooke1999,farhi_quantum_2001,Reichardt:2004}.
Here we report on experimental results that probe the possibility that a putative quantum annealer may be capable of speeding up the solution of certain carefully designed optimization problems. We refer to this either as a \emph{limited} or as a \emph{potential} quantum speedup, since we study the possibility of an advantage relative to a portfolio of classical algorithms that either ``correspond" to the quantum annealer (in the sense that they implement a similar algorithmic approach running on classical hardware), or implement a specific, specialized classical algorithm \cite{speedup}. 
In addition, for technical reasons detailed below we must operate the putative quantum annealer in a suboptimal regime. With these caveats in mind, we push the experimental boundary in searching for a quantum advantage over a class of important classical algorithms, which includes simulated classical and quantum annealing \cite{kirkpatrick_optimization_1983,RevModPhys.80.1061}, using quantum hardware. We achieve this by designing Ising model problems that exhibit frustration, a well-known feature of classically-hard optimization problems \cite{Binder86,Nishimori-book}. In doing so we go beyond the random spin-glass problems 
of earlier studies \cite{q108,speedup}, by ensuring that the problems we study exhibit a degree of hardness that we can tune, and have at least one ``planted" ground state that we know in advance.

The putative quantum annealer used in our work is the D-Wave Two (DW2) device \cite{Bunyk:2014hb}. This device is designed to solve optimization problems by evolving a known initial configuration --- the ground state of a transverse field $H_X = \sum_{i\in V} \sigma_i^x$, where $\sigma_i^x$ is the Pauli spin-$1/2$ matrix acting on spin $i$ --- towards the ground state of a classical Ising-model Hamiltonian which serves as a cost function that encodes the problem that is to be solved:
\beq 
\HI =\sum_{(i,j)\in E} J_{ij} \sigma_i^z \sigma_j^z + \sum_{i\in V} h_i \sigma_i^z \ .
\label{eq:H}
\eeq 
The variables $\{\sigma_i^z\}$ denote either classical Ising-spin variables that take values $\pm1$ or Pauli spin-$1/2$ matrices, the $\{J_{ij}\}$ are programmable coupling parameters, and the $\{h_{i}\}$ are programmable local longitudinal fields. The $N$ spin variables are realized as superconducting flux qubits and occupy the vertices $V$ of a graph $G$ with edge set $E$. Here $G$ is the D-Wave ``Chimera" hardware graph \cite{Bunyk:2014hb,Choi2}. Further details, including a visualization of the Chimera graph and the annealing schedule that interpolates between $H_X$ and $\HI$, are provided in Appendix~\ref{app:methods}.

The dual questions of the computational power and underlying physics of the D-Wave devices --- the $512$-qubit DW2 and its predecessor, the $128$-qubit DW1 --- have generated a fair amount of interest and debate. A major concern is to what extent quantum effects determine the performance of these devices, given that they are inherently noisy and operate at temperatures ($\sim$20 mK) where thermal effects are expected to play a significant role, causing decoherence, excitation and relaxation. Several studies have addressed these concerns \cite{DWave,q-sig,q108,DWave-16q,PAL:13,SSSV,q-sig2,SSSV-comment,PAL:14,Crowley:2014qp,Venturelli:2014nx,Albash:2014if}. Entanglement has been experimentally detected during the annealing process \cite{DWave-entanglement}, and multiqubit tunneling involving up to $8$ qubits has been demonstrated to play a functional role in determining the output of a DW2 device programmed to solve a simple non-convex optimization problem \cite{Boixo:2014yu}. However, the role of quantum effects in determining the computational performance of the D-Wave devices on hard optimization problems involving many variables remains an open problem. A direct approach to try to settle the question is to demonstrate a quantum speedup. Such a demonstration has so far been an elusive goal, possibly because the random Ising problems chosen in previous benchmarking tests \cite{q108,speedup} were too easy to solve for the classical algorithms against which the D-Wave devices were compared; namely such problems exhibit a spin-glass phase only at zero temperature \cite{2014Katzgraber}. The Sherrington-Kirkpatrick model with random $\pm 1$ couplings, exhibiting a positive spin-glass critical temperature, was tested on a DW2 device, but the problem sizes considered were too small (due to the need to embed a complete graph into the Chimera graph) to test for a speedup \cite{Venturelli:2014nx}. The approach we outline next allows us to directly probe for a quantum speedup using frustrated Ising spin glass problems with a tunable degree of hardness, though we do not know whether these problems exhibit a positive-temperature spin-glass phase.

\begin{figure}[t]
\begin{center}
\subfigure{\includegraphics[width=\columnwidth]{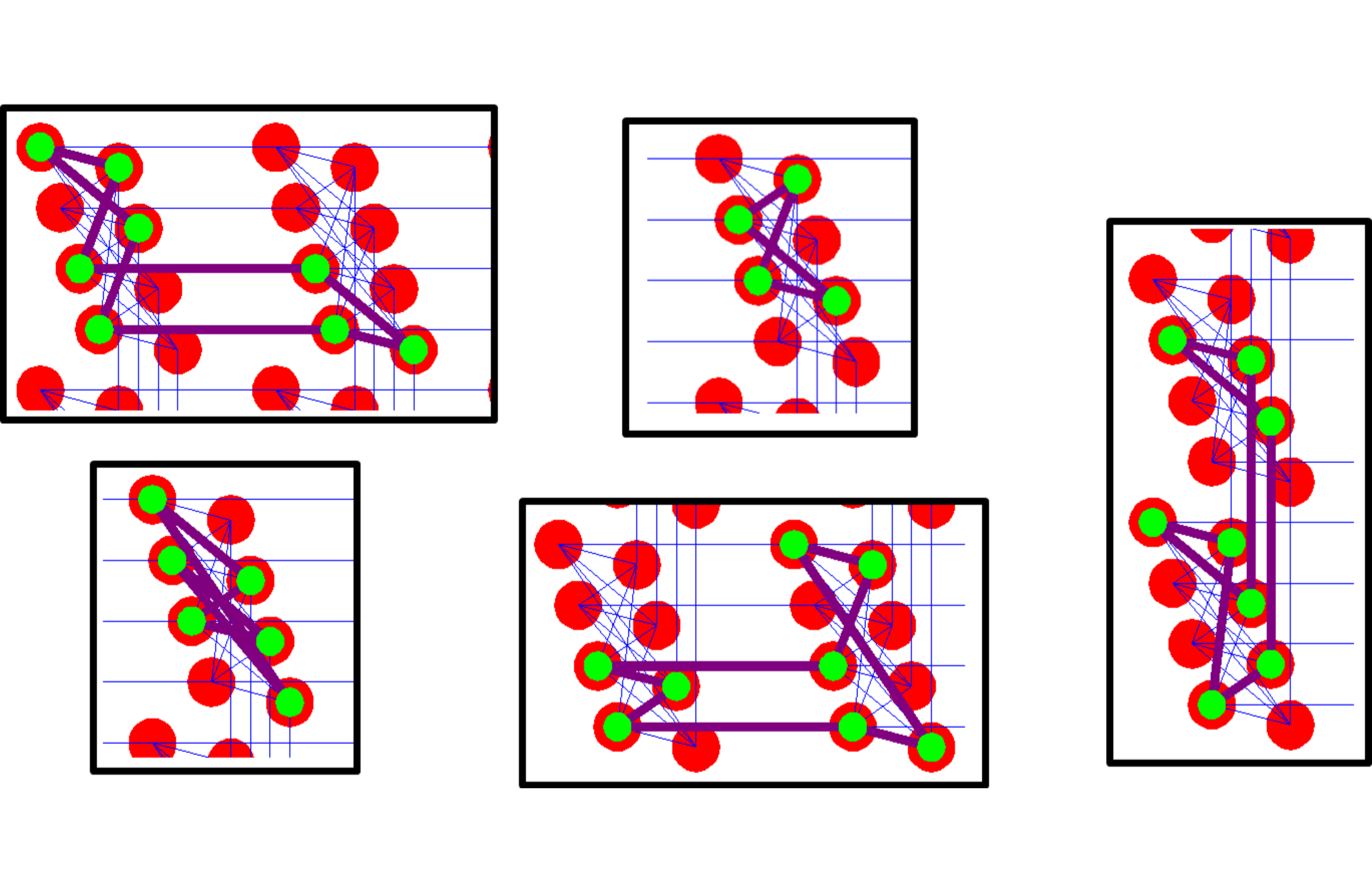}}
\subfigure{\includegraphics[width=\columnwidth]{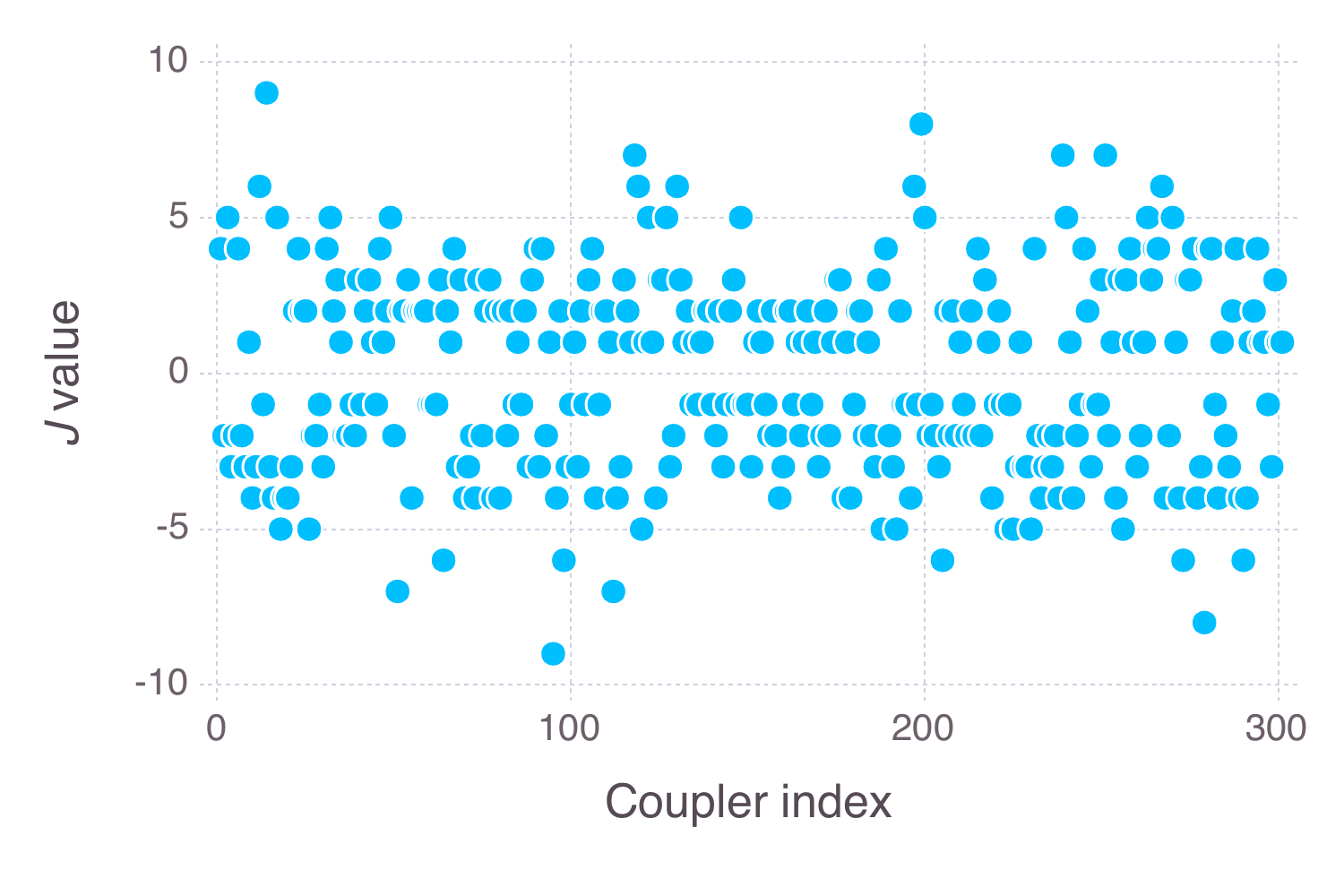}}
\caption{\textbf{Examples of randomly generated loops and couplings on the DW2 Chimera graph.} Top: qubits and couplings participating in the loops are highlighted in green and purple, respectively. Only even-length loops are embeddable on the Chimera graph. 
Bottom: distribution of $J$ values for a sample problem instance with $N=126$ spins and edges, and $101$ loops. It is virtually impossible to recover the loop-Hamiltonians $H_j$ from a given $\HI$. The couplings are all eventually rescaled to lie in $[-1,1]$. We always set the local fields $h_i$ to zero as non-zero fields tended to make the problems easier.}
\label{fig:loops}
\end{center}
\end{figure}

\section{Frustrated Ising problems with planted solutions}
\label{sec:II}

In this section we introduce a method for generating families of benchmark problems that have a certain degree of ``tunable hardness", achieved by adjusting the amount of frustration, a well-known concept from spin-glass theory \cite{Parisi:book}.  In frustrated optimization problems no configuration of the variables simultaneously minimizes all terms in the cost function, often causing classical algorithms to get stuck in local minima, since the global minimum of the problem satisfies only a fraction of the Ising couplings and/or local fields. We construct our problems around ``planted solutions'' -- an idea borrowed from constraint satisfaction (SAT) problems~\cite{Barthel:2002tw,Krzakala:2009qo}. The planted solution represents a ground state configuration of Eq.~\eqref{eq:H} that minimizes the energy and is known in advance. This knowledge circumvents the need to verify the ground state energy using exact (provable) solvers, which rapidly become too expensive computationally as the number of variables grows, and which were employed in earlier benchmarking studies \cite{q108,speedup}.
 
To create Ising-type optimization problems with planted solutions, we make use of frustrated ``local Ising Hamiltonians" $H_j$, i.e., Ising problem instances defined on sub-graphs of the Chimera graph in lieu of the clauses appearing in the SAT formulas. The total Hamiltonian for each problem is then of the form $\HI = \sum_{j=1}^M H_j$, where the sum is over the (possibly partially overlapping) local Hamiltonians.  Similarly to SAT problems, the size of these local Hamiltonians, or clauses, does not scale with the size of the problem. Moreover, to ensure that the planted solution is a ground state of the total Hamiltonian, we construct the clauses so that each is minimized by that portion of the planted solution that has support over the corresponding subgraph.%
\footnote{To see that the planted solution minimizes the total Hamiltonian, assume that a configuration $x_*$ is a minimum of $f_j(x)$ for all $j$, where $\{f_j(x)\}$ is a set of arbitrary real-valued functions. Then, by definition, for each $j$, $f_j(x)\geq f_j(x_*)$ for all possible configurations $x$. Let us now define $f(x) \equiv \sum_j f_j(x)$. It follows then that $f(x) \geq \sum_j f_j(x_*)$. Since also $f(x_*) = \sum_j f_j(x_*)$, $x_*$ is a minimizing configuration of $f(x)$.}
The planted solution is therefore determined prior to constructing the local Hamiltonians, by assigning random values to the bits on the nodes of the graph. The above process generates a Hamiltonian with the property that the planted solution is a simultaneous ground state of all the frustrated local Hamiltonians.%
\footnote{Somewhat confusingly from our perspective of utilizing frustration, such Hamiltonians are sometimes called ``frustration-free'' \cite{Bravyi:2009sp}.}

The various clauses $H_j$ can be generated in many different ways.  This freedom allows for the generation of many different types of instances, and here we present one method.  An $N$-qubit $M$-clause instance is generated as follows. 
\begin{enumerate}
\item 
A random configuration of $N$ bits corresponding to the participating spins of the Chimera graph is generated. 
This configuration constitutes the planted solution of the instance. 
\item
$M$ random loops are constructed along the edges of the Chimera graph. The loops are constructed by placing random walkers on random nodes of the Chimera graph, where each edge is determined at random from the set of all edges connected to the last edge. The random walk is terminated once the random walker crosses its path, i.e., when a node that has already been visited is encountered again. If this node is not the origin of the loop the ``tail" of the path is discarded. Examples of such loops are given in Fig.~\ref{fig:loops}. 
We distinguish between ``short loops'' of length $\ell=4,6$, and ``long loops" of length $\ell \geq 8$, as these give rise to peaks in hardness at different loop densities. Here we focus on long loops; results for short loops, which tend to generate significantly harder problem instances, will be presented elsewhere. 
\item 
On each loop, a clause $H_j$ is defined by assigning $J_{ij}=\pm1$ couplings to the edges of the loop in such a way that the planted solution minimizes $H_j$. As a first step, the $J_{ij}$'s are set to the ferromagnetic  $J_{ij}=-s_i s_j$, where the $s_i$ are the planted solution values. One of the couplings in the loop is then chosen at random and its sign is flipped.
This ensures that no spin configuration can satisfy every edge in that loop, and the planted solution remains a global minimum of the loop, but is now a frustrated ground state.%
\footnote{To see that there can be no spin configuration with an energy lower than that of the planted solution, consider a given loop and a given spin in that loop; note that every spin participates in two couplings. Either both couplings are satisfied after the sign flip, or one is satisfied and the other is not. Correspondingly, flipping that spin will thus either raise the energy or leave it unchanged. This is true for all spins in the loop.}
\item
The total Hamiltonian is then formed by adding up the $M$-loop clauses $H_j$. Note that loops can partially overlap, thereby also potentially ``canceling out''  each other's frustration, a useful feature that will give rise to an easy-hard-easy pattern we discuss below. Since the planted solution is a ground state of each of the $H_j$'s, it is also a ground state of the total Hamiltonian $\HI$.
\end{enumerate}

Ising-type optimization problems with planted solutions, such as those we have generated, have several attractive properties that we utilize later on: 
i) Having a ground-state configuration allows us to readily precompute a measure of frustration, e.g., the fraction of frustrated couplings of the planted solution. We shall show that this type of measure correlates well with the hardness of the problem, as defined in terms of the success probability of finding the ground state or the scaling of the time-to-solution. ii) By changing the number of clauses $M$ we can create different classes of problems, each with a ``clause density" $\alpha=M/N$, analogous to problem generation in SAT.%
\footnote{Note that for small values of $M$, the number of spins actually participating in an instance will be smaller than $N$, the number of spins on the graph from which the clauses are chosen.}
The clause density can be used to tune through a SAT-type phase transition \cite{Bollobas:2001}, i.e., it may be used to control the hardness of the generated problems. Here too, we shall see that the clause density plays an important role in setting the hardness of the problems.
Note that when the energy is unchanged under a spin-flip the solution is degenerate, so that our planted solution need not be unique. 

\section{Algorithms and scaling}
A judicious choice of classical algorithms is required to ensure that our test of a limited or potential quantum speedup is meaningful.
We considered (i) simulated annealing (SA), a well-known, powerful and generic heuristic solver \cite{kirkpatrick_optimization_1983}; (ii) simulated quantum annealing (SQA) \cite{finnila_quantum_1994,Kadowaki_quantum_1998,sqa1,Santoro}, a quantum Monte Carlo algorithm that has been shown to be consistent with the D-Wave devices \cite{q108,Albash:2014if}; (iii) the Shin-Smolin-Smith-Vazirani (SSSV) thermal rotor model, that was specifically designed to mimic the D-Wave devices in their classical limit \cite{SSSV}; (iv) the Hamze-Freitas-Selby (HFS) algorithm \cite{hamze:04,Selby:2014tx}, an algorithm that is fine-tuned for the Chimera graph and appears to be the most competitive at this time.
Of these, the HFS algorithm is the only one that is designed to exploit the scaling of the treewidth of the Chimera graph (see Appendix~\ref{app:methods})
, which renders it particularly efficient in a comparison against the D-Wave devices \cite{selby:13b}.

The D-Wave devices and all the algorithms we considered are probabilistic, and return the correct ground state with some probability of success. 
We thus perform many runs of the same duration $\tau$ for a given problem instance, and estimate the success probability empirically as the number of successes divided by the number of runs. This is repeated for many instances at a given clause density $\alpha$ and number of variables $N$, and generates a distribution of success probabilities. Let $p(\lambda)$ denote the success probability for a given set of parameters $\lambda=\{N,\alpha,q\}$, where $q$ denotes the $q$th percentile of this distribution; e.g., half the instances for given $N$ and $\alpha$ have a higher empirical success probability than the median $p(N,\alpha,0.5)$. The \emph{number of runs} required to find the ground state at least once with a desired probability $p_d$ is \cite{q108,speedup}%
\footnote{We prefer to define $r$ in this manner, rather than rounding it as in \cite{q108,speedup}, as this simplifies the extraction of scaling coefficients.}
\beq
r(\lambda)=\frac{\log (1 - p_d)}{\log (1 - p(\lambda))}\ ,
\label{eq:r-def}
\eeq
and henceforth we set $p_d=0.99$. Correspondingly, the \emph{time-to-solution} is $\TTS(\lambda) = r(\lambda)\tau$, where for the D-Wave device $\tau$ signifies the annealing time $t_a$ (at least $20\mu$s), while for SA, SQA, and SSSV $\tau$ is the number of Monte Carlo sweeps $s$ (a complete update of all spins) multiplied by the time per sweep $\tau_X$ for algorithm $X$.%
\footnote{In our simulations $\tau_{\textrm{SA}} = 3.54\mu$s, $\tau_{\textrm{SQA}} = 9.92\mu$s, and $\tau_{\textrm{SSSV}} = 10.34\mu$s.}
For SA, we further distinguish between using it as a \emph{solver} (SAS) or as an \emph{annealer} (SAA): in SAS mode we keep track of the energies found along the annealing schedule (which we take to be linear in the inverse temperature $\beta$) and take the lowest, while in SAA mode we always use the final energy found. Thus SAA can never be better than SAS, but is a more faithful model of an analog annealing device. A similar distinction can be made for SQA (i.e., SQAA and SQAS), but we primarily consider the annealer version since it too more closely mimics the operation of DW2 (note that SAA and SQAA were  also the modes used in Refs.~\cite{q108,speedup}; SQAS results are shown in Appendix~\ref{app:additional}). For the HFS algorithm, $\TTS(\lambda)$ is calculated directly from the distribution of runtimes obtained from $10^5$ identical, independent executions of the algorithm. Further timing details are given in Appendix~\ref{app:methods}.

\begin{figure}
\begin{center}
\subfigure[\ DW2]{\includegraphics[width=\columnwidth]{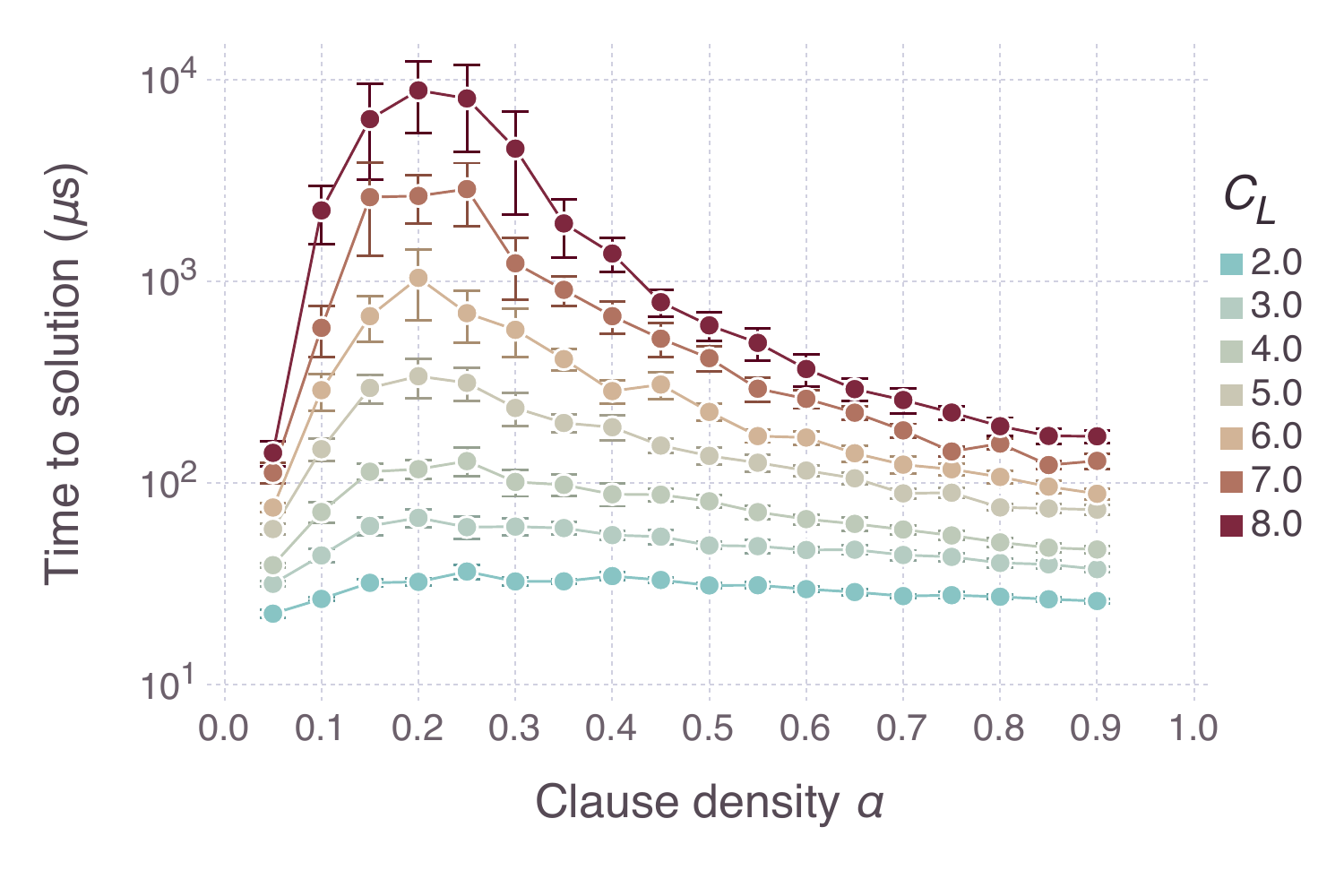}\label{fig:DwaveVsSelby50-a}}
\subfigure[\ HFS]{\includegraphics[width=\columnwidth]{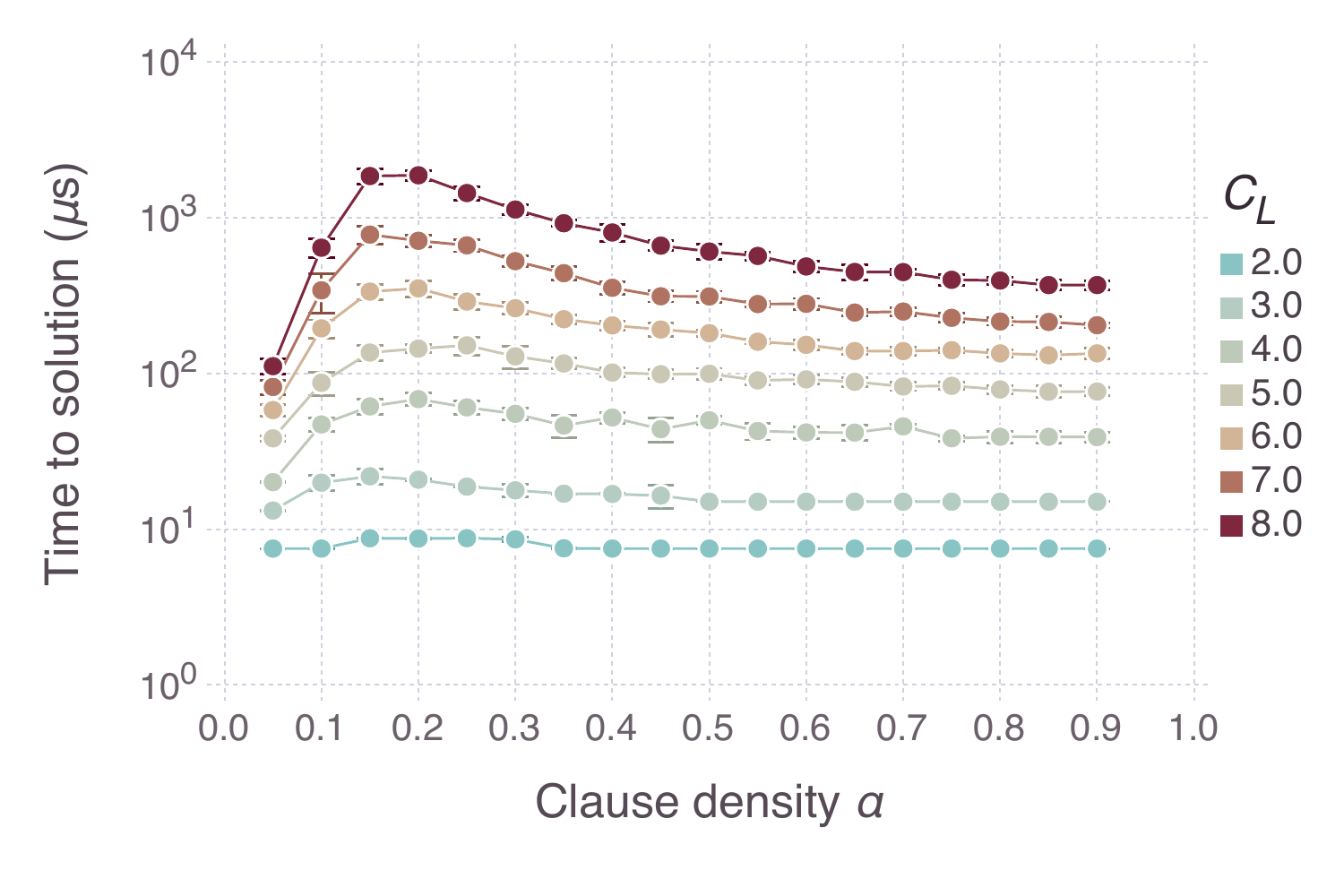}\label{fig:DwaveVsSelby50-b}}
\caption{\textbf{Time to solution as a function of clause density.} Shown is $\TTS(L,\alpha,0.5)$ (log scale) for (a) DW2 and (b) HFS, as a function of the clause density. The different colors represent the different Chimera subgraph sizes, which continue to $L=12$ in the HFS case.  In both cases there is a clear peak. From the HFS results we can identify the peak position as being at $\alpha = 0.17\pm0.01$, which is consistent with the peak position in the DW2 results. Error bars represent $2\sigma$ confidence intervals. 
}
\label{fig:DwaveVsSelby50}
\end{center}
\end{figure}

\begin{figure}[t]
\begin{center}
\includegraphics[width=\columnwidth]{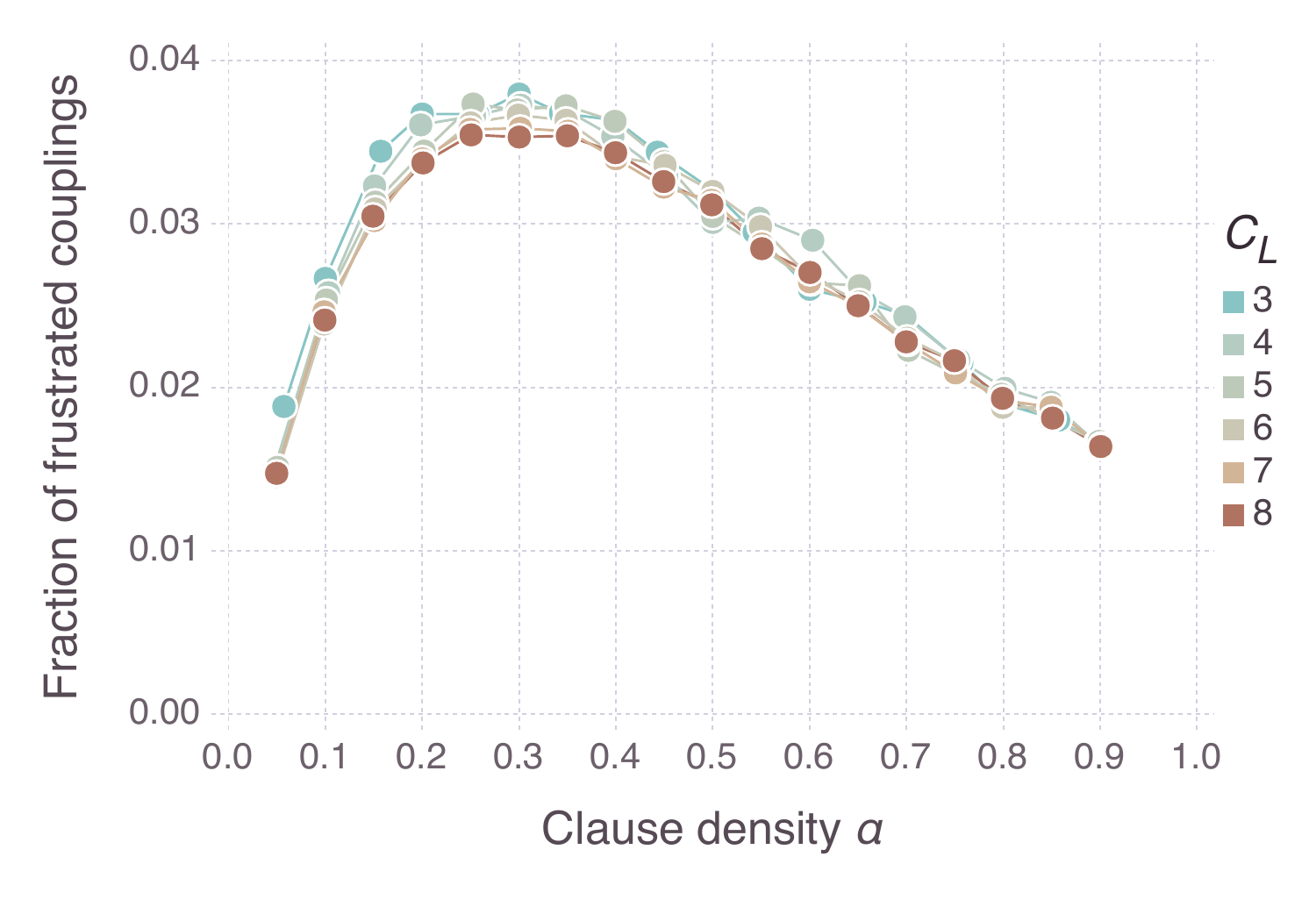}
\caption{\textbf{Frustration fraction.} Shown is the fraction of frustrated couplings (the number of frustrated couplings divided by the total number of couplings, where a frustrated coupling is defined with respect to the planted solution) as a function of clause density for different Chimera subgraphs $C_L$, in the case of  loops of length $\geq 8$, averaged over the $100$ instances for each given $\alpha$ and $N$. There is a broad peak at $\alpha \approx 0.25$. This is the clause density at which there is the largest fraction of frustrated couplings, and is near where we expect the hardest instances to occur, in good agreement with Fig.~\ref{fig:DwaveVsSelby50}. 
}
\label{fig:frustRat}
\end{center}
\end{figure}

\begin{figure}[t]
\begin{center}
\includegraphics[width=\columnwidth]{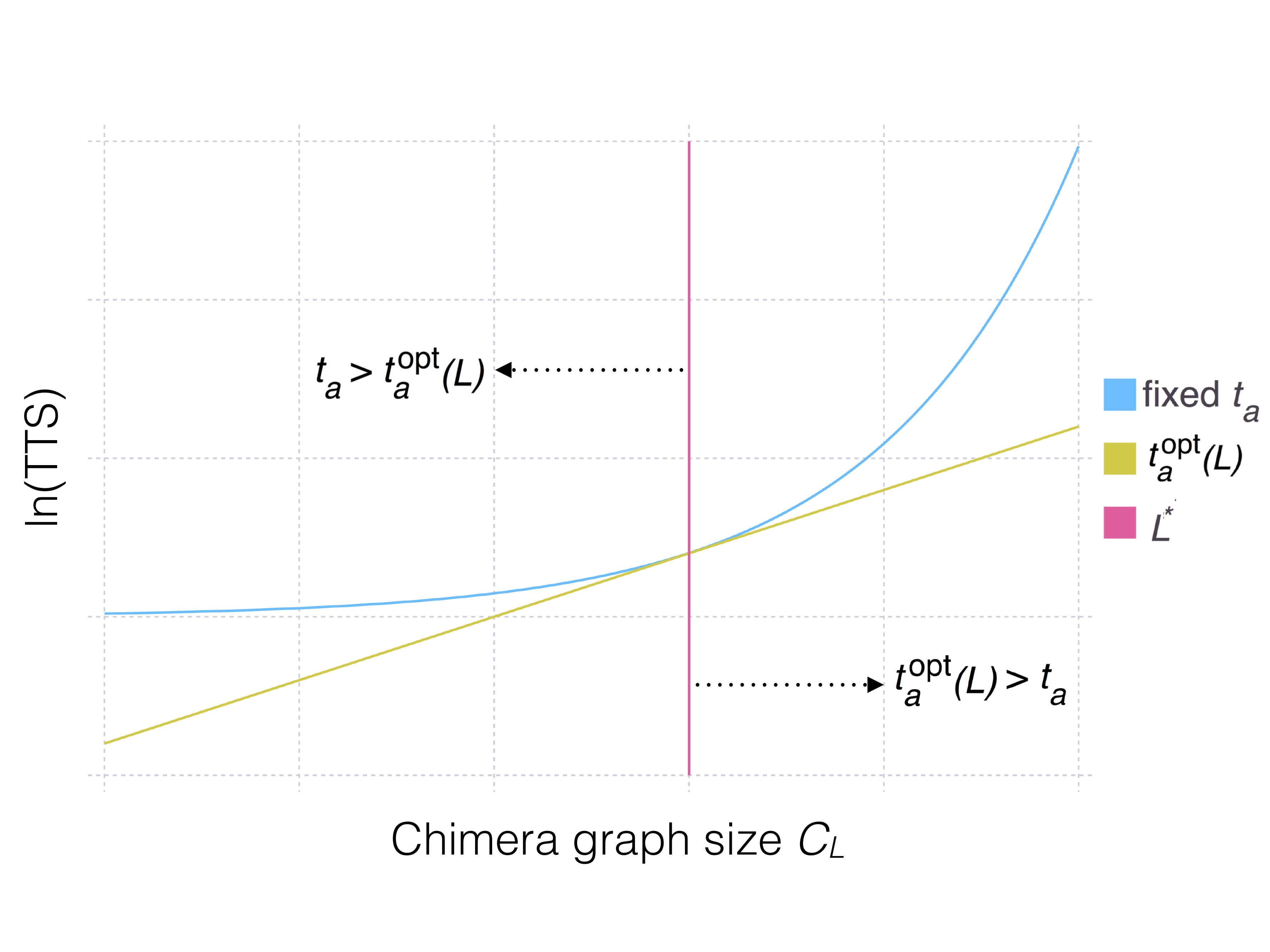}
\caption{\textbf{Sketch of the relation between the $\log(\TTS)$ curves for optimal and suboptimal annealing times.} The  annealing time needs to be optimized for each problem size. Blue represents the TTS with a size-independent annealing time $t_a$.  Red represents the optimal TTS corresponding to having an optimal size-dependent annealing time $t_a^{\mathrm{opt}}(L)$, i.e. the lower envelope of the full series of fixed annealing time TTS curves.  This curve need not be linear as depicted, though we expect it to be linear for NP-hard problems. The blue line upper bounds the red line since by definition $\TTS_{\textrm{DW2}}(\lambda,t_a^{\mathrm{opt}}(L)) \leq \TTS_{\textrm{DW2}}(\lambda,t_a)$. The vertical dotted line represents the problem size $L^*$ at which $t_a = t_a^\textrm{opt}(L^*)$. To the left of this line $t_a > t_a^{\mathrm{opt}}$ and the slope of the fixed-$t_a$ TTS curve lower-bounds the slope of the optimal TTS curve, since for very small problem sizes a large $t_a$ results in insensitivity to problem size, and the success probability is essentially constant. The opposite happens to the right of this line, where $t_a < t_a^{\mathrm{opt}}$, and where the success probability rapidly drops with $L$ at fixed $t_a$. }
\label{fig:TTS_Illustration}
\end{center}
\end{figure}

\begin{figure*}[t]
\begin{center}
\includegraphics[width=\textwidth]{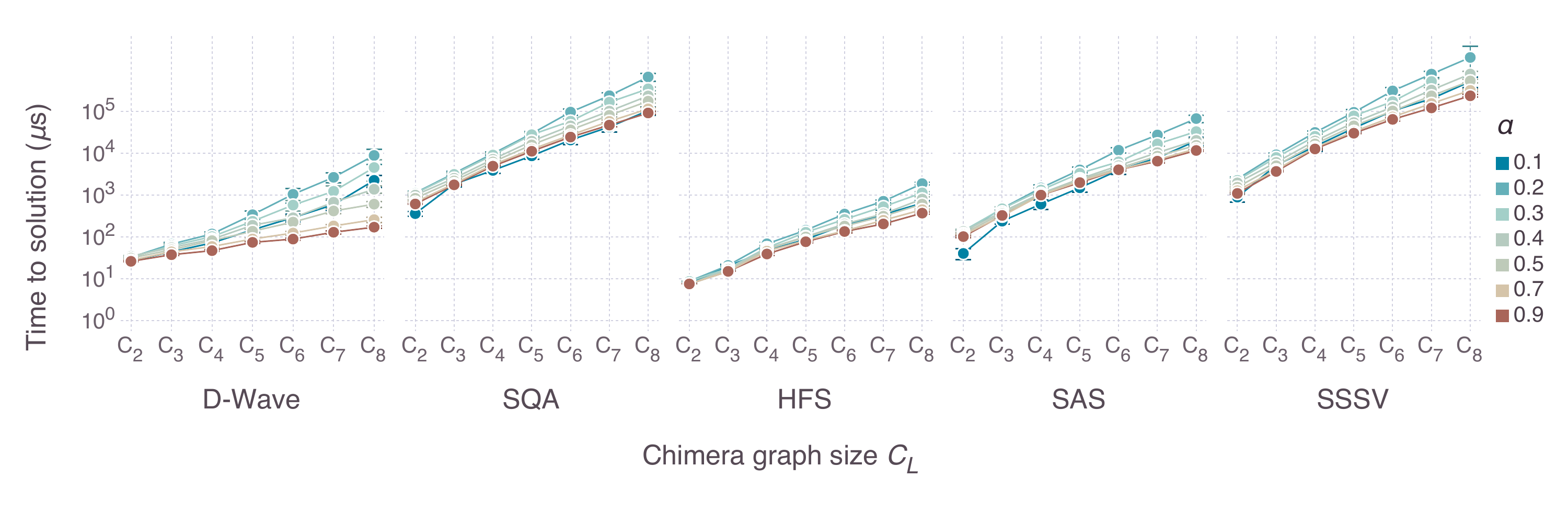}
\caption{\textbf{Median time-to-solution over instances.} Plotted is $\TTS(L,\alpha,0.5)$ (log scale) as a function of the Chimera subgraph size $C_L$, for a range of clause densities and for all solvers we tested. Note that only the scaling matters and not the actual TTS, since it is determined by constant factors that vary from processor to processor, compiler options, etc.  All algorithms' timing reflects the result after accounting for parallelism, as described in Appendix~\ref{app:methods}. Error bars represent $2\sigma$ confidence intervals. The DW2 annealing time is $t_a=20\mu$s.  }
\label{fig:TTSscalingbasic50}
\end{center}
\end{figure*}

\begin{figure*} [t]
\begin{center}
\includegraphics[width=\textwidth]{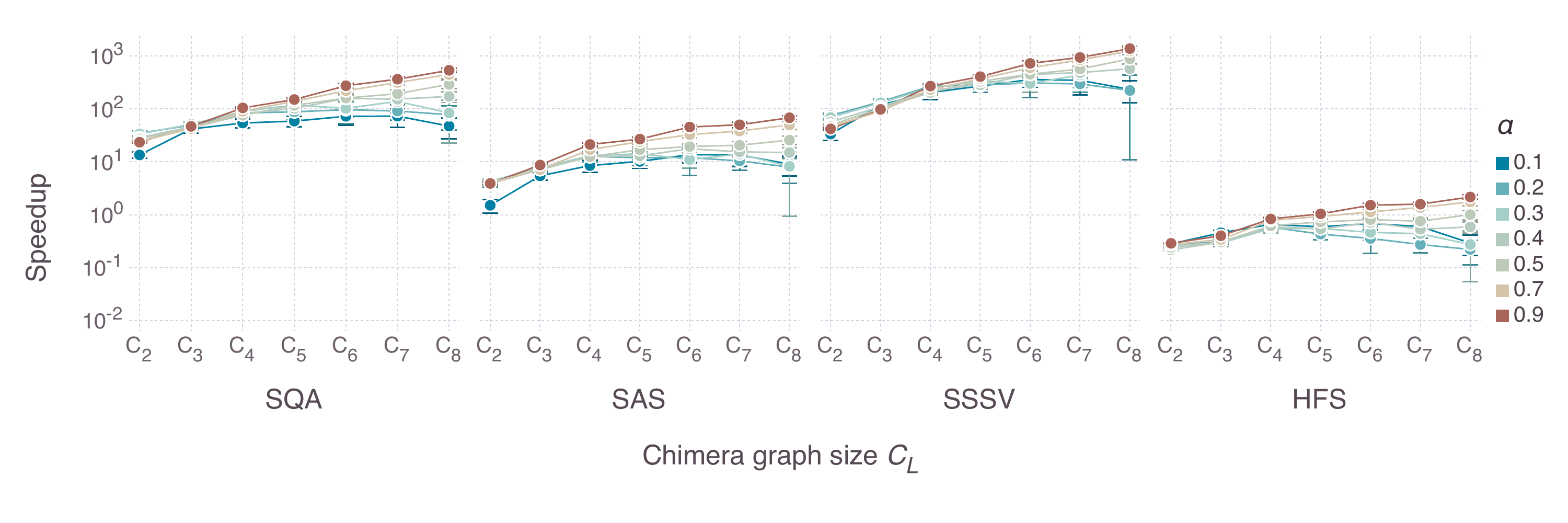}
\caption{\textbf{Speedup ratio.} Plotted is the median speedup ratio $S_X(L,\alpha,0.5)$ (log scale) as defined in Eq.~\eqref{eq:speedup} for all algorithms tested. A negative slope indicates a definite slowdown for the DW2. A positive slope indicates the possibility of an advantage for the DW2 over the corresponding classical algorithm. This is observed for $\alpha > 0.4$ in the comparison to SAS, SQA, SSSV, and HFS (see Fig.~\ref{fig:DiffDWslope} for a more detailed analysis). Error bars represent $2\sigma$ confidence intervals. }
\label{fig:Speedupscalingbasic50}
\end{center}
\end{figure*}

We next briefly discuss how to properly address resource scaling \cite{speedup}. The D-Wave devices use $N$ qubits and $O(N)$ couplers to compute the solution of the input problem. Thus, it uses resources which scale linearly in the size of the problem, and we should give our classical solvers the same opportunity. Namely, we should allow for the use of linearly more cores and memory for our classical algorithms as we scale the problem size. 
For annealers such as SA, SQA, and SSSV, this is trivial as we can exploit the bipartite nature of the Chimera graph to perform spin updates in parallel so that each sweep takes constant time and a linear number of cores and memory. 
The HFS algorithm is not perfectly parallelizable but as we explain in more detail in Appendix~\ref{app:methods}, we take this into account as well. Finally, note that dynamic programming can always find the true ground state in a time that is exponential in the Chimera graph treewidth, i.e., that scales as $\exp(c\sqrt{N})$ \cite{Choi2}. The natural size scale in our study is the square Chimera subgraph $C_L$ comprising $L^2$ unit cells of $8$ qubits each, i.e., $N = 8L^2$. Therefore, henceforth we replace $N$ by $L$ so that $\lambda = \{L,\alpha,q\}$.


\section{Probing for a quantum speedup}
We now come to our main goal, which is to probe for a limited or potential quantum speedup on our frustrated Ising problem set. We reserve the term ``limited speedup" for our comparisons with SA, SQA, and SSSV, while the term ``potential speedup" refers to the comparison with the HFS algorithm, which unlike the other three algorithms, does not implement a similar algorithmic approach to a quantum annealer. 

\subsection{Dependence on clause density}
We first analyze the effect of the clause density. This is shown in Fig.~\ref{fig:DwaveVsSelby50}, where we plot $\TTS(L,\alpha,0.5)$ for the DW2 and the HFS algorithm.%
\footnote{In this study we focus mostly on the median since with $100$ instances per setting, higher quantiles tend be noise-dominated.}
We note that the worst-case TTS of $\sim\!10$ms is smaller than that observed for random Ising problems in Ref.~\cite{speedup} [$\sim\!100$ms for range $7$, $C_8$, and $q=0.5$ (median)]. However, as we shall demonstrate, the classical algorithms against which the DW2 was benchmarked in Ref.~\cite{speedup} (SA and SQA) scale significantly less favorably in the present case, i.e., whereas in Ref.~\cite{speedup} no possibility of a limited speedup against SA was observed for the median, here we will find that such a possibility remains. In this sense, the problem instances considered here are relatively harder for the classical solvers than those of Ref.~\cite{speedup}.

For our choice of random loop characteristics, the time-to-solution peaks at a clause density $\alpha\approx 0.17$, reflecting the hardness of the problems in that regime. To correlate the hardness of the instances with their degree of frustration, we plot the frustration fraction, defined as the ratio of the number of unsatisfied edges with respect to the planted solution to the total number of edges on the graph, as a function of clause density. The frustration fraction curve, shown in Fig.~\ref{fig:frustRat}, has a peak at $\alpha \approx 0.25$, confirming that frustration and hardness are indeed correlated.%
\footnote{Note that in our definition of frustration fraction, frustration is measured with respect to all edges of the graph from which clauses are chosen, similarly to the way clause density is defined in SAT problems.}
The hardness peak is reminiscent of the analogous situation in SAT, where the clause density 
can be tuned through a phase transition between satisfiable and unsatisfiable phases \cite{Bollobas:2001}. The peak we observe may be interpreted as a finite-size precursor of a phase transition. This interpretation is corroborated below by time-to-solution results of all other tested algorithms which will also find problems near the critical point the hardest. Indeed, all the  algorithms we studied exhibit qualitatively similar behavior to that seen in Fig.~\ref{fig:DwaveVsSelby50} (see Fig.~\ref{fig:pt_25_75} in Appendix~\ref{app:additional}),
with an easy-hard-easy pattern separated at $\alpha\approx 0.2$. This is in agreement with previous studies, e.g., for MAX 2-SAT problems on the DW1~\cite{MAX2SAT}, and for $k$-SAT with $k > 2$, where a similar pattern is found 
for backtracking solvers \cite{monasson}. It is important to note that we do not claim that this easy-hard-easy transition coincides with a spin-glass phase transition; we have not studied which phases actually appear in our problem set as we tune the clause density.

A qualitative explanation for the easy-hard-easy pattern is that when the number of loops (and hence $\alpha$) is small they do not overlap and thus each loop becomes an easy optimization problem. In the opposite limit many loops pass through each edge, thus tending to reduce frustration, since each loop contributes either a ``frustrated" edge with small probability $1/\ell$ (where $\ell$ is the loop length) or an ``unfrustrated" edge with probability $1-1/\ell$.
The hard problems thus lie in between these two limits, where a constant fraction (bounded away from $0$ and $1$) of loops overlap.  

\begin{figure}[t]
\begin{center}
\subfigure[]{\includegraphics[width=\columnwidth]{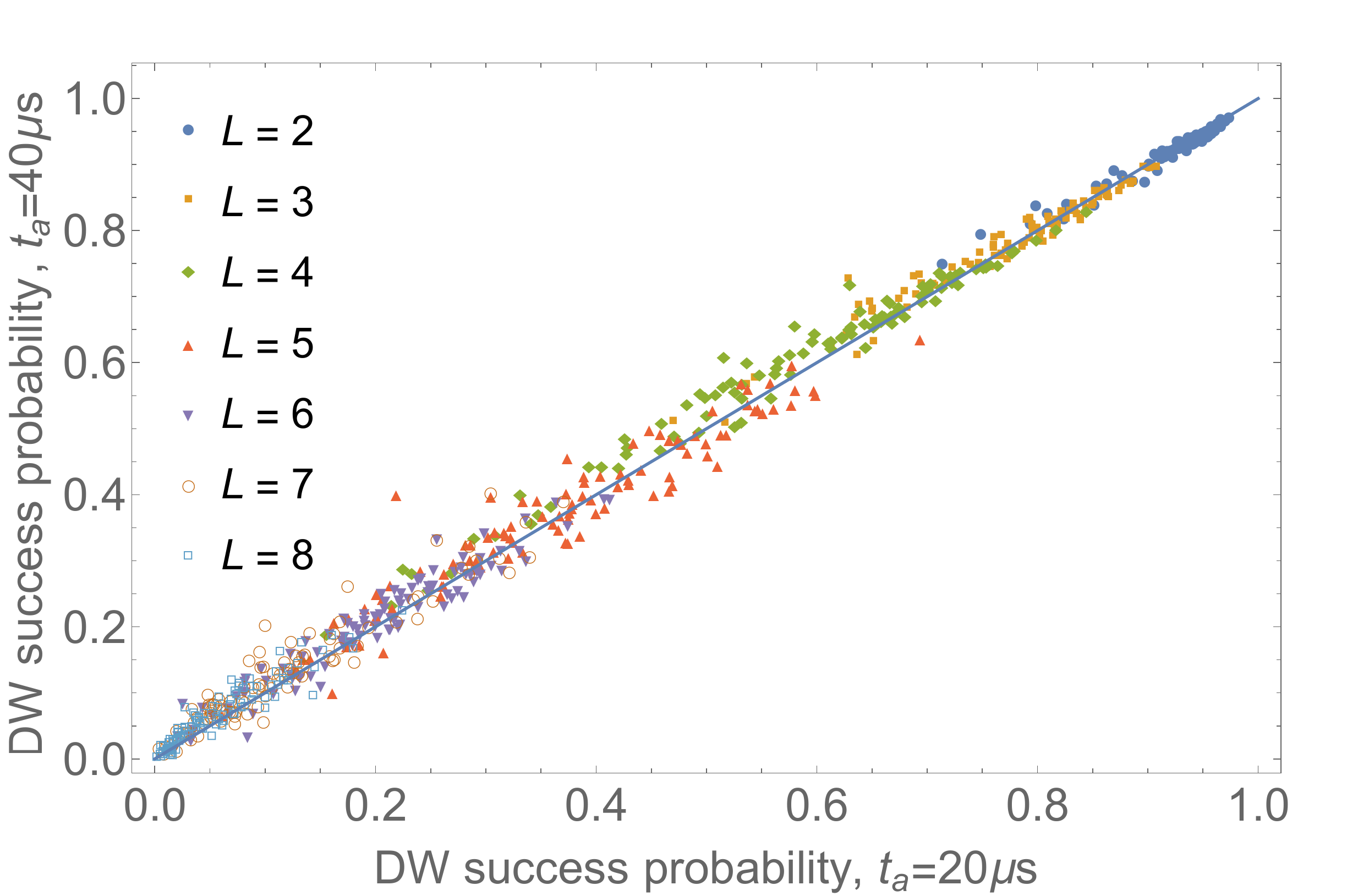}\label{fig:DW-DW-corr}}
\subfigure[]{\includegraphics[width=\columnwidth]{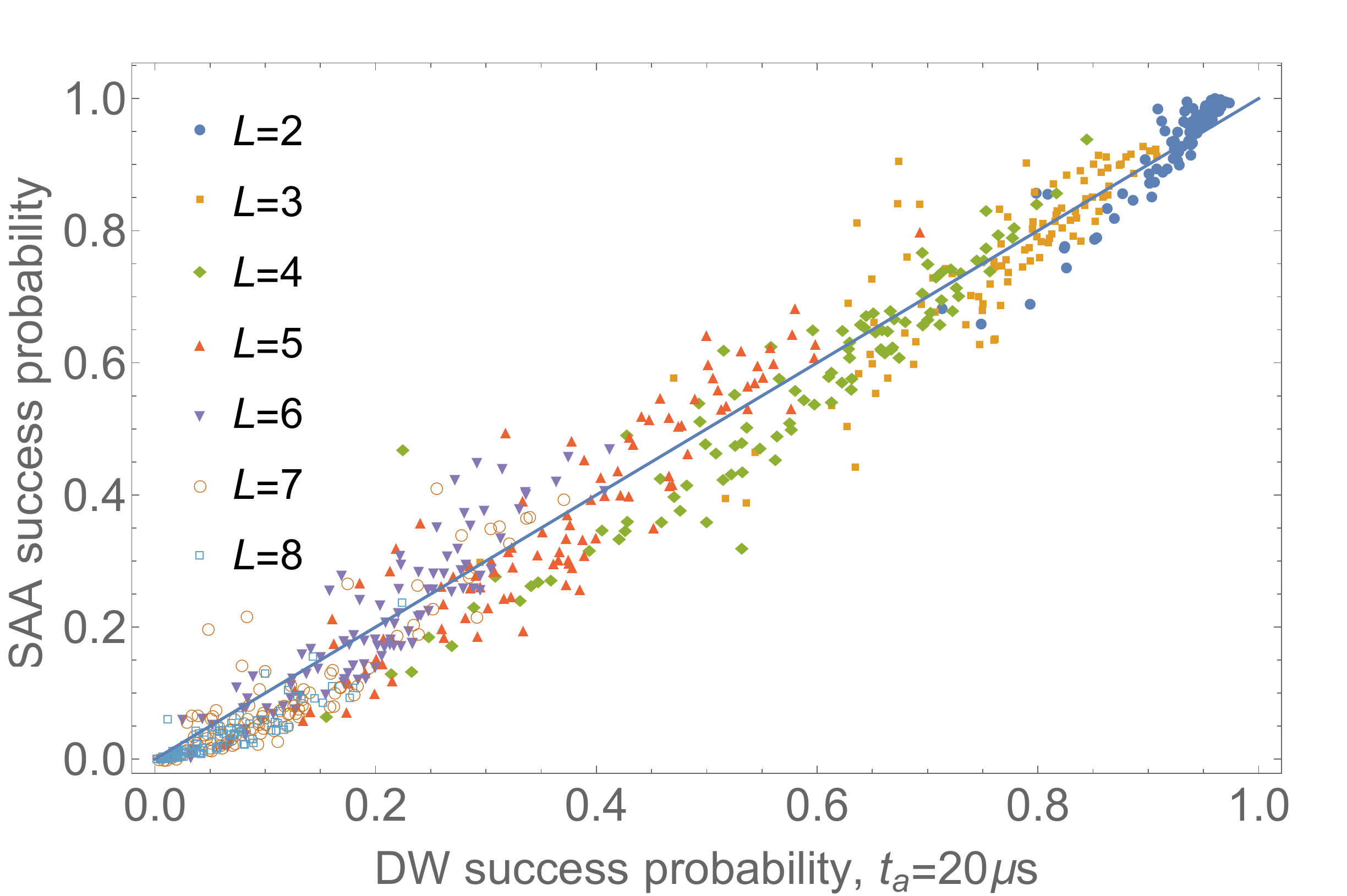}\label{fig:DW-SAA-corr}}
\caption{\textbf{Success probability correlations.} The results for all instances at $\alpha=0.35$ are shown. Each datapoint is the success probability for the same instance, (a) for DW2 at $t_a=20\mu$s and $t_a=40\mu$s, (b) for DW2 at $t_a=20\mu$s and SAA at $50,\!000$ sweeps and $\beta_f=5$ [in dimensionless units, such that $\max(J_{ij}) = 1$]. Perfect correlation means that all data points would fall on the diagonal, and a strong correlation is observed in both cases. The data is colored by the problem size $L$ and shows a clear progression from high success probabilities at small $L$ to large success probabilities at small $L$. Qualitatively similar results are seen for $t_a=20\mu$s \vs $t_a=40\mu$s at all $\alpha$ values, and for DW2 \vs SAA at intermediate $\alpha$ values (see Figs.~\ref{fig:corr-DWvsDW-all-alphas}
 and \ref{fig:corr-SAAvsDW-all-alphas} in Appendix~\ref{app:additional}).}
\label{fig:DW-DW-SAA-corr}
\end{center}
\end{figure}

\subsection{General considerations concerning scaling and speedup}
 
Of central interest is the question of whether there is any scaling advantage (with $L$) in using a quantum device to solve our problems. Therefore, we define a quantum speedup ratio for the DW2 relative to a given algorithm $X$ as \cite{speedup}
\beq 
\label{eq:speedup}
S_X(\lambda,t_a) = \frac{\TTS_X(\lambda)}{\TTS_{\mathrm{DW2}}(\lambda,t_a)} \ .
\eeq
Since this quantity is specific to the DW2 we refer to it simply as the empirical ``DW2 speedup ratio" from now on, though of course it generalizes straightforwardly for any other putative quantum annealer or processor against which algorithm $X$ is compared. 

We must be careful in using $S_X(\lambda,t_a)$ in assessing a speedup, since the annealing time must be optimized for each problem size $L$ in order to avoid the pitfall of a fake speedup \cite{speedup}. Let us denote the (unknown) optimal annealing time by $t_a^{\mathrm{opt}}(L)$. By definition, $\TTS_{\textrm{DW2}}(\lambda,t_a^{\mathrm{opt}}(L)) \leq \TTS_{\textrm{DW2}}(\lambda,t_a)$, where $t_a$ is a fixed annealing time. 
We now define the \emph{optimized speedup ratio} as 
\begin{align}
\label{eq:speedup-opt}
S_X^{\textrm{opt}}(\lambda,t_a^{\mathrm{opt}}(L)) = \frac{\TTS_X(\lambda)}{\TTS_{\textrm{DW2}}(\lambda,t_a^{\mathrm{opt}}(L))}
\end{align}
and clearly $
S_X(\lambda,t_a) \leq  S_X^{\textrm{opt}}(\lambda,t_a^{\mathrm{opt}}(L))$, i.e., the speedup ratios computed using Eq.~\eqref{eq:speedup} are lower bounds on the optimized speedup. However, 
what matters for a speedup is the \emph{scaling} of the speedup ratio with problem size. Thus, we are interested not in the numerical value of the speedup ratio but rather in the slope $dS/dL$ (recognizing that this is a formal derivative since $L$ is a discrete variable). A positive slope would indicate a DW2 speedup, while a negative speedup slope would indicate a slowdown. 

We thus define the \emph{DW2 speedup regime} as the set of problem sizes $\mathcal{L}^+$ where $\frac{d}{dL} S_X(\lambda,t_a^{\mathrm{opt}}(L)) > 0$ for all $L\in \mathcal{L}^+$. Likewise, the \emph{DW2 slowdown regime} is the set of problem sizes $\mathcal{L}^-$ where $\frac{d}{dL} S_X(\lambda,t_a^{\mathrm{opt}}(L)) < 0$ for all $L\in \mathcal{L}^-$.

From a computational complexity perspective one is ultimately interested in the asymptotic performance, i.e., the regime where $L$ becomes arbitrarily large. In this sense a \emph{true} speedup would correspond to  the observation that $\mathcal{L}^+ = [L^+_{\min},L^+_{\max}]$, with $L^+_{\min}$ a positive constant and $L^+_{\max} \rightarrow \infty$. Of course, such a definition is meaningless for a physical device such as the DW2, for which $L^+_{\max}$ is necessarily finite. Thus the best we can hope for is an observation that $\mathcal{L}^+$ is as large as is consistent with the device itself, which in our case would imply that $\mathcal{L}^+ =[1,8]$. However, as we shall argue, we can in fact only rule out a speedup, while we are unable to confirm one. I.e., we are able to identify $\mathcal{L}^- = [L^-_{\min},L^-_{\max}]$, but not $\mathcal{L}^+$.

The culprit, as in earlier benchmarking work \cite{speedup} and as we establish below, is the fact that the DW2 minimum annealing time of $t_a=20 \mu$s is too long (see also Appendix~\ref{app:additional}). This means that the smaller the problem size the longer it takes to solve the corresponding instances compared to the case with an optimized annealing time, and hence the observed slope of the DW2 speedup ratio should be interpreted as a \emph{lower bound} for the optimal scaling. This is illustrated in Fig.~\ref{fig:TTS_Illustration}. Without the ability to identify $t_a^{\mathrm{opt}}(L)$ we do not know of a way to infer, or even estimate $\mathcal{L}^+$. However, as we now demonstrate, under a certain reasonable assumption we can still \emph{bound} $\mathcal{L}^-$.

The assumption is that if $t_a > t_a^{\mathrm{opt}}$ then  $\frac{d}{dL} \TTS_{\mathrm{DW2}}(\lambda,t_a) \leq \frac{d}{dL}\TTS_{\mathrm{DW2}}(\lambda,t_a^{\mathrm{opt}}(L))$ for all $L < L^*$, the problem size for which $t_a = t_a^{\mathrm{opt}}(L^*)$. This assumption is essentially a statement that the TTS is monotonic in $L$ \footnote{Individual instances may not satisfy this assumption \cite{Amin:2015qf,Albash:2015nx}, but we are not aware of any cases where averaging over an ensemble of instances violates this assumption.}, as illustrated in Fig.~\ref{fig:TTS_Illustration}. Next we consider the (formal) derivatives of Eqs.~\eqref{eq:speedup} and \eqref{eq:speedup-opt}:
\bes
\begin{align}
&\frac{\frac{d}{dL} S_X(\lambda,t_a)}{S_X(\lambda,t_a)} =  \frac{\partial_L \TTS_\mathrm{X}(\lambda)}{\TTS_\mathrm{X}(\lambda)}   -\frac{\partial_L \TTS_{\mathrm{DW2}}(\lambda,t_a)}{\TTS_{\mathrm{DW2}}(\lambda,t_a)} \\
&\frac{\frac{d}{dL} S_X(\lambda,t_a^{\mathrm{opt}}(L))}{S_X(\lambda,t_a^{\mathrm{opt}}(L))} = 
\frac{\partial_L \TTS_\mathrm{X}(\lambda)}{\TTS_\mathrm{X}(\lambda)} \notag \\
&\qquad\qquad\qquad\qquad - \frac{\frac{d}{dL} \TTS_{\mathrm{DW2}}(\lambda,t_a^{\mathrm{opt}}(L))}{\TTS_{\mathrm{DW2}}(\lambda,t_a^{\mathrm{opt}}(L))} \ .
\end{align}
\ees
Collecting these results we have
\beq
\label{eq:rulingout}
 \frac{{S_X(\lambda,t_a)}}{S_X(\lambda,t_a^{\mathrm{opt}})} \frac{d}{dL} S_X(\lambda,t_a^{\mathrm{opt}})< {\frac{d}{dL} S_X(\lambda,t_a)} \ \mathrm{if} \ t_a > t_a^{\mathrm{opt}}(L). \notag \\
\eeq
Therefore, if we find that $\frac{d}{dL} S_X(\lambda,t_a)< 0$ in the suboptimal regime where $t_a > t_a^{\mathrm{opt}}(L)$, then it follows that $\frac{d}{dL} S_X(\lambda,t_a^{\mathrm{opt}}(L)) < 0$. In other words, \emph{a DW2 speedup is ruled out if we observe a slowdown using a suboptimal annealing time.}

\subsection{Scaling and speedup ratio results}
In Fig.~\ref{fig:TTSscalingbasic50} we show the scaling of the median time-to-solution for all algorithms studied, for a representative set of clause densities. All curves appear to match the general dynamic programming scaling for $L\gtrsim 4$, i.e., $\TTS(\lambda) \sim \exp[b(\alpha) L]$, but the scaling coefficient $b(\alpha)$ clearly varies from solver to solver. This scaling is similar to that observed in previous benchmarking studies of random Ising instances \cite{q108,speedup}.

In Fig.~\ref{fig:Speedupscalingbasic50} we show the median scaling of $S_X$ for the same set of clause densities as shown in Fig.~\ref{fig:TTSscalingbasic50}. 
We  observe that in all cases there is a strong dependence on the clause density $\alpha$, with a negative slope of the DW2 speedup ratio for the lower clause densities, corresponding to the harder, more frustrated problems. In this regime the DW2 exhibits a scaling that is worse than the classical algorithms and by Eq.~\eqref{eq:rulingout} there is no speedup. The possibility of a DW2 speedup remains open for the higher clause densities, where a positive slope is observed, i.e., the DW2 appears to find the easier, less frustrated problems easier than the classical solvers. This apparent advantage is most pronounced for $\alpha\geq 0.4$, where we observe the possibility of a potential speedup even against the highly fine-tuned HFS algorithm (this is seen more clearly in Fig.~\ref{fig:DiffDWslope}). Moreover, the DW2 speedup ratio against HFS improves slightly at the higher percentiles (see Fig.~\ref{fig:speedup_25_75} in Appendix~\ref{app:additional}), which is encouraging from the perspective of a potential quantum speedup.

\begin{figure}[t]
\begin{center}
\includegraphics[width=\columnwidth]{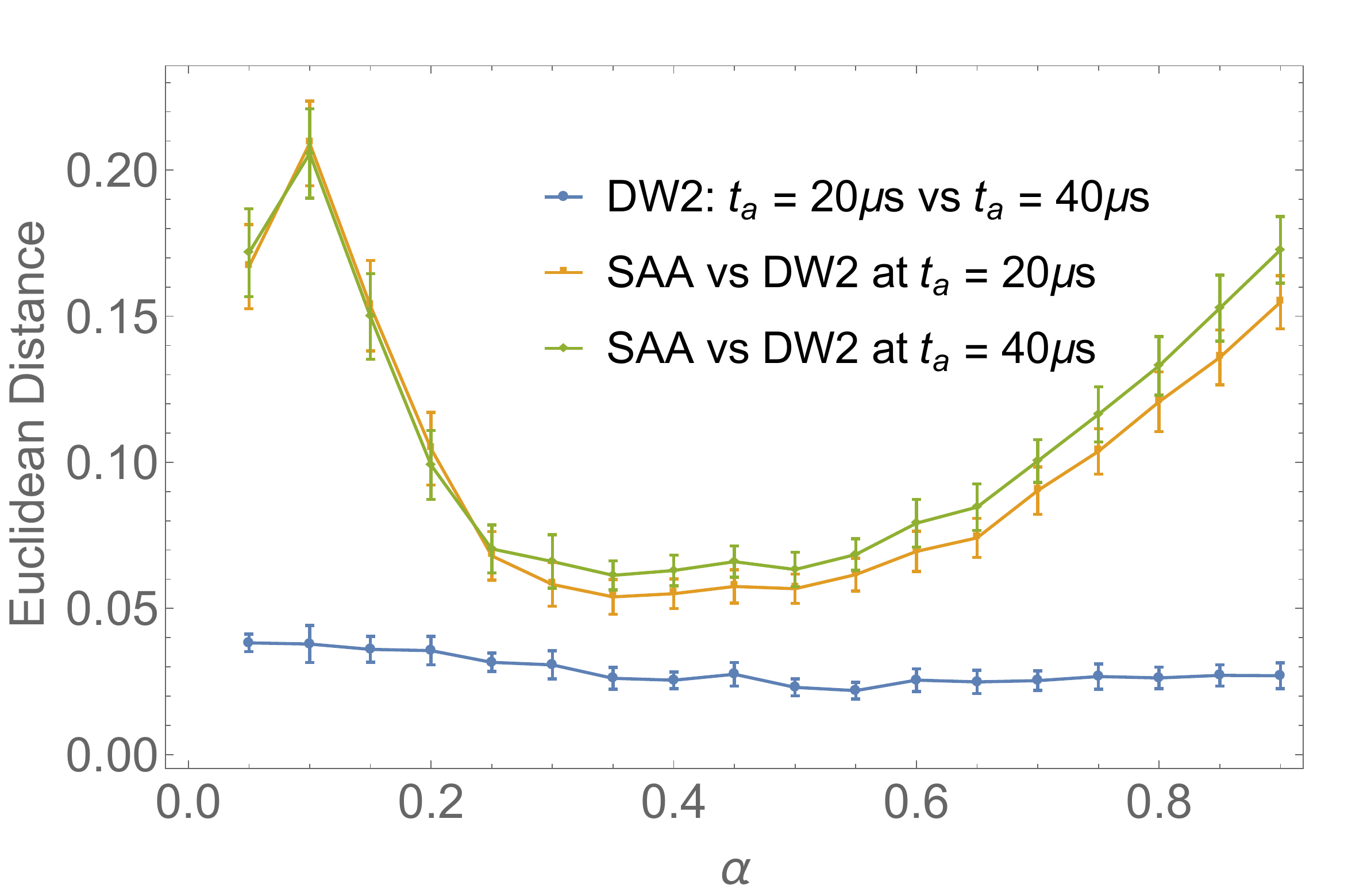}
\caption{\textbf{Correlation between the DW2 data for two different annealing times and SAA.} Plotted is the normalized Euclidean distance $D(\vec{p}_1,\vec{p}_2)$ for $\vec{p}_1$ and $\vec{p}_2$ being, respectively, the ordered success probability 
for DW2 at $t_a=20\mu$s and $t_a=40\mu$s (blue circles), DW2 at $t_a=20\mu$s and SAA (yellow squares), DW2 at $t_a=40\mu$s and SAA (green diamonds).  For comparison, the Euclidean distance between two random vectors with elements $\in [0,1]$  is $\sim 0.4$. SAA data is for $50,\!000$ sweeps and $\beta_f=5$. 
The correlation with SAA degrades slightly for $t_a=40\mu$s. Error bars represent $2\sigma$ confidence intervals and were computed using bootstrapping (see Appendix~\ref{app:error-estimation} for details). In each comparison, to construct $\vec{p}_1$ and $\vec{p}_2$ we fixed $\alpha$ and used half the instances (for bootstrapping purposes) for $L\in[2,8]$. 
}
\label{fig:Euclid-dist}
\end{center}
\end{figure}

\subsection{Scaling coefficient results}
To test the dependence on $t_a$, we repeated our DW2 experiments for $t_a \in [20,40]\mu$s, in intervals of $2\mu$s. Figure~\ref{fig:DW-DW-corr} is a success probability correlation plot between $t_a=20\mu$s and $t_a=40\mu$s, at $\alpha=0.35$. The correlation appears strong, suggesting that the device might already have approached the asymptotic regime where increasing $t_a$ does not modify the success probabilities. To check this more carefully let us first define the normalized Euclidean distance between two length-$M$ vectors of probabilities $\vec{p}_1$ and $\vec{p}_2$ as 
\beq
D(\vec{p}_1,\vec{p}_2):=\frac{1}{M}\|\vec{p}_1-\vec{p}_2\|
\label{eq:D}
\eeq
(where $\|\vec{p}\|=\sqrt{\vec{p}\cdot\vec{p}}$); clearly, $0\leq D(\vec{p}_1,\vec{p}_2) \leq 1$. The result for the DW2 data with $\vec{p}_1$ and $\vec{p}_2$ being the ordered sets of success probabilities for all 
instances with given $\alpha$ at $t_a=20\mu$s and $t_a=40\mu$s respectively, is shown as the blue circles in Fig.~\ref{fig:Euclid-dist}. The small distance for all $\alpha$ suggests that for $t_a \geq 20\mu$s the distribution of ground state probabilities has indeed nearly reached its asymptotic value. 

\begin{figure}[t]
\begin{center}
\includegraphics[width=\columnwidth]{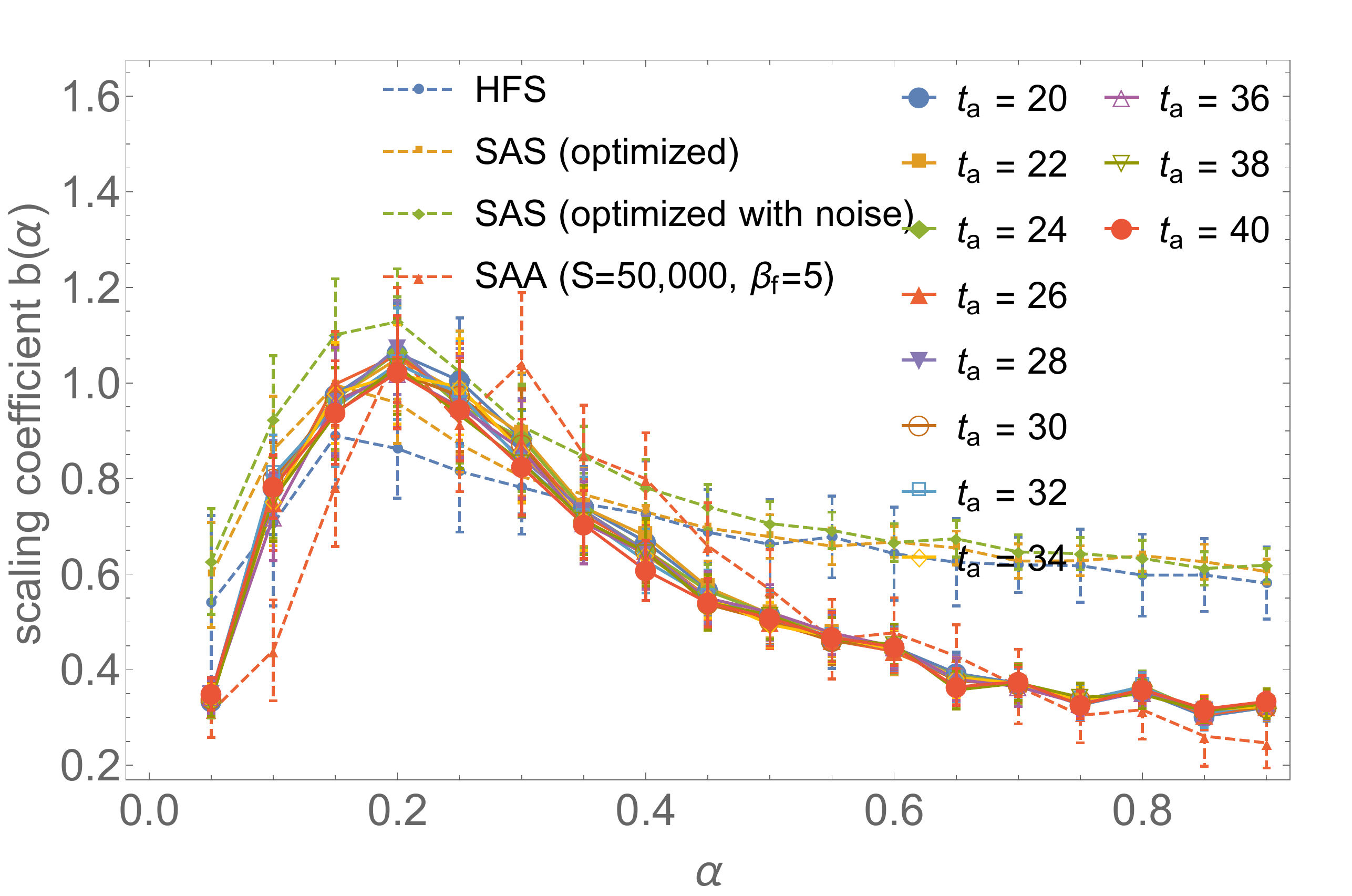}
\caption{\textbf{Scaling coefficients of the number of runs.} Plotted here is $b(\alpha)$ [Eq.~\eqref{eq:fit-exp-r}] for the DW2 at all annealing times (overlapping solid lines and large symbols), for the HFS algorithm, for SAS with an optimal number of sweeps, for SAS with noise and an optimal number of sweeps, and for SAA with a large enough number of sweeps that the asymptotic distribution has been reached at $\beta_f=5$. The scaling coefficients of HFS and of optimized SAS each set an upper bound for a DW2 speedup against that particular algorithm.  In terms of the scaling coefficient the DW2 result is statistically indistinguishable (except at $\alpha=0.1$) from SAA run at $S=50,\!000$ and $\beta_f=5$. The coefficients shown here are extracted from fits with $L\geq 4$ (see Fig.~\ref{fig:DW-fit} in Appendix~\ref{app:additional}). Error bars represent $2\sigma$ confidence intervals.}
\label{fig:DWslope}
\end{center}
\end{figure}

\begin{figure}[t]
\begin{center}
\includegraphics[width=\columnwidth]{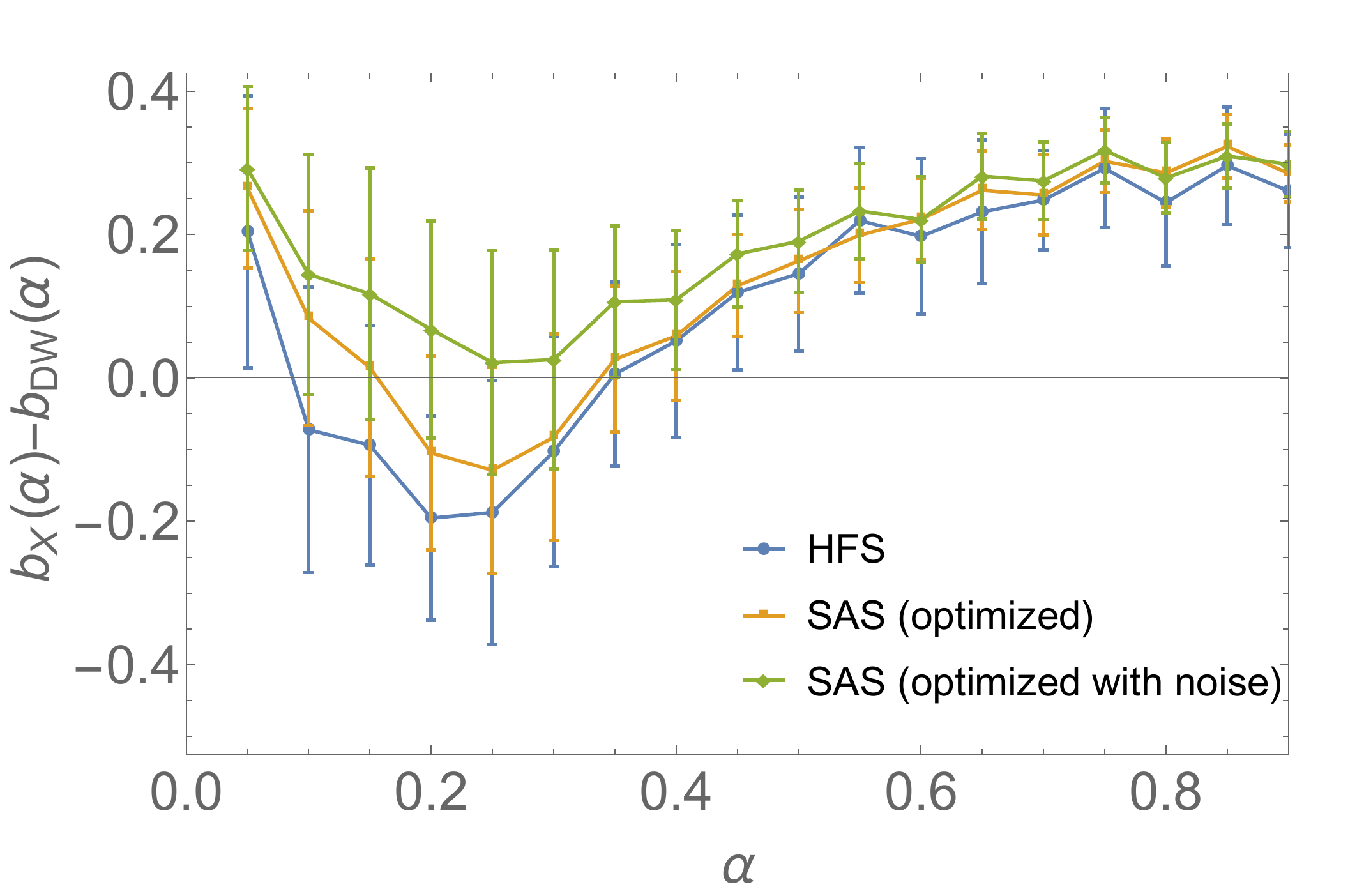}
\caption{\textbf{Difference between the scaling coefficients.} Plotted here is the difference between the scaling coefficients in Fig~\ref{fig:DWslope}, $b_{\textrm{X}}(\alpha) - b_{\textrm{DW2}}(\alpha)$, where $X$ denotes the HFS algorithm, SAS with an optimal number of sweeps, SAS with noise and an optimal number of sweeps, or SAA with $S=50,000$ and $\beta_f=5$. When the difference is non-positive there can be no speedup since optimizing $t_a$ can only increase $b_{\textrm{DW2}}(\alpha)$; conversely, when the difference is positive a speedup is still possible, i.e., not accounting for the error bars, for $\alpha\leq 0.05$ and $\alpha\geq 0.4$ for HFS, and for $\alpha \leq 0.15$ and $\alpha \geq 0.35$ for SAS without noise. These ranges shrink if the error bars are accounted for, but notably, for most $\alpha$ values SAS with noise does not disallow a limited speedup, suggesting that control noise may play an important factor in masking a DW2 speedup.  Error bars represent $2\sigma$ confidence intervals.}
\label{fig:DiffDWslope}
\end{center}
\end{figure}

This result means that the number of runs $r(\lambda)$ [Eq.~\eqref{eq:r-def}] does not depend strongly on $t_a$ either. To demonstrate this we first fit the number of runs to
\beq
r(L,\alpha,0.5) = e^{a(\alpha)+b(\alpha)L},
\label{eq:fit-exp-r}
\eeq
in accordance with the abovementioned expectation of the scaling with the treewidth, and find a good fit across the entire range of clause densities (see Fig.~\ref{fig:DW-fit} in Appendix~\ref{app:additional}). The corresponding scaling coefficient $b(\alpha)$ is plotted in Fig.~\ref{fig:DWslope} for all annealing times (the constant $a(\alpha)$ is shown in Fig.~\ref{fig:DW-scaling-constant} in Appendix~\ref{app:additional} and is well-behaved); the data collapses nicely, showing that the scaling coefficient $b(\alpha)$ has already nearly reached its asymptotic value. Also plotted in Fig.~\ref{fig:DWslope} is the scaling coefficient $b(\alpha)$ for HFS and for SAS with an optimized numbers of sweeps at each $\alpha$ and $L$, as extracted from the data shown in Fig.~\ref{fig:TTSscalingbasic50}. 

\emph{By Eq.}~\eqref{eq:rulingout}, \emph{where} $b_{\textrm{DW2}}(\alpha) \geq b_{\textrm{X}}(\alpha)$ \emph{there is no DW2 speedup $(\mathcal{L}^- = [4,8])$, whereas where} $b_{\textrm{DW2}}(\alpha) < b_{\textrm{X}}(\alpha)$, \emph{a DW2 speedup over algorithm $X$ is still possible.} 

We thus plot the difference in the scaling coefficients in Fig.~\ref{fig:DiffDWslope}. Figures~\ref{fig:DWslope} and \ref{fig:DiffDWslope} do allow for the possibility of a speedup against both HFS and SAS, at sufficiently high clause densities. However, we stress once more that the smaller DW2 scaling coefficient may be a consequence of $t_a=20\mu$s being excessively long, and that we cannot rule out that the observed regime of a possible speedup would have disappeared had we been able to optimize $t_a$ by exploring annealing times shorter than $20\mu$s. Note that an optimization of the SAS annealing schedule might further improve its scaling, but the same cannot be said of the HFS algorithm, 
and it seems unlikely that it could be outperformed even by the fully optimized version of SAS.

\subsection{DW2 \vs SAA}
Earlier work has ruled out SAA as a model of the D-Wave devices for random Ising model problems \cite{q108}, as well as for certain Hamiltonians with suppressed and enhanced ground states for which quantum annealing and SAA give opposite predictions \cite{q-sig,q-sig2}. The observation made above, that the DW2 success probabilities have nearly reached their asymptotic values, suggests that for the problems studied here the DW2 device may perhaps be described as a thermal annealer. While we cannot conclude on the basis of ground state probabilities alone that the DW2 state distribution has reached the Gibbs state, we can compare the ground state distribution to that of SAA in the regime of an asymptotic number of sweeps, and attempt to identify an effective final temperature for the classical annealer that allows it to closely describe the DW2 distribution. We empirically determine $\beta_f=5$ to be the final temperature for our SAA simulations that gives the closest match, and $S=50,\!000$ (corresponding to $150$ms, much larger than the DW2's $t_a = 20\mu$s) to be large enough for the SAA distribution to have become stationary (see Fig.~\ref{fig:slope-DW_all_ta-SAA_many_sweeps} in Appendix~\ref{app:additional} for results with additional sweep numbers confirming this). The result for the Euclidean distance measure is shown in Fig.~\ref{fig:Euclid-dist} for the two extremal annealing times, and the distance is indeed small. To more closely assess the quality of the correlation we select $\alpha=0.35$, the value that minimizes the Euclidean distance as seen in Fig.~\ref{fig:Euclid-dist}, and present the correlation plot in Fig.~\ref{fig:DW-SAA-corr}. Considering the results for each size $L$ separately, it is apparent that the correlation is good but also systematically skewed, i.e., for each problem size the data points approximately lie on a line that is not the diagonal line. Additional correlation plots are shown in Fig.~\ref{fig:corr-SAAvsDW-all-alphas} of Appendix~\ref{app:additional}. 

As a final comparison we also extract the scaling coefficients $b(\alpha)$ and compare DW2 to SAA in Fig.~\ref{fig:DWslope}. It can be seen that in terms of the scaling coefficients the DW2 and SAA results are essentially statistically indistinguishable. However, we stress that since the SAA number of sweeps is not optimized, one should refrain from concluding that the equal scaling observed for the DW2 and SAA rules out a DW2 speedup.

\begin{figure}[t]
\begin{center}
\includegraphics[width=\columnwidth]{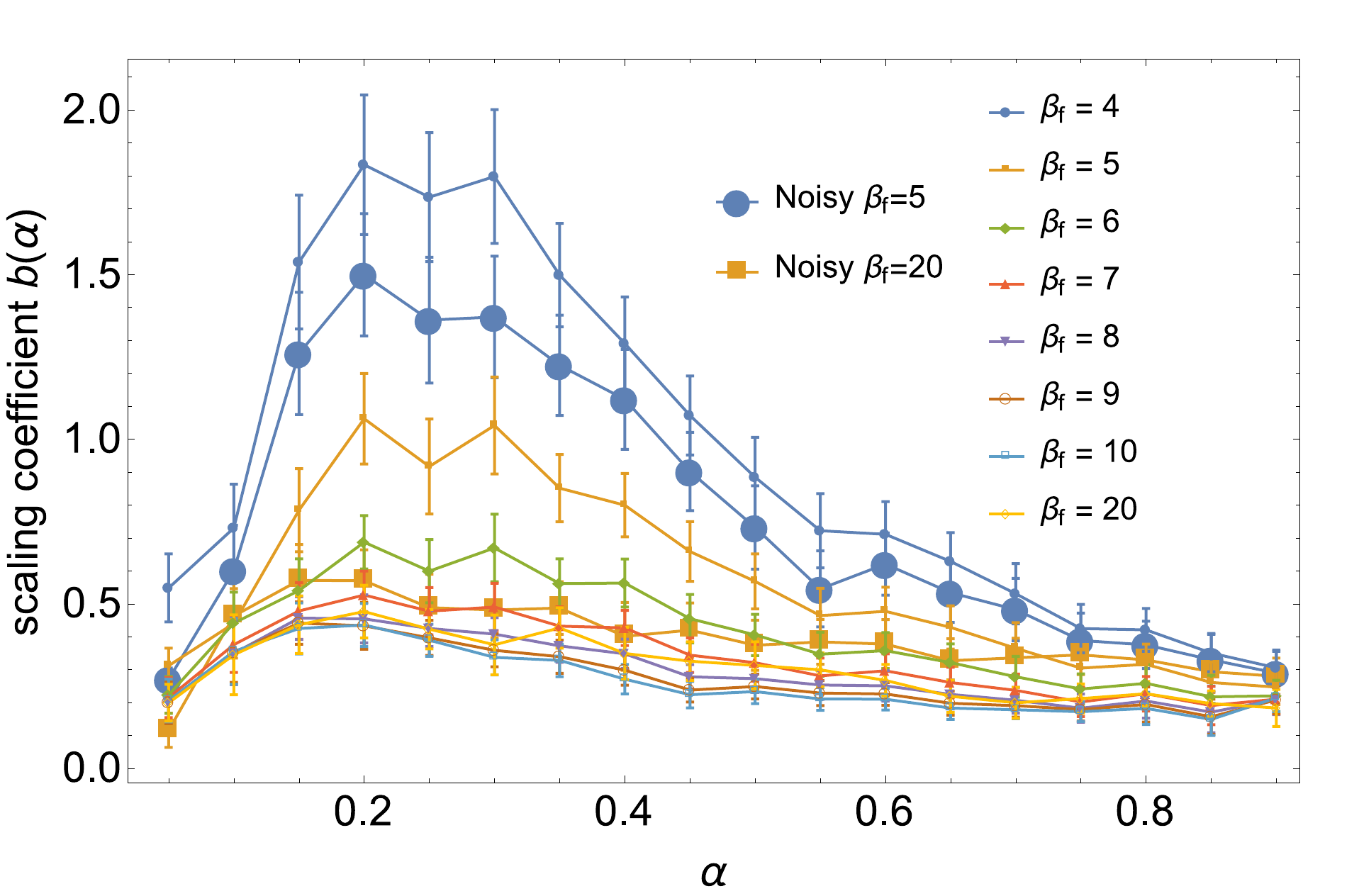}
\caption{\textbf{Scaling of SAA at different final temperatures.} SAA is run at $S=50,\!000$ and various final inverse temperatures. The peak at $\alpha\sim 0.2$ remains a robust feature. The data marked ``Noisy" is with $5\%$ random noise added to the couplings $J_{ij}$.}
\label{fig:SAA-slope}
\end{center}
\end{figure}



\section{Discussion}
\label{sec:conclusions}
In this work we proposed and implemented a method for generating problems with a range of hardness, tuned by the clause density, mimicking the phase structure observed for SAT problems. 
By comparing the DW2 device to a number of classical algorithms we delineated where there is no DW2 speedup and where it might still be possible, for this problem set. No advantage is observed for the low clause densities corresponding to the hardest optimization problems, but a speedup remains possible for the higher clause densities. In this sense these results are more encouraging than for the random Ising problems studied in Ref.~\cite{speedup}, where only the lowest percentiles of the success probability distribution allowed for the possibility of a DW2 speedup. In our case there is in fact a slight improvement for the higher percentiles (see Fig.~\ref{fig:speedup_25_75} in Appendix~\ref{app:additional}). 
In the same sense our findings are also more encouraging than very recent theoretical results showing that quantum annealing does not provide a speedup relative to simulated annealing on 2SAT problems with a unique ground state and a highly degenerate first excited state \cite{Neuhaus:2014mz}.

The close match between the DW2 and SAA scaling coefficients seen in Fig.~\ref{fig:DWslope} suggests that SAA in the regime of an asymptotic number of sweeps can serve as a model for the expected performance of the D-Wave device as its temperature is lowered. Thus we plot in Fig.~\ref{fig:SAA-slope} the scaling coefficient for SAA at various final inverse temperatures, at a fixed (and still asymptotic) value of $S=50,\!000$. Performance improves steadily as $\beta_f$ increases, suggesting that SAA does not become trapped at $50,\!000$ sweeps for the largest problem sizes we have studied (this may also indicate that even at the hardest clause densities these problems do not exhibit a positive-temperature spin-glass phase). 
We may infer that a similar behavior can be expected of the D-Wave device if its temperature were lowered.

An additional interesting feature of the fact that the asymptotic DW2 ground state probability observed here is in good agreement with that of a thermal annealer, is that it gives the ground state with a similar probability as expected from a Gibbs distribution. This result is consistent with the weak-coupling limit that underlies the derivation of the adiabatic Markovian master equation \cite{ABLZ:12-SI}, i.e., it is consistent with the notion that the system-bath coupling is weak and \emph{decoherence occurs in the energy eigenbasis} \cite{childs_robustness_2001}. This rules out the possibility that decoherence occurs in the computational basis, as this would have led to a singular-coupling limit master equation with a ground state probability drawn not from the Gibbs distribution but rather from a uniform distribution, if the single-qubit decoherence time is much shorter than the total annealing time \cite{ABLZ:12-SI}. This is important, as the weak-coupling limit is compatible with decoherence between eigenstates with different energies while maintaining ground state coherence, a necessary condition for a speedup via quantum annealing. In contrast, in the singular-coupling limit no quantum effects survive and no quantum speedup of any sort is possible.

We note that an important disadvantage the DW2 device has over all the classical algorithms we have compared it with, is control errors in the programming of the local fields and couplings \cite{King:2014uq,perdomo:15a,perdomo:15b,Martin-Mayor:2015dq}.  As shown in Appendix~\ref{app:additional} (Fig.~\ref{fig:precision}), many of the rescaled couplings $J_{ij}$ used in our instances are below the single-coupler control noise specification, meaning that with some probability the DW2 is giving the right solution to the wrong problem. We are unable to directly measure the effect of such errors on the DW2 device, but their effect is demonstrated in Figs.~\ref{fig:DiffDWslope} and \ref{fig:SAA-slope}, where both SAS and SAA with $\beta_f=5,20$, respectively, are seen to have substantially increased scaling coefficients after the addition of noise that is comparable to the control noise in the DW2 device. In fact, the DW2 scaling coefficient is smaller than the scaling coefficient of optimized SAS with noise over almost the entire range of $\alpha$, suggesting that a reduction in such errors will extend the upper bound for a DW2 speedup against SAS over a wider range of clause densities. The effect of such control errors can be mitigated by improved engineering, but also emphasizes the need for the implementation of error correction on putative quantum annealing devices. The beneficial effect of such error correction has already been demonstrated experimentally \cite{PAL:13,PAL:14} and theoretically  \cite{Young:2013fk}, albeit at the cost of a reduction in the effective number of qubits and reduced problem sizes. 

To summarize, we believe that at least three major improvements will be needed before it becomes possible to demonstrate a conclusive (limited or potential) quantum speedup using putative quantum annealing devices: (1) harder optimization problems must be designed that will allow the annealing time to be optimized, (2) decoherence and control noise must be further reduced, and (3) error correction techniques must be incorporated. Another outstanding challenge is to theoretically design optimization problems that can be unequivocally shown to benefit from quantum annealing dynamics.  

Finally, the methods introduced here for creating frustrated problems with tunable hardness should serve as a general tool for the creation of suitable benchmarks for quantum annealers. Our study directly illustrates the important role that frustration plays in the optimization of spin-glass problems for classical algorithms as well as for putative quantum optimizers. It is plausible that different, perhaps more finely-tuned choices of clauses to create novel types of benchmarks, may be used to establish a clearer separation between the performance of quantum and classical devices.
These may eventually lead to the demonstration of the coveted
experimental annealing-based quantum speedup. 

\textit{Note added}. Work on problem instances similar to the ones we studied here, but with a range of couplings above the single-coupler control noise specification, appeared shortly after our preprint \cite{King:2015zr}. The DW2 scaling results were much improved, supporting our conclusion that such errors have a strong detrimental effect on the performance of the DW2.

\acknowledgments
Part of the computing resources were provided by the USC Center for High Performance Computing and Communications.  This research used resources of the Oak Ridge Leadership Computing Facility at the Oak Ridge National Laboratory, which is supported by the Office of Science of the U.S. Department of Energy under Contract No. DE-AC05-00OR22725. I.H. and D.A.L. acknowledge support under ARO grant number W911NF-12-1-0523. The work of J.J., T.A. and D.A.L. was supported under ARO MURI Grant No. W911NF-11-1-0268, and the Lockheed Martin Corporation. M.T. and T.F.R. acknowledges support by the Swiss National Science Foundation through the National Competence Center in Research NCCR QSIT, and by the European Research Council through ERC Advanced Grant SIMCOFE. We thank Mohammad Amin, Andrew King, Catherine McGeoch, and Alex Selby and  for comments and discussions. I.H. would like to thank Gene Wagenbreth for assistance with the implementation of solution-enumeration algorithms.  J.J. would like to thank the developers of the Julia programming language \cite{Bezanson:2014}, which was used extensively for data gathering and analysis. 

\appendix

\section{Methods}
\label{app:methods}
In this section we describe various technical details regarding out methods of data collection and analysis.

\subsection{Experimental details} 
\begin{figure}[t]
\begin{center}
\includegraphics[width=\columnwidth]{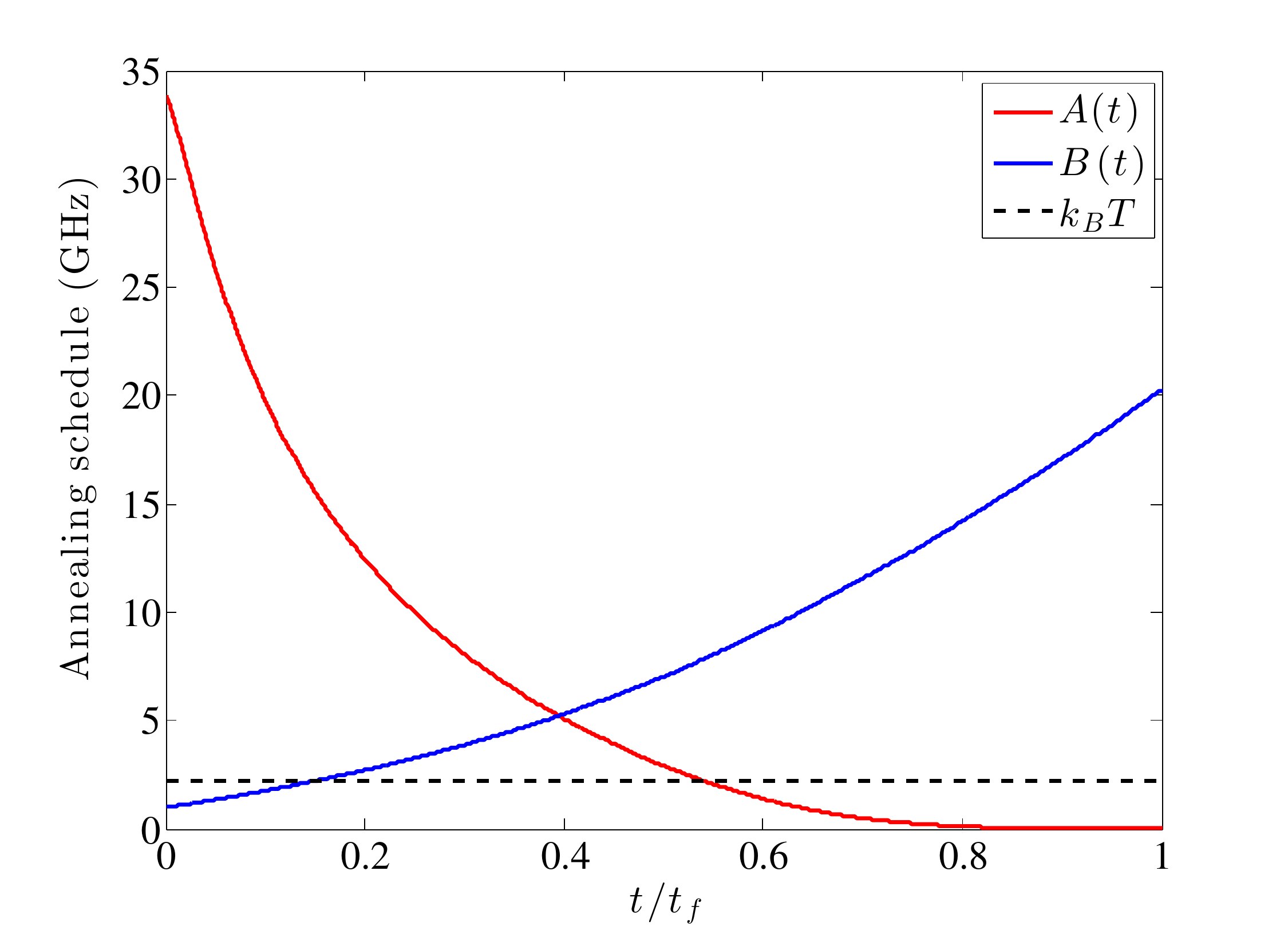}
\caption{\textbf{Annealing schedule of the DW2.}  The annealing curves $A(t)$ and $B(t)$ are calculated using rf-SQUID models with independently calibrated qubit parameters. Units of $\hbar = 1$.  The operating temperature of $17$mK is also shown.}
\label{fig:schedule}
\end{center}
\end{figure}

\begin{figure}[t]
\begin{center}
\includegraphics[width=\columnwidth]{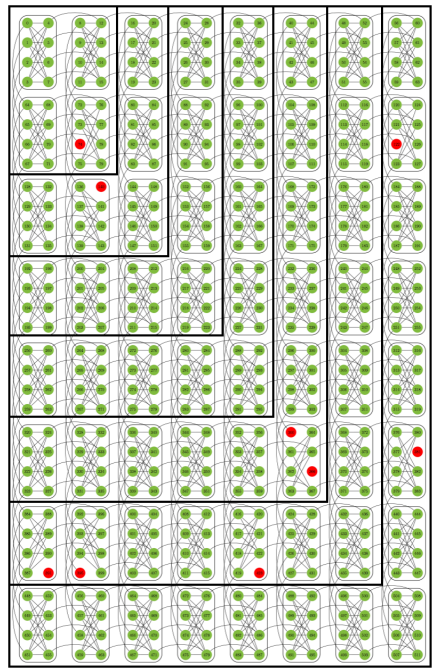}
\vspace{-1.2cm}
\caption{\textbf{The DW2 Chimera graph.} The qubits or spin variables occupy the vertices (circles) and the couplings $J_{ij}$ are along the edges. 
Of the $512$ qubits, $503$ were operative in our experiments (green circles) and $9$ were not (red circles). 
We utilized subgraphs comprising $L\times L$ unit cells, denoted $C_L$, indicated by the solid black lines. There were $31, 70, 126, 198,284, 385, 503$ qubits in our $C_2,\dots,C_8$ graphs, respectively.}
\label{fig:chimera}
\end{center}
\end{figure}

The DW2 is marketed by D-Wave Systems Inc. as a quantum annealer, which evolves a physical system of superconducting flux qubits according to the time-dependent Hamiltonian
\beq \label{eq:Hquantum}
H(t) = A(t) \sum_{i\in V} \sigma_i^x + B(t) \HI \ , \quad t\in[0,t_a] \ ,
\eeq
with $\HI$ given in Eq.~\eqref{eq:H}. The annealing schedules given by $A(t)$ and $B(t)$ are shown in Fig.~\ref{fig:schedule}.  Our experiments used the DW2 device housed at the USC Information Sciences Institute, with an operating temperature of $17$mK. 
The Chimera graph of the DW2 used in our work is shown in Figure~\ref{fig:chimera}. Each unit cell is a balanced $K_{4,4}$ bipartite graph. In the ideal Chimera graph (of $512$ qubits) the degree of each vertex is $6$ (except for the corner unit cells). In the actual DW2 device we used $503$ qubits were functional.
For the scaling analysis we considered $L\times  L$ square sub-lattices of the Chimera graph, and restricted our simulations and tests on the DW2 to the subset of functional qubits within these subgraphs, denoted $C_L$ (see Figure~\ref{fig:chimera}). For each clause density $\alpha$ and problem size $N$ (or $C_L$) we generated $100$ instances, for a total of $12,\!600$ instances. We performed approximately $990000 \mu s/t_a$ annealing runs (experiments) for each problem instance and for each annealing time $t_a \in [20,40]\mu$s, in steps of 2$\mu$s, for a total of more than $10^{10}$ experiments. No gauge averaging \cite{q108} was performed because we are not concerned with the timing data for a single instance but a collection of instances, and the variation over instances is larger than the variation over gauges. 


\subsection{Algorithms}
\label{app:algorithms}

\subsubsection{HFS algorithm}
The HFS algorithm is due to Hamze \& Freitas \cite{hamze:04} and Selby \cite{selby:13b}. This is a tree-based optimization algorithm, which exploits the sparsity and local connections of the Chimera graph to construct very wide induced trees
and repeatedly optimizes over such trees until no more improvement is likely. 

\begin{figure}
\begin{center}
\includegraphics[width=\columnwidth]{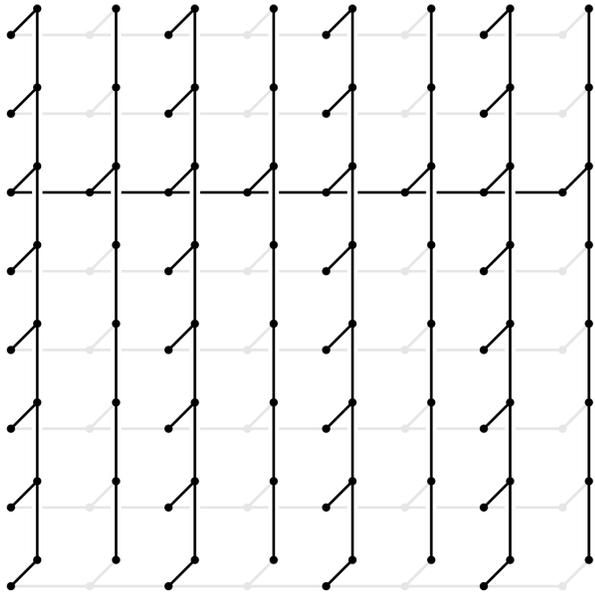}
\caption{An example of the view of the Chimera graph in the HFS algorithm. Each vertex is one half of a unit cell, and is $2^4$-dimensional. The algorithm repeatedly finds the minimum energy configuration of the graph over trees (like the one highlighted) given the state of the remaining vertices (in gray).}
\label{fig:SelbyTree}
\end{center}
\end{figure}

We briefly discuss the tree construction of the HFS algorithm.  It considers each part of the bipartite unit cell as a single $2^4$-dimensional vertex instead of four distinct $2$-dimensional vertices, resulting in a graph as depicted in Fig.~\ref{fig:SelbyTree}. The tree represented by the dark vertices covers $78\%$ of the graph, and such trees will cover $75\%$ of the graph in the limit of infinitely large Chimera graphs. Finding the minimum energy configuration of such a tree conditioned on the state of the rest of the graph can be done in $\mathcal{O} (N)$ time where $N$ is the number of vertices. Since the tree encodes so much of the graph, optimizing over such trees can quickly find a low-lying energy state. This is the strategy behind Prog-QUBO \cite{selby:13a}. Since each sample takes a variable number of trees, we estimate time to solution as as the average number of trees multiplied by the number of operations per tree, with a time constant of $0.5\mu s$ per operation.

Prog-QUBO runs in a  serial manner on a single core of a CPU. Since we allow the DW2 to use $\mathcal{O} (N)$ resources, we should also parallelize the HFS algorithm. In principle, HFS proceeds in a series of steps --- at each step, one shears off each of the leaves of the tree (i.e., the outermost vertices) and then proceeds to the next step until the tree has collapsed to a point. Each of the leaves can be eliminated separately, however each elimination along a particular branch is dependent on the previous eliminations. As a result, we must perform between $L+2$ and $\frac32 L+2$ (average $\frac54 L+2$) irreducibly serial operations when reducing the trees on an $L\times L$ unit cell Chimera graph (depending on which of the trees we use and the specifics for how we do the reduction). Since we deal only with square graphs and are concerned with asymptotic scaling, we ignore the constant steps and use $\frac54 L$ basic operations per tree (see Sec.~\ref{subsec:HFS} for additional details).

We ran Prog-QUBO in mode ``-S3'' \cite{selby:13a}, meaning that we used maximal induced trees (treewidth $1$ in our case). Other options enable reduction over lines or treewidth $2$ subgraphs, but we did not use those options here. In principle, one could define a new solver by optimizing the treewidth of subgraphs over which to reduce for each problem size, which will likely perform better than only using treewidth $1$ for arbitrary problem sizes. However, this was not investigated in this work.

\subsubsection{Simulated Annealing}
Our SA algorithm uses a single spin-flip Metropolis update method.  In a single sweep, each spin is updated once according to the Metropolis rule: the spin is flipped, the change in energy $\Delta E$ is calculated.  If the energy is lowered, the flip is accepted, and if not, it is accepted with a probability given by the Metropolis probability:
\beq
P_{\mathrm{Met}} = \min \left( 1, \exp (- \beta \Delta E) \right)
\eeq
A linear annealing schedule in $\beta$ is used; Starting at $\beta_i = 0.01$, we increment $\beta$ in steps of  $\delta\beta = (\beta_f-\beta_i)/(S-1)$ where $S$ is the number of sweeps, up to $\beta_f$.  Each instance is repeated $10^4$ times.

\subsubsection{SSSV}
The SSSV model was first proposed \cite{SSSV} as a classical model that reproduced the success probabilities of the DW1 device studied in Ref.~\cite{q108}, although there is growing evidence that this model fails to capture the behavior of the device for specific instances \cite{q-sig2,Albash:2014if,Boixo:2014yu}.  The model can be understood as describing coherent single qubits interacting incoherently by replacing qubits by O(2) rotors; the Hamiltonian can be generated by replacing $\sigma_i^x \mapsto \sin \theta_i$ and $\sigma^z_i \mapsto \cos \theta_i$.  The system is then ``evolved'' by Monte Carlo updates on the angles $\theta_i \in [0, \pi]$.  Although SSSV is not designed to be a fast solver, we studied it here as a potential classical limit of the DW2 and checked what the scaling of such a classical limit would be.  The DW2 annealing schedule in Fig.~\ref{fig:schedule} was used, and the temperature was kept constant at $10.56$mK. Each instance was repeated $10^4$ times.  \\

\subsubsection{Simulated Quantum Annealing}
SQA \cite{sqa1,Santoro} is an annealing algorithm based on discrete-time path-integral quantum Monte Carlo simulations of the transverse field Ising model but using Monte Carlo dynamics instead of the open system evolution of a quantum system. This amounts to sampling the world line configurations of the quantum Hamiltonian \eqref{eq:Hquantum} while slowly changing the couplings. SQA has been shown to be consistent with the input/output behavior of the DW1 for random instances \cite{q108}, and we accordingly used a discrete-time quantum annealing algorithm.  We always used $64$ Trotter slices and the code was run at $\beta=10$ (in dimensionless units, such that $\max(|J_{ij}|) = 1$) with linearly decreasing and increasing $A(t)$ and $B(t)$, respectively.  Cluster updates were performed only along the imaginary time direction.  A single sweep amounts to the following: for each space-like slice, a random spin along the imaginary time direction is picked.  The neighbors of this spin are added to the cluster (assuming they are parallel) according to the Wolff algorithm \cite{PhysRevLett.62.361} with probability $1 - e^{-2 J_{\perp}}$, where $J_\perp = -0.5 \ln \left[ \tanh A(t) \right]$ is the spin-spin coupling along the imaginary time direction.  When the cluster construction terminates, the cluster is flipped according to the Metropolis probability using the change in energy along the space-like direction associated with flipping the cluster.  Therefore a single sweep involves a single cluster update for each space-like slice.
For every problem instance, we performed $10^3$ repetitions in order to estimate the instance success rates.  

\subsubsection{Constraint Solver Algorithm}

Here we outline the algorithm we developed and used to search for, and list, solutions to a specified planted-solution Hamiltonian, taking advantage of the fact that the total cost function is a sum of easily-solvable cost functions each defined on a finite number of bits. Since in our case each term in the Hamiltonian is a frustrated loop, it is easy to list the constraints each loop induces on any potential solution of the total Hamiltonian. The algorithm is an exhaustive search and is an implementation of the Bucket Elimination Algorithm (see, e.g., Ref.~\cite{dechter1999bucket}).

The problem structure consists of a set of values (or bits) where each value may be $+1$ or $-1$.  A set of constraints restricts subsets of the bits to certain values. 
The problem is to assign values to all the bits so as to satisfy all the constraints.

Each constraint (in this case, the optimizing configurations of loop Hamiltonians) applies to a subset of the bits, and contains a list of allowed settings for that subset of bits. 
The constraint is met if the values of the bits in the subset matches one of the allowed settings.

The Bucket Elimination Algorithm as applied to this problem consists of the following steps.
First is the constraint elimination stage: (i) Select a bit to eliminate; (ii)
Save constraints which contain selected bit; (iii)
Combine all constraints which contain selected bit; (iv) 
Generate new constraints without selected bit; (v) If combined constraints contain a contradiction, exit; (vi)  
Repeat until all bits have been addressed.

The second part of the algorithm tackles the enumeration of the solutions. 
It involves using the tables created in the constraint elimination stage to grow solution sets one bit at a time.  
The last table created in the constraint satisfaction stage contains tables for a single bit (the last to be eliminated). 
If it has no entries, no solutions are possible for the processed constraint set. 
If there is a single entry, it indicates the value that the bit must be $-1$, or $+1$. 
If there are two entries, then both $-1$ and $+1$ are possible values. 
A solution set is built listing the values of the final bit that was eliminated.  
The table created for the final two bits is processed next. It contains all the legal values for the final two bits. 
For each allowed value of the final bit, a solution is generated for the final two bits, resulting in a new solution set for the final two bits. 
The bit solution tables generated during the elimination phase are processed in reverse order, each time adding a single bit, 
combining the solutions from the previous step with the single bit, generating a new solution set with a bit added. 
Ultimately all of the bit solution tables generated during the elimination phase are processed and solutions with values 
for all bits are generated. The logic driving the elimination phase ensures that it is always possible to combine the 
set of solutions with the next bit. 

This algorithm is guaranteed to find all solutions, but may exceed time and memory constraints. 
All steps are well defined except for the step to ``select a bit to eliminate''. 
The order in which bits are eliminated dramatically affects the time and memory required. 
Determination of the optimal order to eliminate bits is known to be NP complete. 
A more detailed description of the algorithm (including examples) will be published elsewhere as part of another study in the near future. 

\subsection{Error estimation}
\label{app:error-estimation}

\subsubsection{Annealers}

For the annealers, the time to solution can be trivially found by applying a function $T(p)$ to the probability of success $p$. In the main text, the $T(p)$ used is the time required to find the ground state at least once with a probability $0.99$: 
\beq
T(p)=\frac{\log (1 - 0.99)}{\log (1 - p)}\ ,
\eeq
For the purposes of the discussion below this function can in fact be totally arbitrary and may depend on system size.  

We imagine our annealing algorithms as essentially a sequence of binary trials, where each run is either a success or a failure. Because we do not have infinitely many observations, there is some uncertainty in our estimate of the true probability of success of our binary trial. Since our data was collected as a series of $r$ independent trials (which varies by solver) for each solver and instance, the annealers' results are described by a binomial distribution. 

In order to determine what the probability of success is, we assumed a Bayesian stance --- we do not know what the probability is currently and so must assume some prior distribution for our belief, and will update our belief based on the evidence. Which prior should we choose? The obvious choice is a $\beta$ distribution as it is the conjugate prior for the binomial distribution, giving us a closed form for our posterior distribution \cite{fink:97}. We have chosen the Jeffreys prior for the binomial distribution, $\beta(\frac12,\frac12)$ for two reasons: 1) It is invariant under reparameterization of the parameter space, 2) it is the prior which maximizes the mutual information between the sample and the parameter over all continuous, positive priors. In other words, it yields the same prior no matter how we parameterize our space and $\beta(\frac12,\frac12)$ maximizes the amount of information gained by learning the data. The other obvious prior is the uniform distribution or $\beta(1,1)$ but it is not invariant under reparameterization and learns less from the data than $\beta(\frac12,\frac12)$ \cite{clarke:94}. For these reasons, $\beta(\frac12,\frac12)$ tends to be the standard choice of prior for binomial distributions \cite{bernardo:11}.

After Bayesian updating by our prior, our probability distribution for $p$ is $\beta(x+\frac12,r-x+\frac12)$, where $x$ is the number of successes and $r$ the number of runs. 

There may be a concern that our solvers do not fully conform to a binomial distribution: if there are correlations between successive runs then our empirical success probability $\frac xr$ would be inflated for hard problems. However, since we did not observe an advantage for the DW2 over the classical algorithms for the hardest problems, and the only potential advantage we observed is for the easier problems at high clause density, we do not believe this issue is affecting our conclusions.

Thus, for all instances $\mathcal{I}_i \in \{\mathcal{I}\}$ in a particular range of interest (for example, a particular problem size and clause density), we have some distribution $\beta_i$ for the probability of success for that instance. To generate our error bars we used the bootstrap method, which we describe next.

If we have $M$ instances in our set of interest, then we resample with replacement a ``new'' set of instances $\{\mathcal{I}_i\}_j$, also of length $M$, from our set $\mathcal{I}$. For each instance $\mathcal{I}_{i,j}$ from this new set we sample a value for its probability from its corresponding distribution $\beta_{\mathcal{I}_{i,j}}$ to get a set of probabilities $\{p_{i,j}\}$. We then calculate whatever function $F_j = f(\{p_{i,j}\})$ we wish on these probabilities, e.g., the median over the set of instances. We repeat this process a large number of times (in our case, $1000$), to have many values of our function $\{F_j\}$. We then take the mean and standard deviation over the set $\{F_j\}$ to get a value $\bar{F}$ and a standard deviation $\sigma_F$, which form our value and error bar for that size and clause density.

When we take the ratio of two algorithms $A$ and $B$, for each pair of corresponding data points for the two algorithms (i.e., each size and clause density) we characterize each point with a normal distributions $\mathcal{N}(\mu_A,\sigma_A)$ and $\mathcal{N}(\mu_B,\sigma_B)$. We then resample from each distribution a large number of times ($1000$) to get two sets of values for the function we wish to plot $F$:  $\{F_A\}$ and $\{F_B\}$. We take the ratio of the corresponding elements in the two sets $S_i = F_{A,i} / F_{B,i}$ and take the mean and standard deviation of the set $\{S_i\}$ to get our data point and error bar for the ratio.

\subsubsection{HFS algorithm}
\label{subsec:HFS}
For the HFS algorithm, time to solution for each problem is computed as the mean number of trees per sample multiplied by $0.625\mu s \times L$ for a $C_L$ problem. The number $0.625$ is chosen to approximate the times on a standard laptop, and the scaling with $L$ is due to the fact that, in a parallel setting, the number of steps to reduce a tree is linear in $L$ (exactly, it is $\frac 54 L+2$, but the $2$ serves only to mask asymptotic scaling and is thus not included in our TTS or speedup plots or the scaling analysis).
 
\subsubsection{Euclidean distance}
In order to generate the error bars in Fig.~\ref{fig:Euclid-dist}, instead of calculating the Euclidean distance over the total number of instances at a given $\alpha$ ($700$  total), we calculated the Euclidean distance over half the number of instances ($350$  total).  We were then able to perform $100$ bootstraps over the instances, i.e., we picked $350$ instances at random for each Euclidean distance calculation.  For each bootstrap, we calculated the Euclidean distance, and the data points in Fig.~\ref{fig:Euclid-dist} are the mean of the Euclidean distances over the bootstraps, while the error bars are twice the standard deviation of the Euclidean distances.

\subsubsection{Difference in slope}
In order to generate the error bars in Fig.~\ref{fig:DiffDWslope}, we used the data points and the error bars in Fig.~\ref{fig:DWslope} as the mean and twice the standard deviation of a Gaussian distribution.  We then took $1000$ samples from this distribution and calculated their differences.  The means of the differences are the data points in Fig.~\ref{fig:DiffDWslope}, and the error bars are twice the standard deviation of the differences.

\section{Additional results}
\label{app:additional}
In this section we collect additional results in support of the main text.

\subsection{Degeneracy-hardness correlation}
\begin{figure}[t]
\begin{center}
\includegraphics[width=\columnwidth]{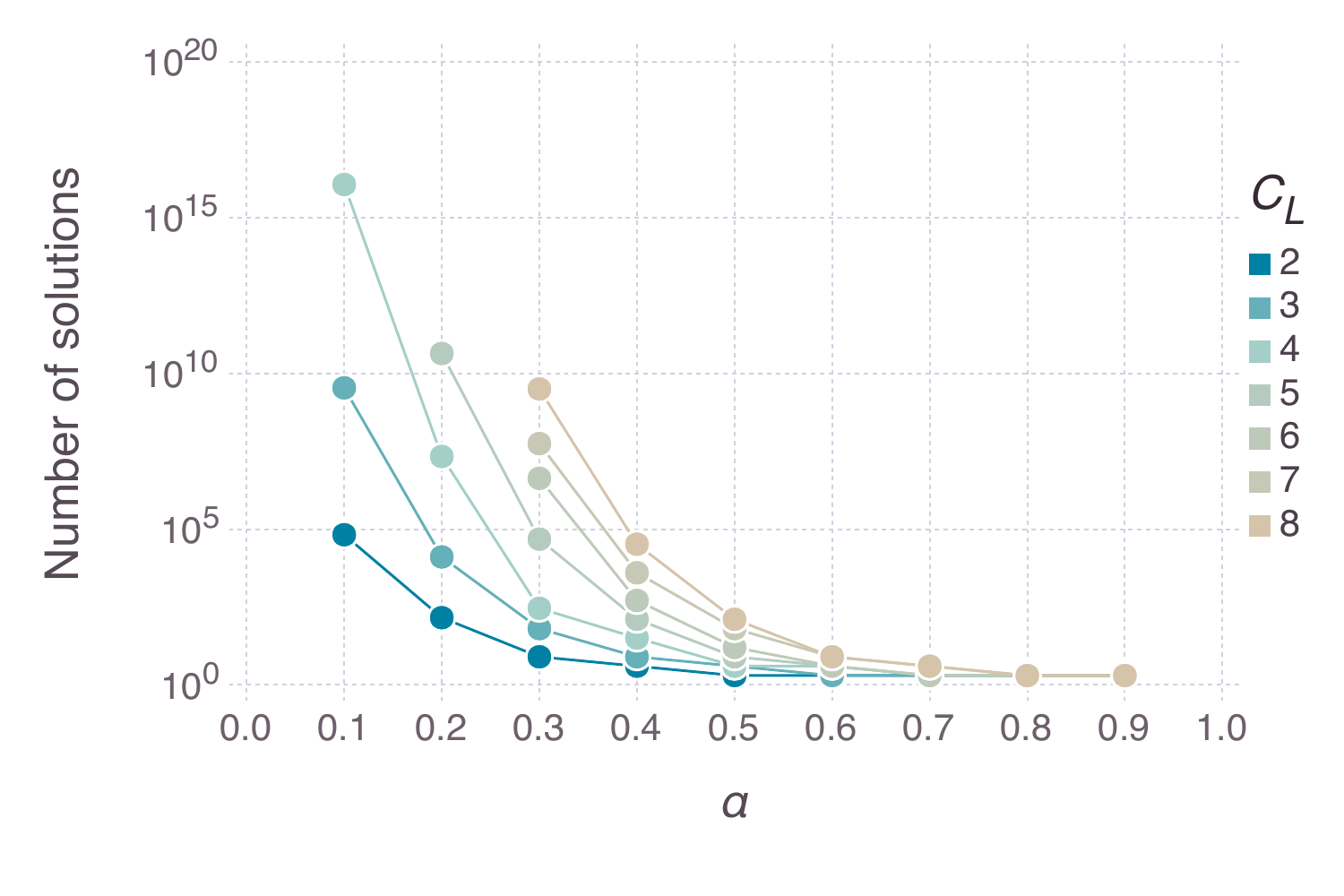}
\caption{\textbf{Ground state degeneracy}.
Number of unique solutions as a function of clause density for different Chimera subgraph sizes $C_L$.
As the clause density is increased, the number of unique solutions found decreases to one (up to the global bit flip symmetry).  
Shown is the median degeneracy, i.e., we sort the degeneracies of the $100$ instances for each value of $L$ and $\alpha$, and find the median. Our procedure counts the degenerate solutions and stops when it reaches $10^5$ solutions. If the median has $10^5$ solutions then we assume that not all solutions were found and hence the degeneracy for that value of $L$ and $\alpha$ is not plotted.
These are solutions on the used qubits (e.g., there are many instances for each $\alpha$ at $L=8$ that use $< 503$ qubits); to account for the $n_{uq}$ unused qubits we multiply the degeneracy by $2^{n_{uq}}$.}
 \label{fig:numDiffSols}
\end{center}
\end{figure}

\begin{figure}
\begin{center}
{\includegraphics[width=0.48\textwidth]{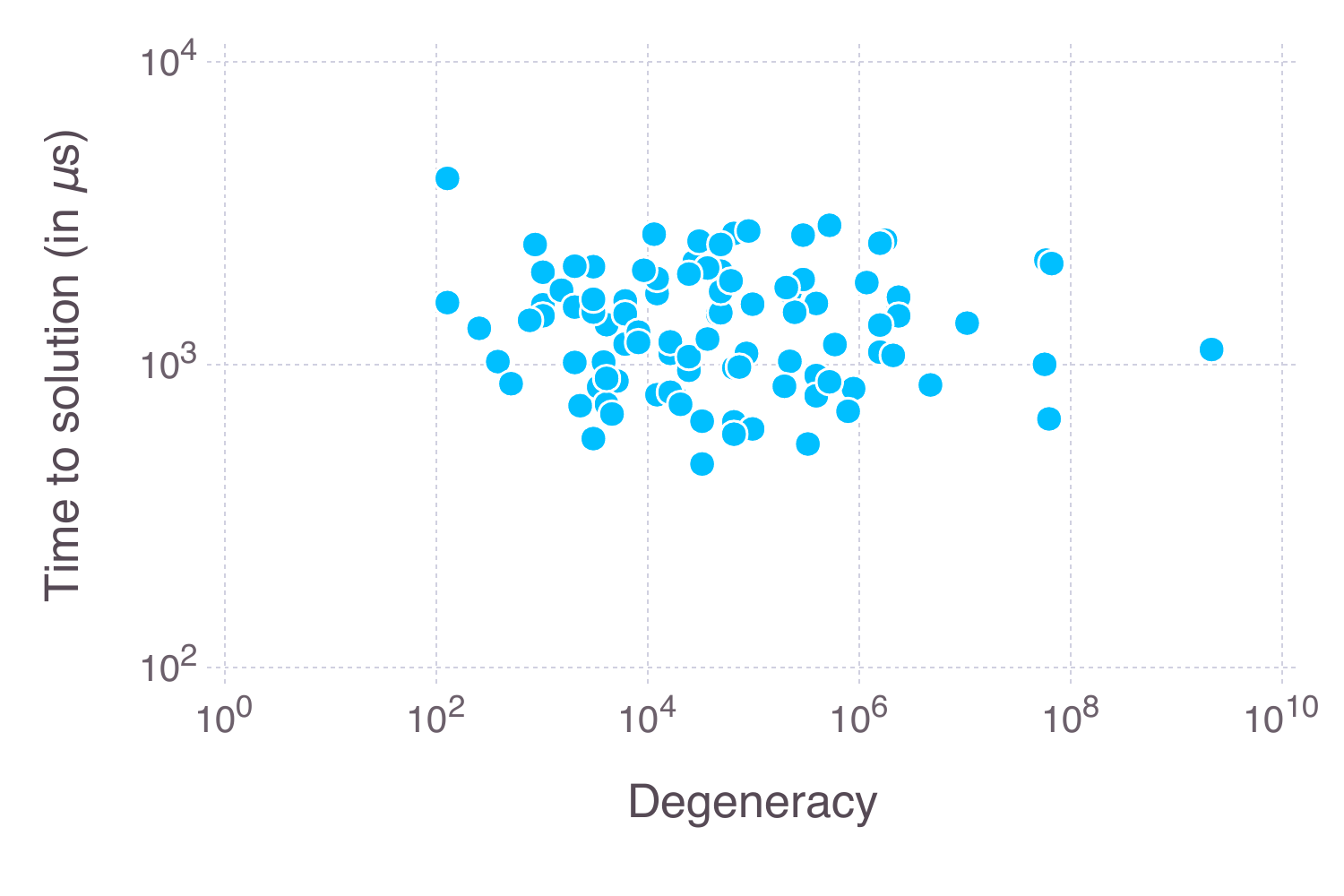}}
\caption{Scatter plot of the TTS for the HFS algorithm and the degeneracy at $L=8$ and $\alpha = 0.4$ ($100$ instances total). 
  Even though there is a wide range of degeneracy over several orders of magnitude, we do not observe any trend in the TTS.  
The degeneracy accounts for the fact that some qubits are not coupled into the problem (e.g., if $n$ qubits are not specified for that particular problem, then the degeneracy is $2^n$ times the directly counted degeneracy). The Pearson correlation coefficient is $-0.046$.}
\label{fig:DegHardness}
\end{center}
\end{figure}

It is known that a non-degenerate ground state along with an exponentially (in problem size) degenerate first excited state leads to very hard SAT-type optimization problems \cite{Neuhaus:2014kx}. Here we focus on the ground state degeneracy and ask whether it is correlated with hardness. We show the ground state degeneracy in Fig.~\ref{fig:numDiffSols}. It decays rapidly as $\alpha$ grows, and for sufficiently large $\alpha$ (that depends on the problem size) the ground state found is unique (up to the global $\mathbf{Z}_2$ symmetry). This suggests that degeneracy is not necessarily correlated with hardness, since in the main text we found that hardness peaks at $\alpha \sim 0.25$. To this test directly  we restrict ourselves to $L=8$ and $\alpha = 0.4$.  We then bin the $100$ instances at this $\alpha$ according to their degeneracy and study their median TTS using the HFS algorithm.  We show the results in Fig.~ \ref{fig:DegHardness}, where we see no correlation between degeneracy and hardness for a fixed $\alpha$.  We find a similar result when we use the success probability of the DW2 as the metric for hardness.

%


\subsection{Additional easy-hard-easy transition plots} \label{sec:phasetransition}
\begin{figure*}
\begin{center}
\subfigure{\includegraphics[width=0.48\textwidth]{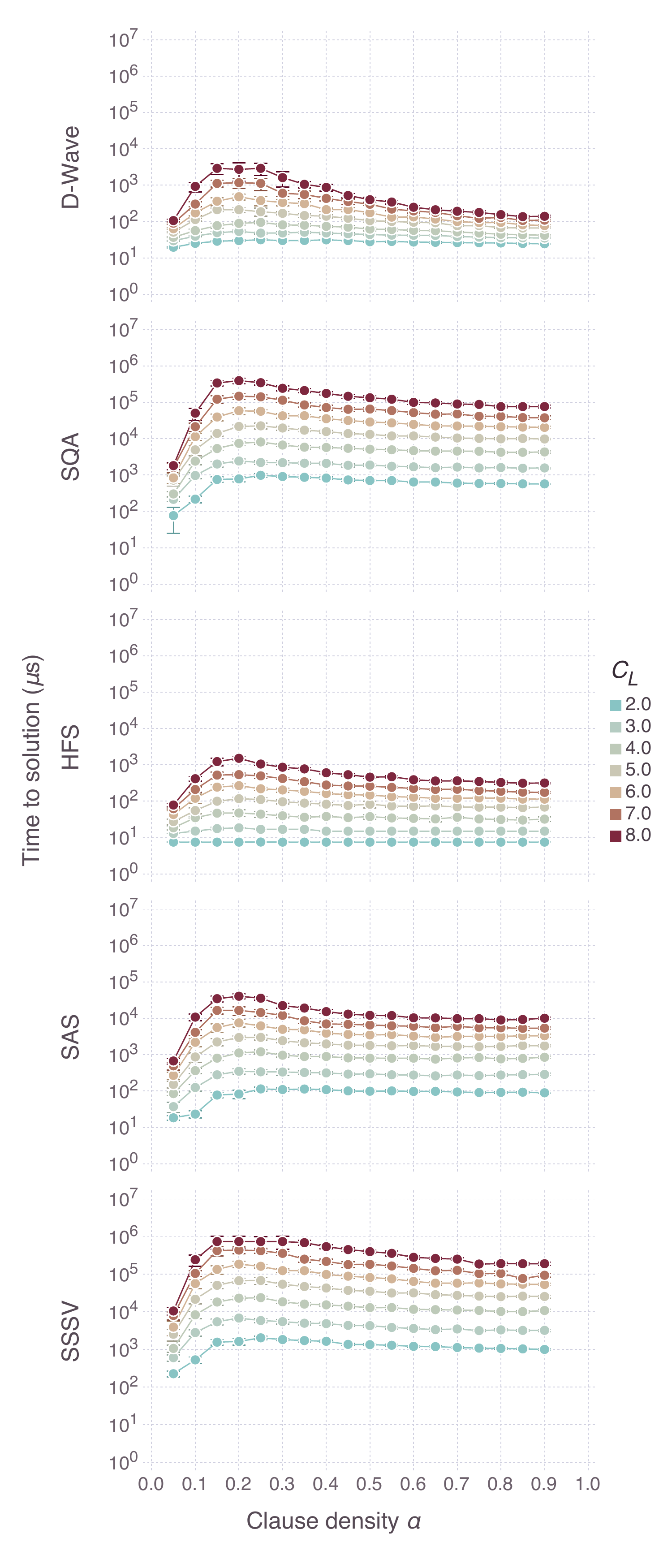}}
\subfigure{\includegraphics[width=0.48\textwidth]{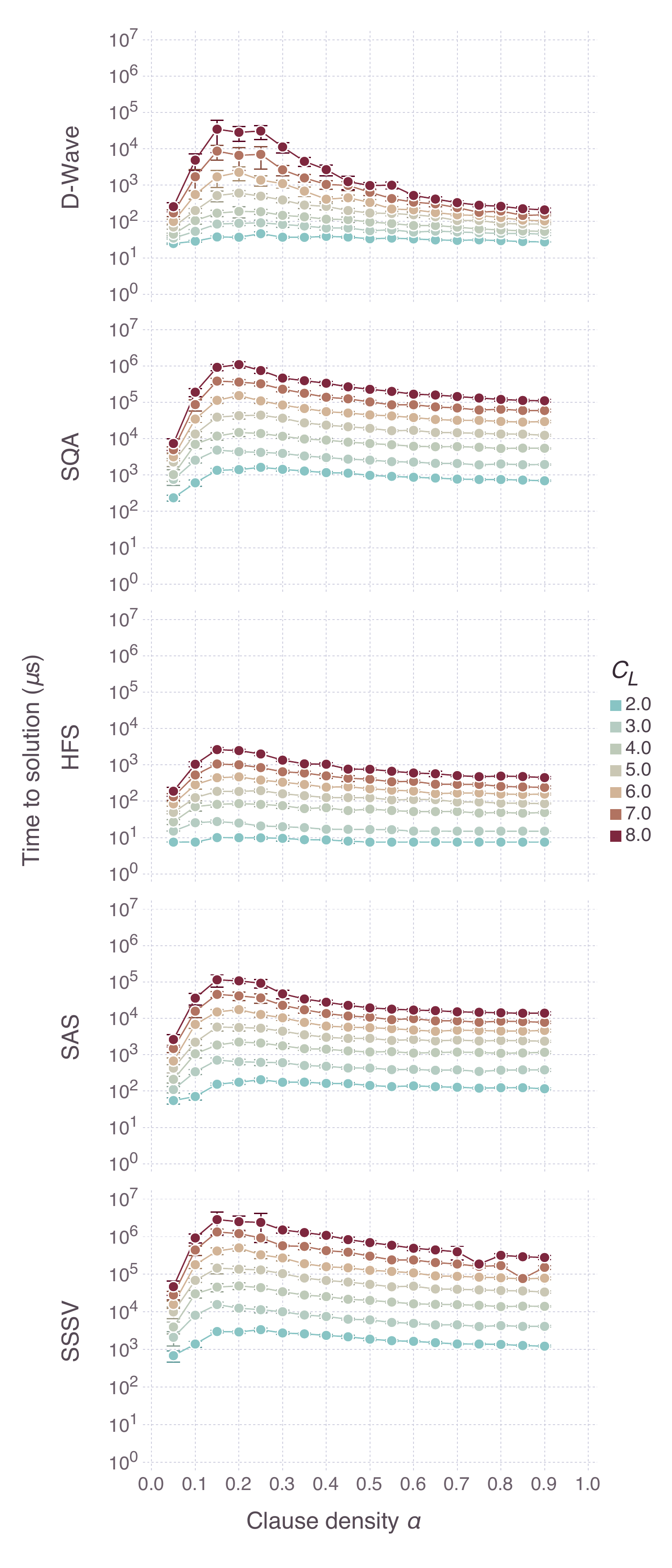}}
\caption{
Comparison of the 25th (left) and 75th (right) percentiles of the TTS (log scale) for all algorithms as a function of clause density $\alpha$. The different colors represent the different Chimera sizes tested.  All solvers show a peak at the same density value of $\alpha \approx 0.2$.}
\label{fig:pt_25_75}
\end{center}
\end{figure*}

The universal nature of the scaling behavior can be seen in Fig.~\ref{fig:pt_25_75}, complementing Fig.~\ref{fig:DwaveVsSelby50} with results for the $25$th and $75$th percentiles respectively. The peak in the TTS near $\alpha=0.2$ is a feature shared by all the solvers we considered. 

\subsection{Optimality plots}
\label{sec:optimality}

\begin{figure*}[t]
\subfigure[\ DW2]{\includegraphics[width=0.98\columnwidth]{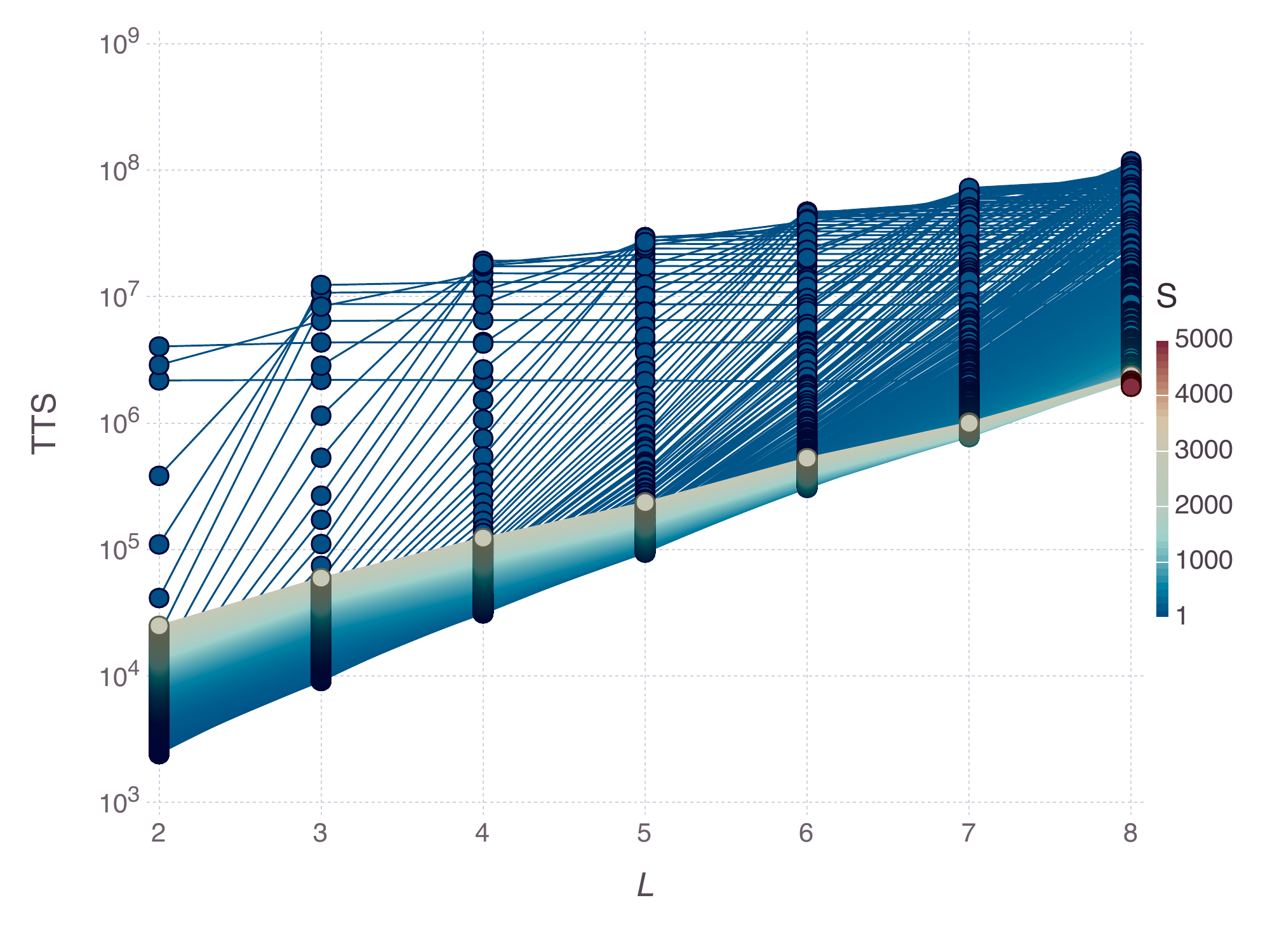}\label{fig:DW-opt}}
\subfigure[\ SQA]{\includegraphics[width=0.98\columnwidth]{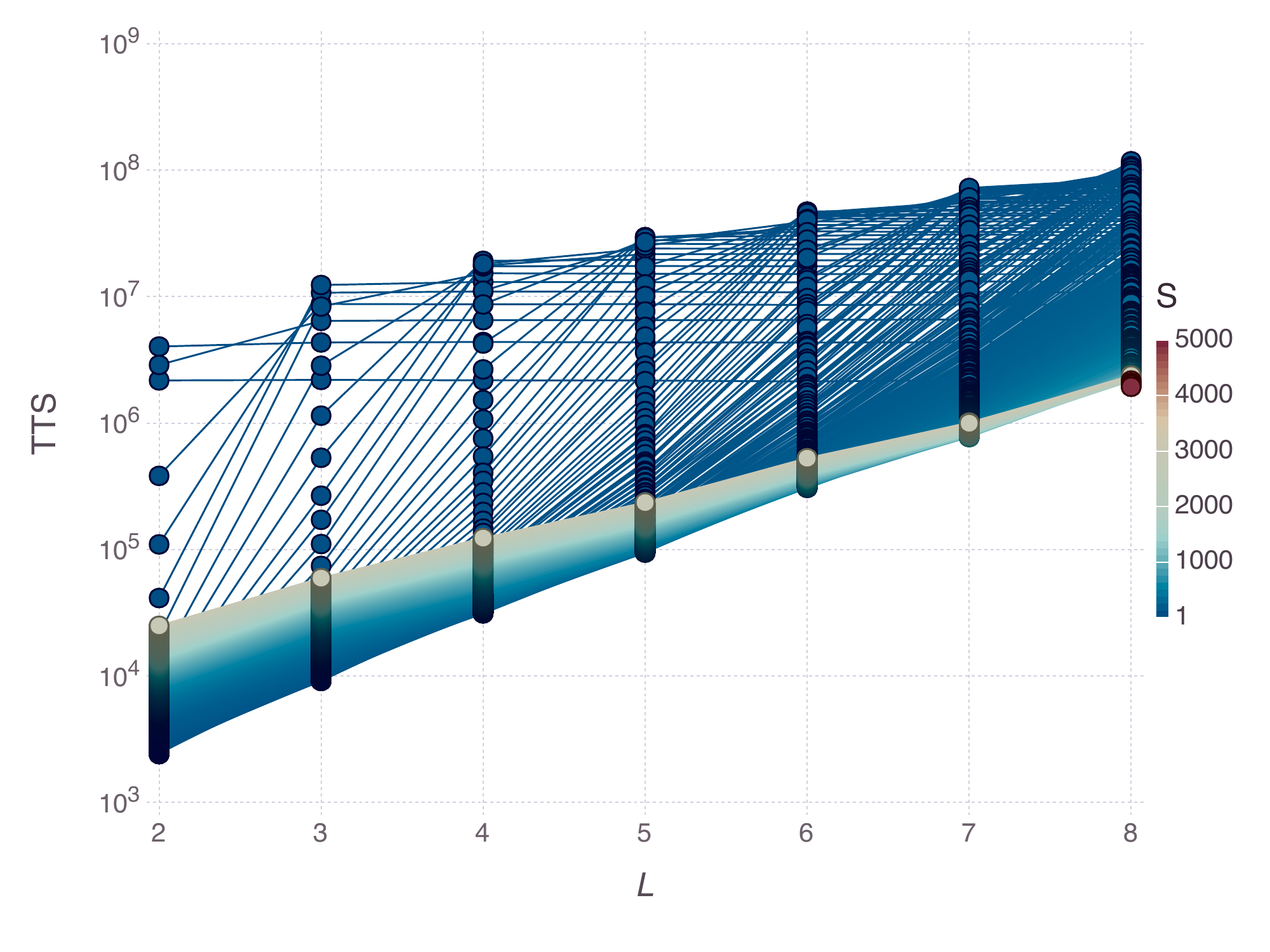}\label{fig:SQA-opt}}
\subfigure[\ SA]{\includegraphics[width=0.98\columnwidth]{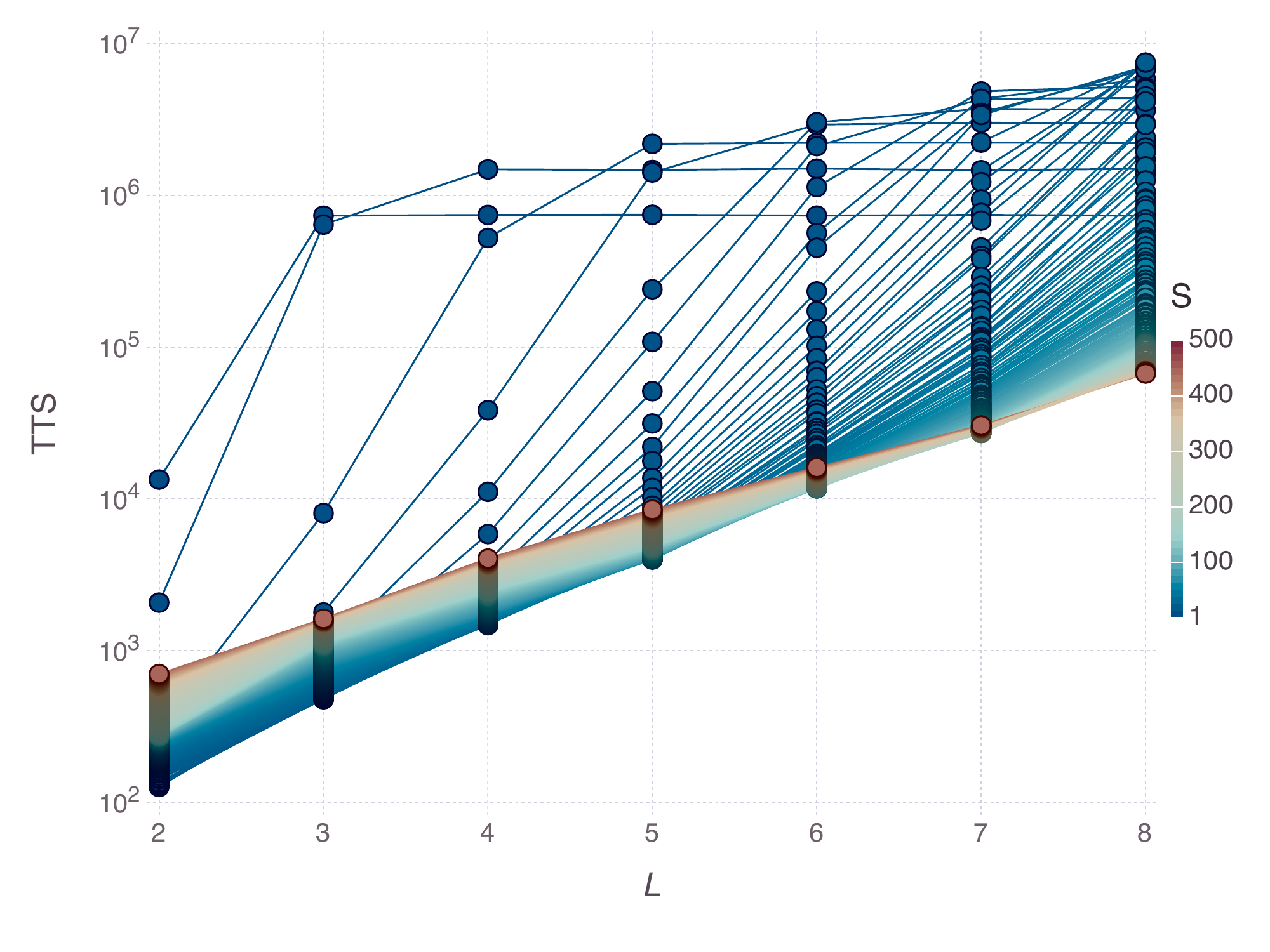}\label{fig:SA-opt}}
\subfigure[\ SSSV]{\includegraphics[width=0.98\columnwidth]{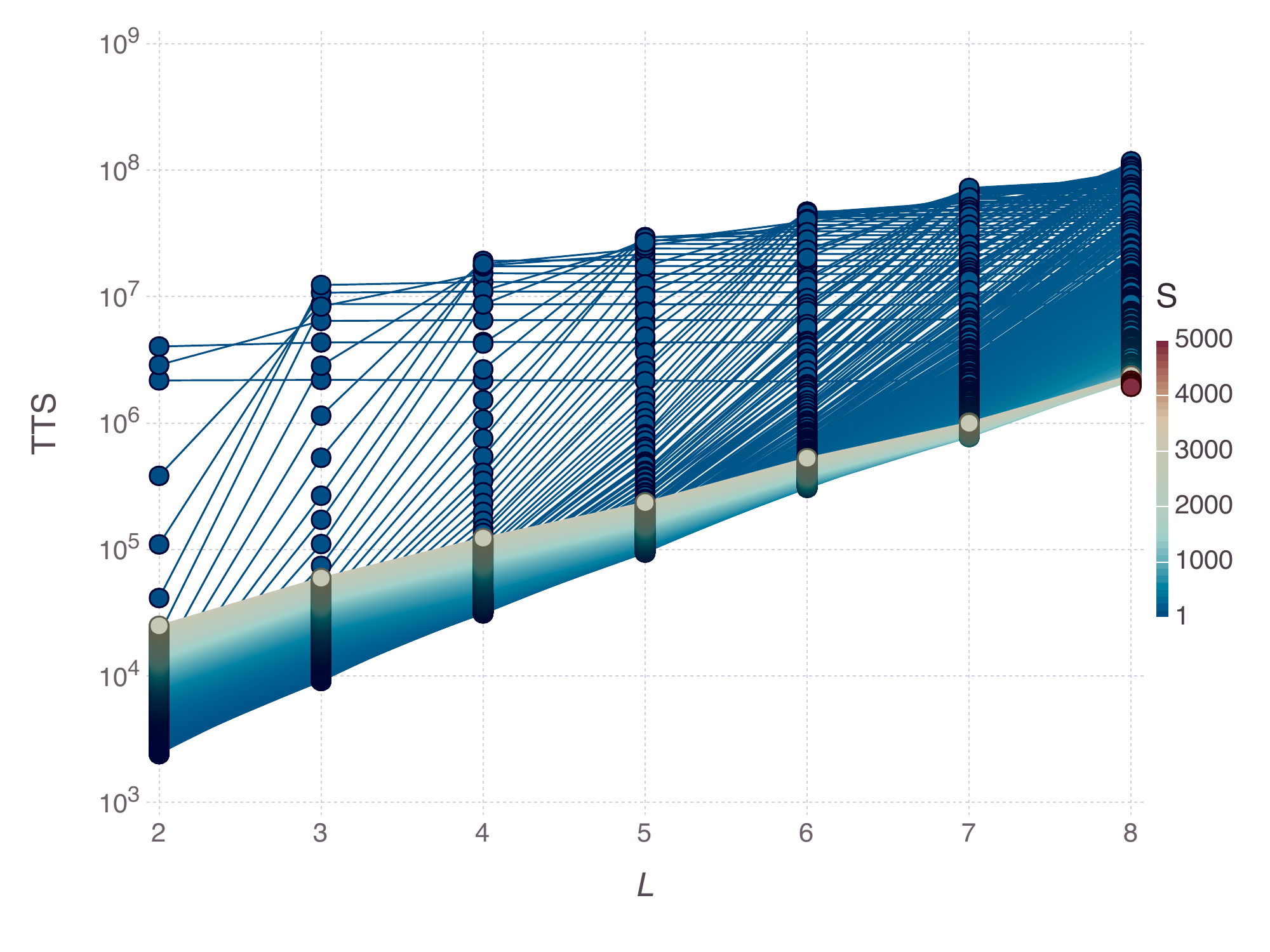}\label{fig:SSSV-opt}}
\caption{\textbf{Suboptimal annealing time and optimal sweeps for $\alpha=0.2$.} Plotted is the TTS (log scale) as a function of size $L$ for (a) the DW2, with all available annealing times, (b) SQA, (c) SA, and (d) SSSV, for many different sweep numbers. The lower envelope gives the scaling curves shown in Fig.~\ref{fig:TTSscalingbasic50} for $\alpha=0.2$. The TTS curves flatten at high $L$ for the following reason: each classical annealer was run $N_X$ times ($N_{\textrm{SA}}=N_{\textrm{SSSV}}=10^4$, $N_{\textrm{SQA}}=10^3$), and our $\beta$ distribution is $\beta(0.5,N_X+0.5)$ for $0$ successes, which has an average value of $\sim 1/(2N_X)$. This reflects the (Bayesian) information acquired after $N_X$ runs with $0$ successes (one would not expect the probability to be $0$). The flattening has no impact on the scale of the optimal number of sweeps.}
\label{fig:all-opt}
\end{figure*}

The absence of an optimal DW2 annealing time was discussed in detail in the main text, along with the optimality of the number of sweeps of the classical algorithms. Figure~\ref{fig:all-opt} illustrates this: a clear lower envelope is formed by the different curves plotted for SQA, SA, and SSSV, from which the optimal number of sweeps at each size can be easily extracted. Unfortunately no such envelope is seen for the DW2 results [Fig.~\ref{fig:DW-opt}], leading to the conclusion that $t_a=20\mu$s is suboptimal. 

\begin{figure*}
\begin{center}
\subfigure[\ DW2]{\includegraphics[width=\columnwidth]{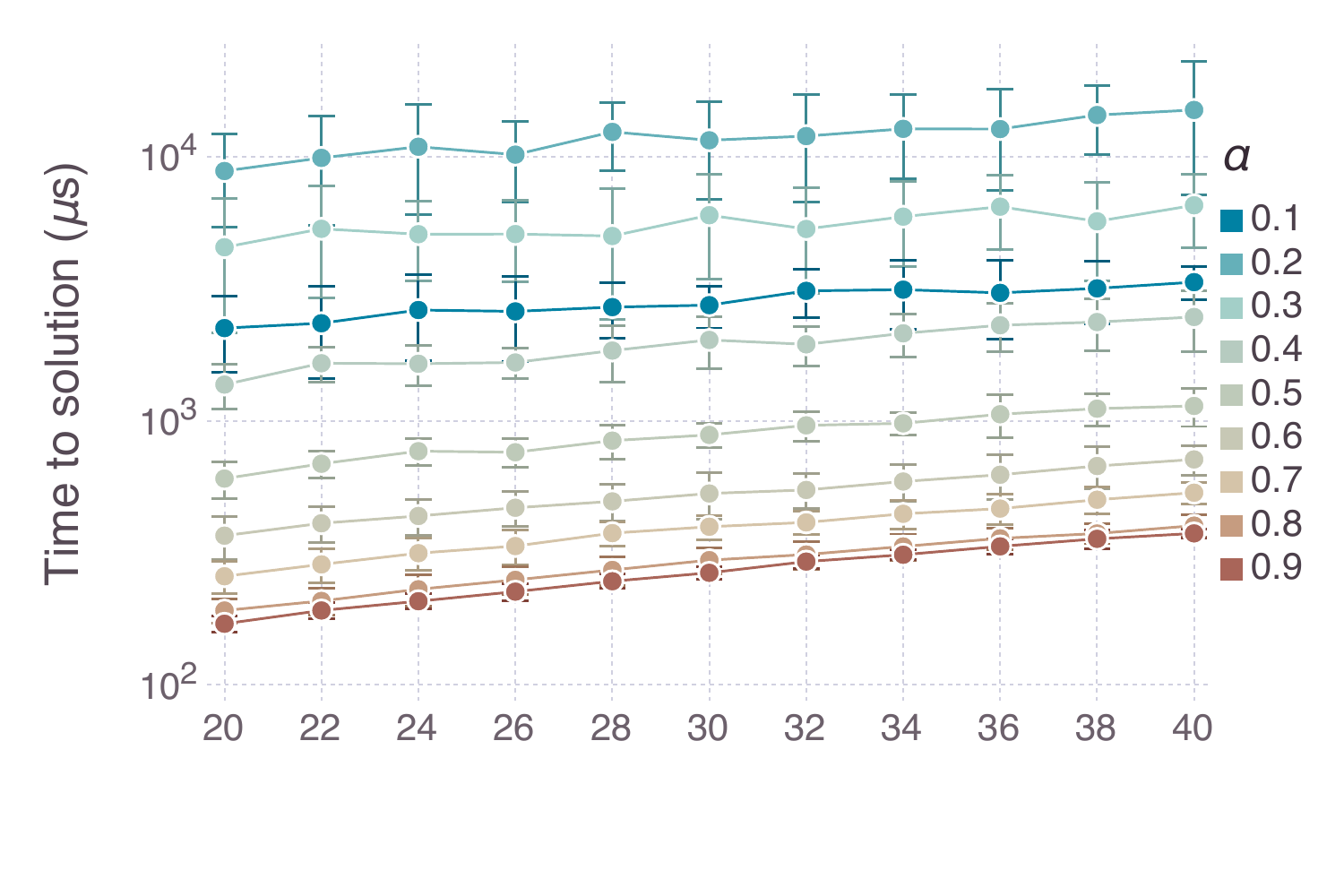}}
\subfigure[\ SQA]{\includegraphics[width=\columnwidth]{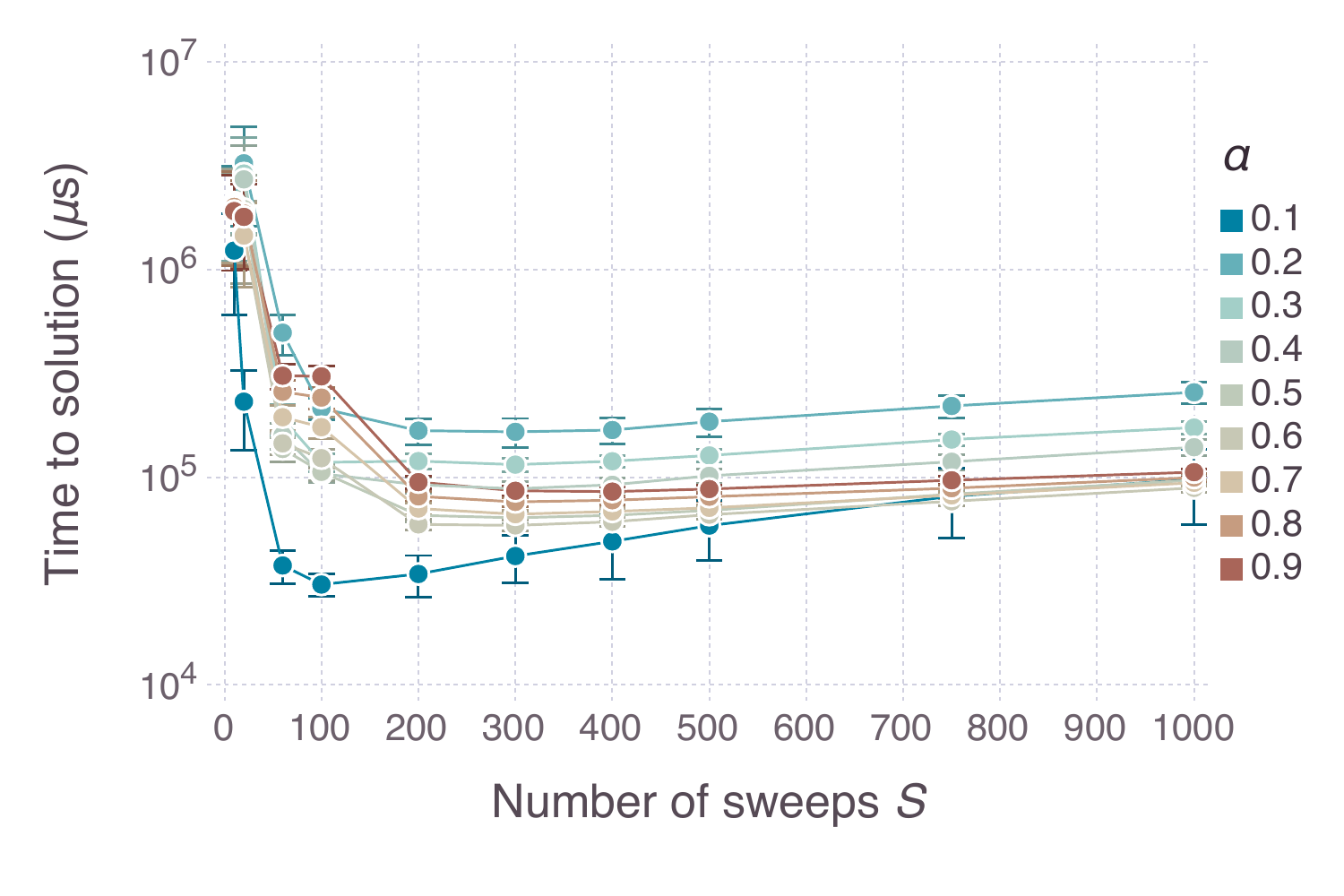}}
\subfigure[\ SA]{\includegraphics[width=\columnwidth]{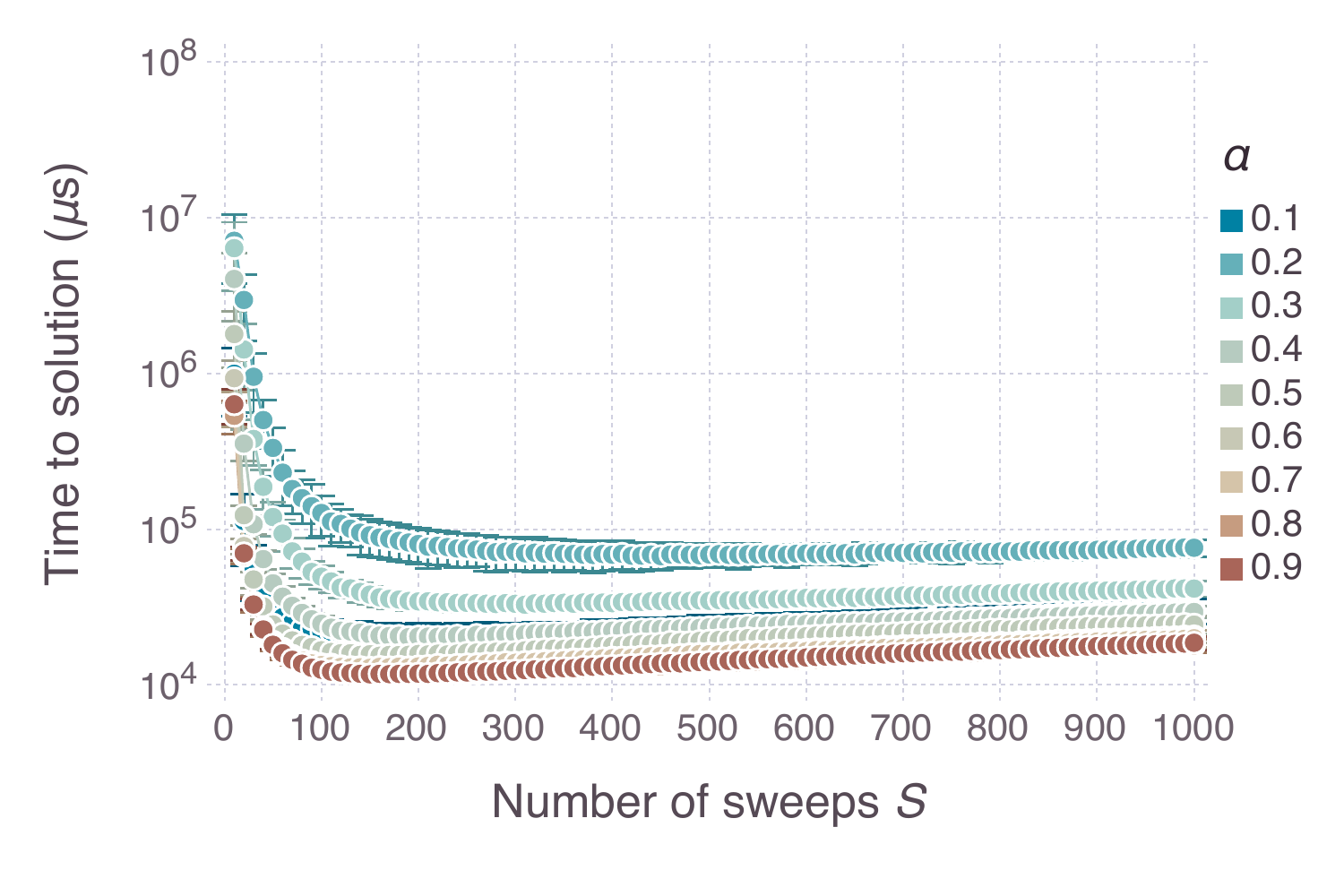}}
\subfigure[\ SSSV]{\includegraphics[width=\columnwidth]{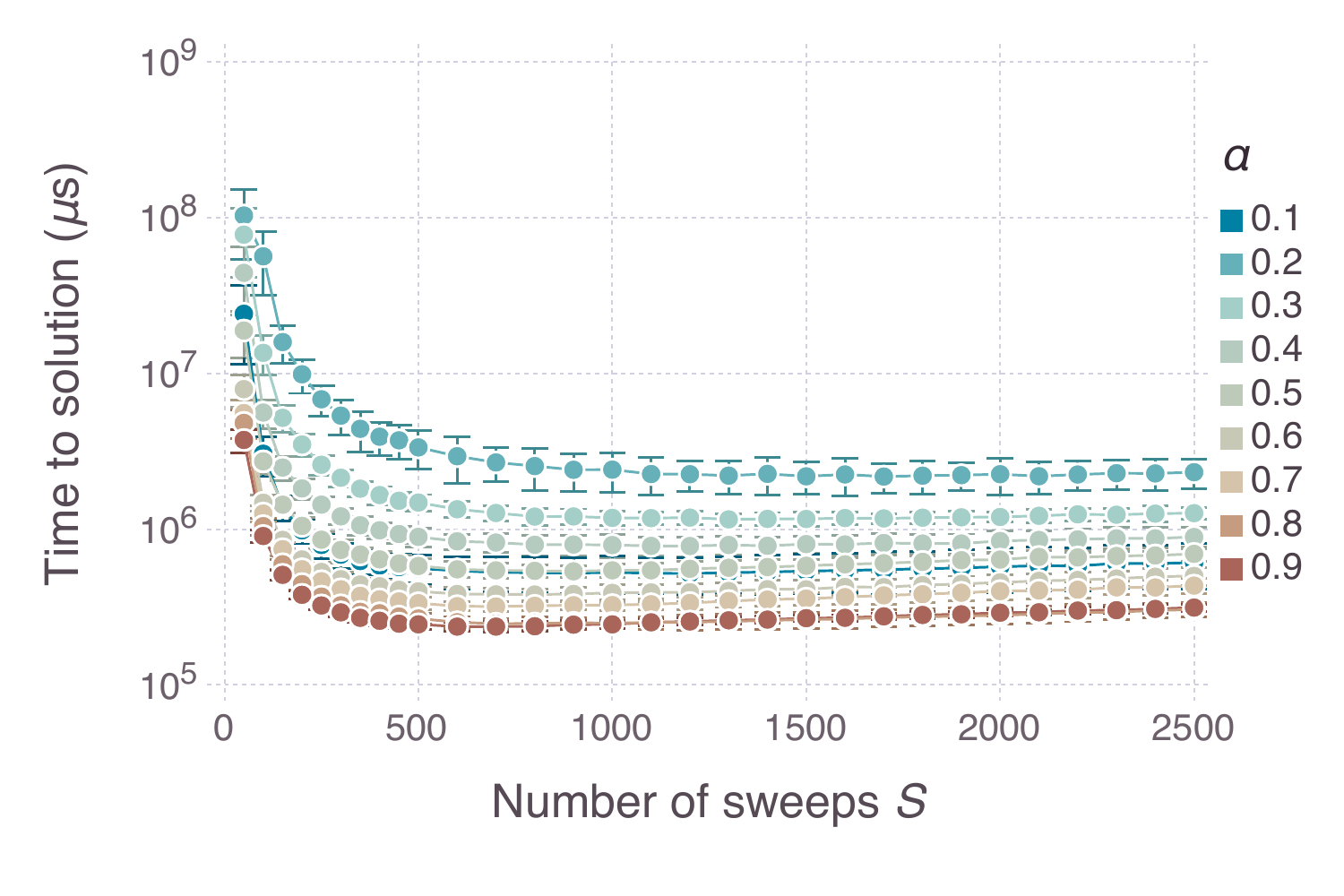}}
\caption{
TTS (log scale) for $L=8$ as a function of number of sweeps for DW2, SQA, SA, and SSSV used to identify the optimal number of sweeps.}
\label{fig:median_envelope}
\end{center}
\end{figure*}

A complementary perspective is given by Fig.~\ref{fig:median_envelope}, where we plot the TTS as a function of the number of sweeps, for a fixed problem size $L=8$. It can be seen that the classical algorithms all display a minimum for each clause density. The DW2 curves all slope upward, suggesting that the minimum lies to the left, i.e., is attained at $t_a < 20\mu$s. We note that in an attempt to extract an optimal annealing time
we tried to fit the DW2 curves to various functions inspired by the shape of the classical curves, but this proved unsuccessful since the DW2 curves essentially differ merely by a factor of $t_a$, as discussed in the main text.

\begin{figure*}[t]
{\includegraphics[width=0.98\columnwidth]{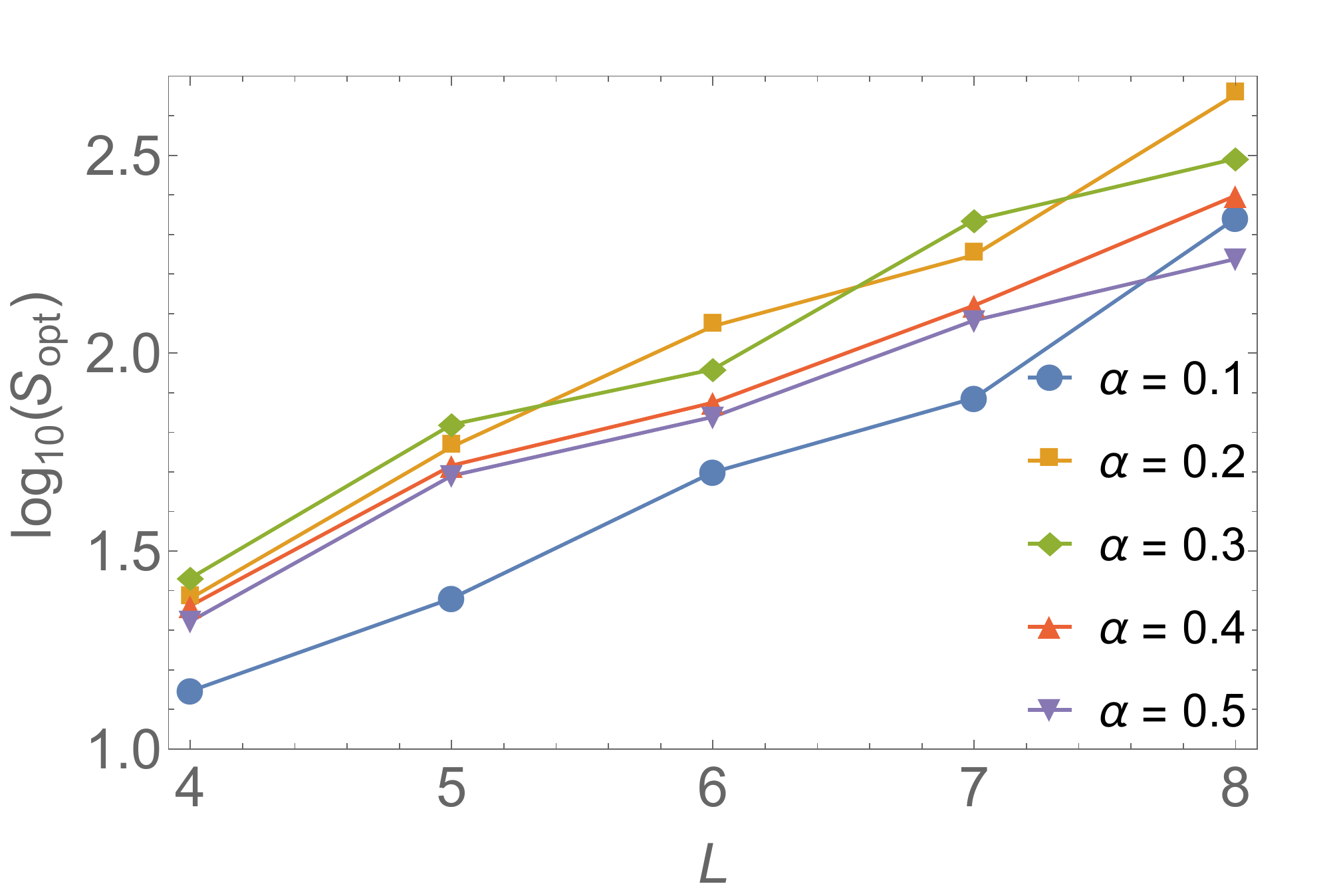}\label{fig:SAA-opt-small}}
{\includegraphics[width=0.98\columnwidth]{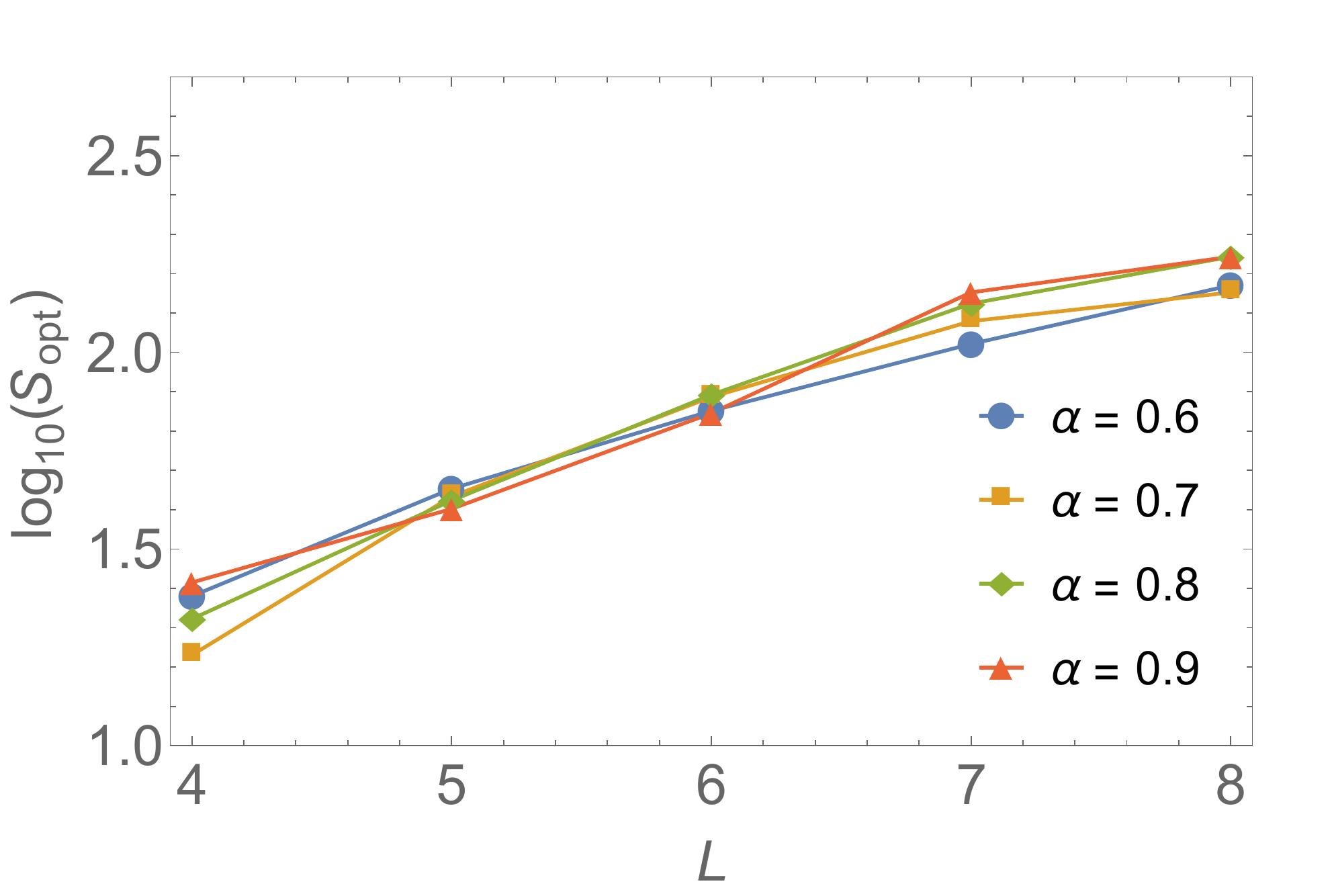}\label{fig:SA-opt-large}}
\caption{\textbf{Scaling of the SAA optimal number of sweeps.} The optimal number of sweeps is extracted for each $L$ from Fig.~\ref{fig:median_envelope}.
The scaling is roughly exponential for the smaller $\alpha$ values, and appears to be close to exponential for the larger $\alpha$ values. Lines are guides to the eye.}
\label{fig:SAA-opt-scaling}
\end{figure*}

We can take this a step further and use these optimal number of sweeps results to demonstrate that the problems we are considering here really are hard. To this end we plot in Fig.~\ref{fig:SAA-opt-scaling} the optimal number of sweeps as a function of problems size for SAA. We observe that certainly for the smaller clause densities the optimal number of sweeps $s_\textrm{opt}$ appears to scale exponentially in $L$, which indicates that the TTS (which is proportional to $s_\textrm{opt}$) will also grow exponentially in $L$.

%

\subsection{Additional speedup ratio plots} \label{sec:speedup}
\begin{figure*}
\begin{center}
\subfigure[\ $25$th percentile]{\includegraphics[width=0.98\textwidth]{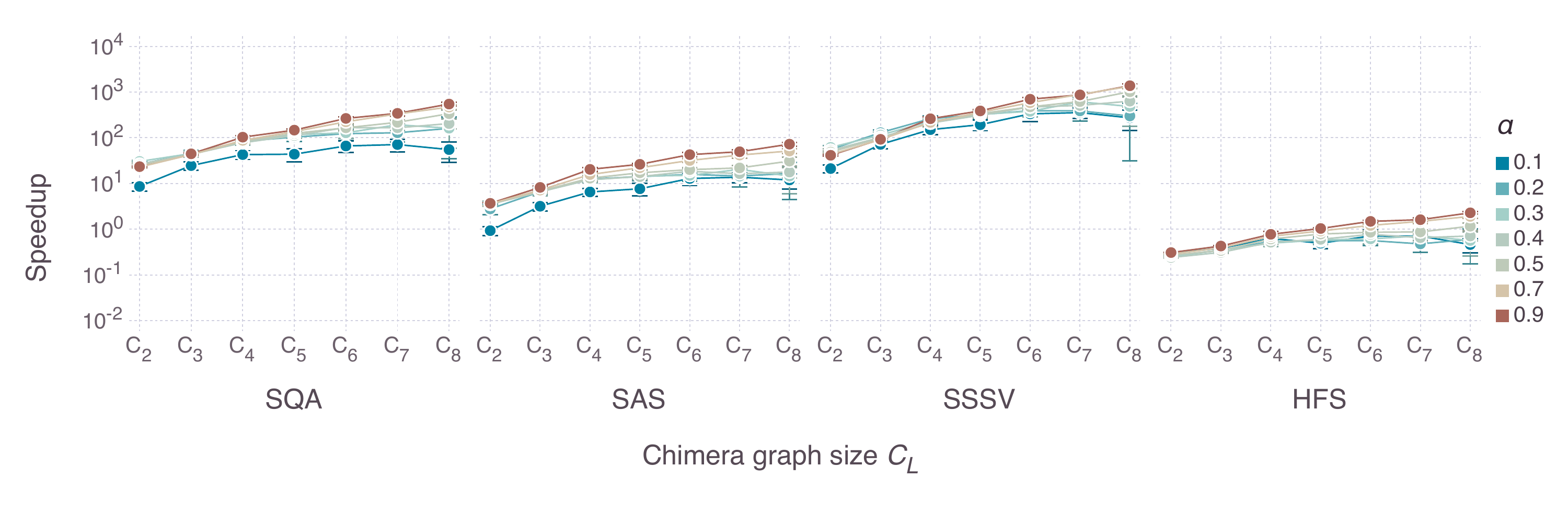}}

\subfigure[\ $75$th percentile]{\includegraphics[width=0.98\textwidth]{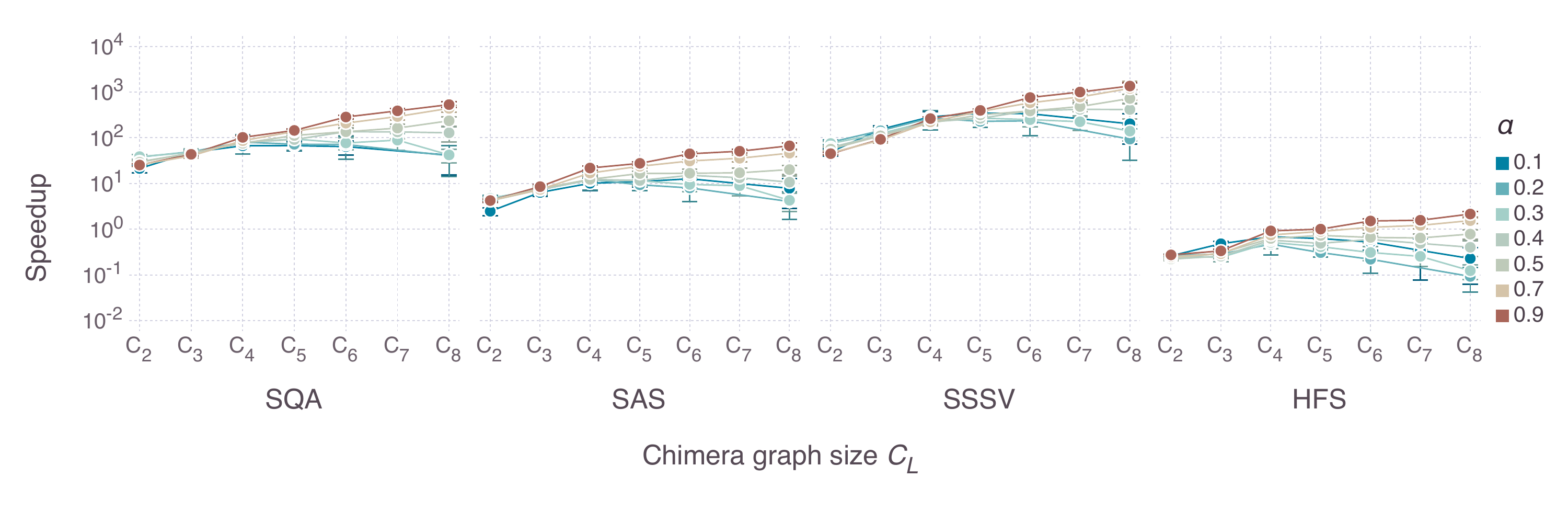}}
\caption{
The speedup ratios (log scale) for the 25th and 75th percentile of time-to-solution as a function of system size for various $\alpha$. The different colors denote a representative sample of clause densities.}
\label{fig:speedup_25_75}
\end{center}
\end{figure*}

To test whether our speedup ratio results depend strongly on the percentile of the success probability distribution, we recreate Fig.~\ref{fig:Speedupscalingbasic50} for the $25$th and $75$th percentiles in Fig.~\ref{fig:speedup_25_75}. The results are qualitatively similar, with a small improvement in the speedup ratio relative to HFS at the higher percentile.

\subsection{Additional correlation plots} \label{sec:correlationplots}
\begin{figure*}
\begin{center}
\includegraphics[width=\textwidth]{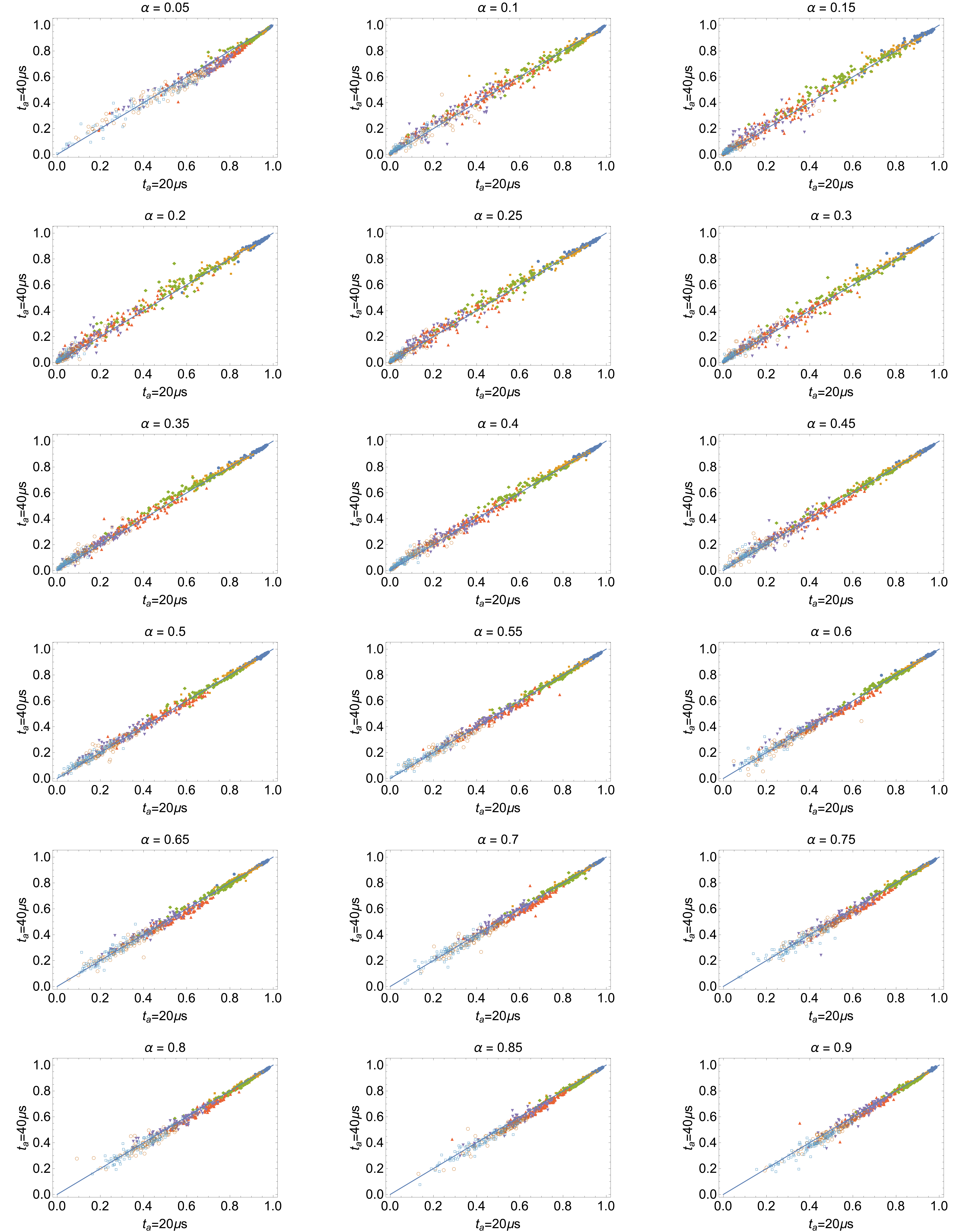}
\caption{\textbf{Success probability correlations.} The results for all instances at all $\alpha$ values are shown for the DW2 data at $t_a=20\mu$s and $t_a=40\mu$s. This complements Fig.~\ref{fig:DW-DW-SAA-corr}; the color scheme corresponds to different sizes $L$ as in that figure.}
\label{fig:corr-DWvsDW-all-alphas}
\end{center}
\end{figure*}

\begin{figure*}
\begin{center}
\includegraphics[width=\textwidth]{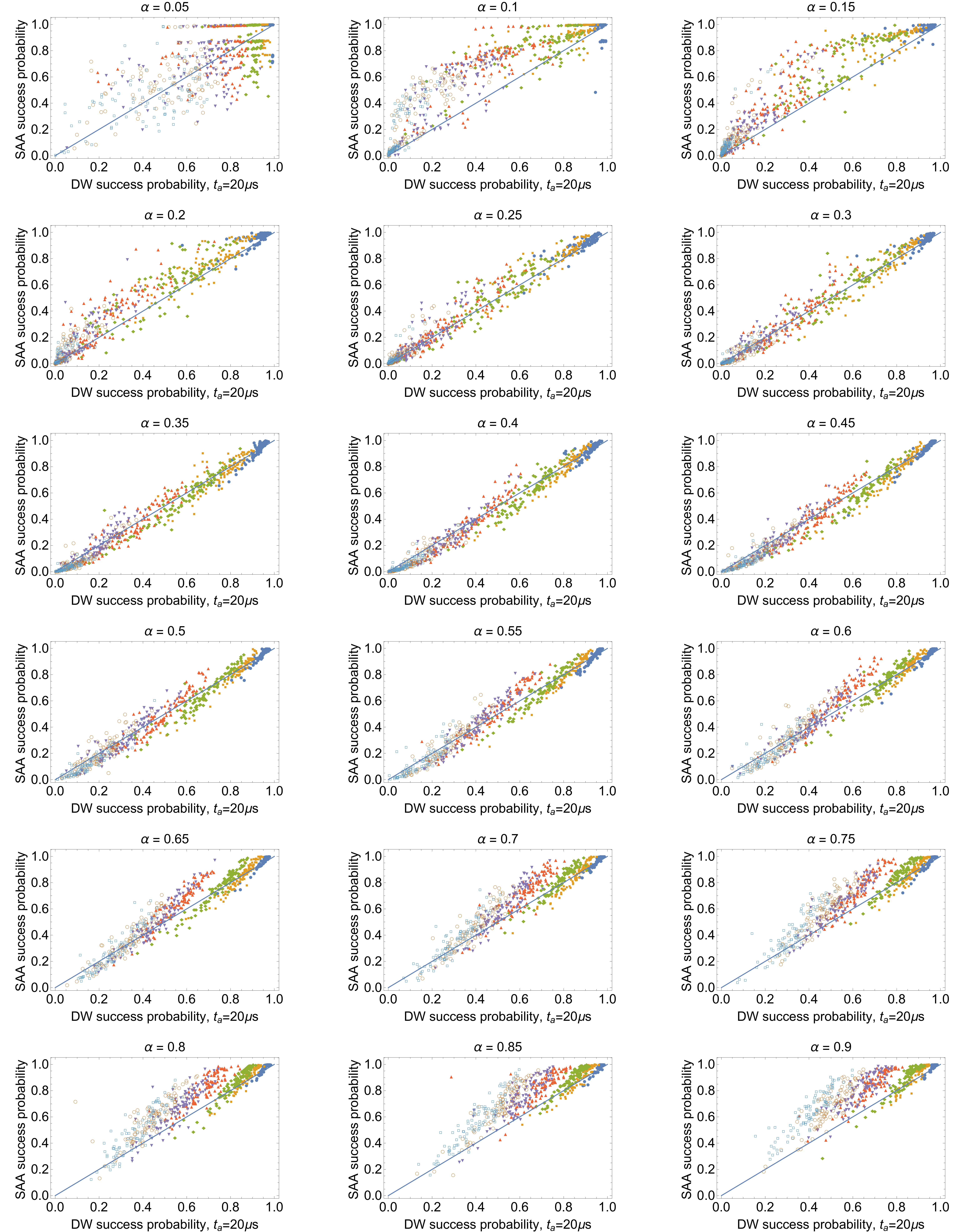}
\caption{\textbf{Success probability correlations.} The results for all instances at all $\alpha$ values are shown for the DW2 data at $t_a=20\mu$s and SAA with $S=50,\!000$ and $\beta_f=5$. This complements Fig.~\ref{fig:DW-DW-SAA-corr}; the color scheme corresponds to different sizes $L$ as in that figure. The correlation gradually improves from poor for the lowest clause densities to strong at $\alpha=0.35$, then deteriorates again.}
\label{fig:corr-SAAvsDW-all-alphas}
\end{center}
\end{figure*}

\begin{figure*}
\begin{center}
\includegraphics[width=\textwidth]{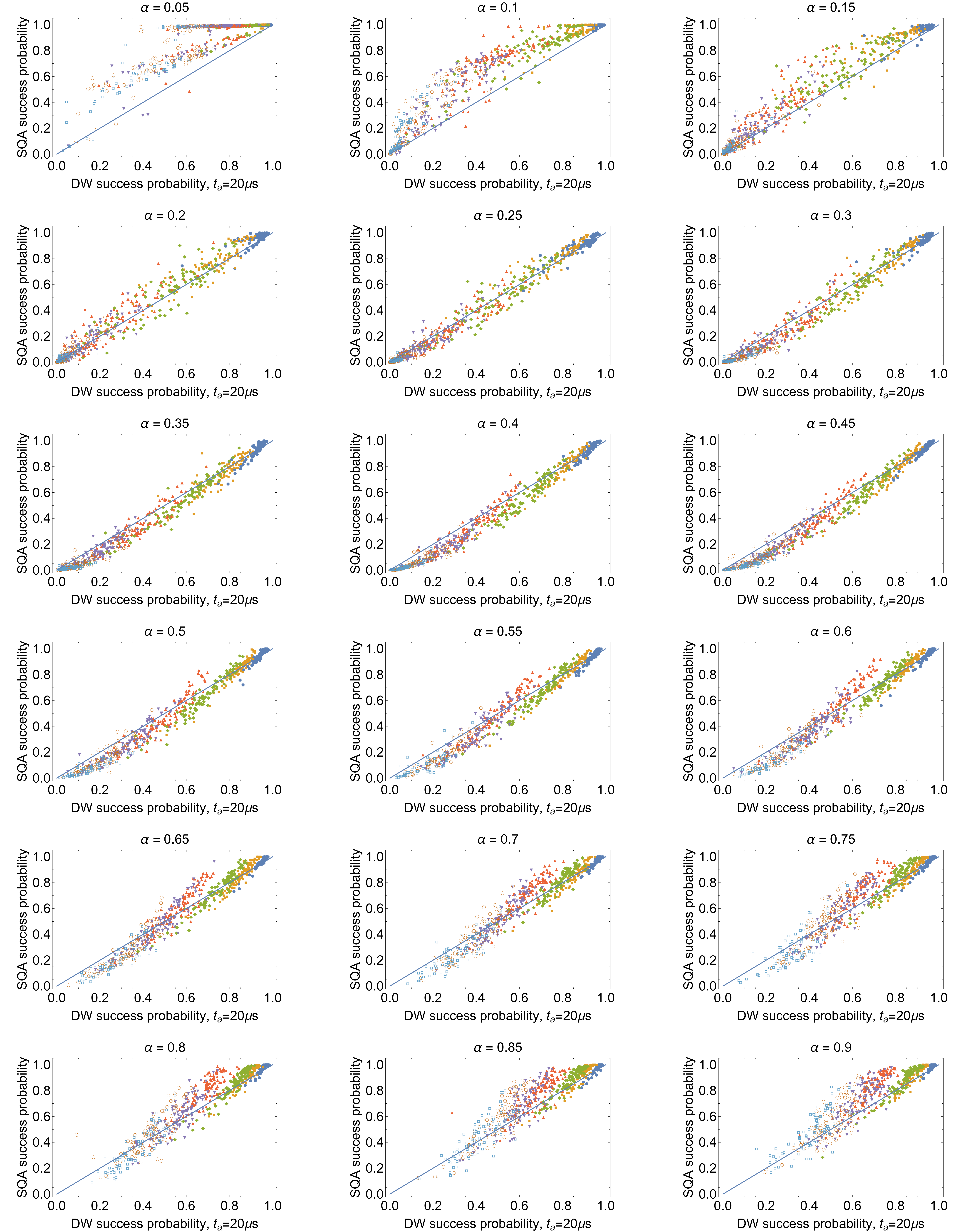}
\caption{\textbf{Success probability correlations.} The results for all instances at all $\alpha$ values are shown for the DW2 data at \protect{$t_a=20\mu$}s and SQAA with $S=10,\!000$ and $\beta=5$. This complements Fig.~\ref{fig:DW-DW-SAA-corr}; the color scheme corresponds to different sizes $L$ as in that figure. The correlation gradually improves from poor for the lowest clause densities to strong at $\alpha=0.35$, then deteriorates again.}
\label{fig:corr-SQAvsDW-all-alphas}
\end{center}
\end{figure*}

\begin{figure*}
\begin{center}
\includegraphics[width=\textwidth]{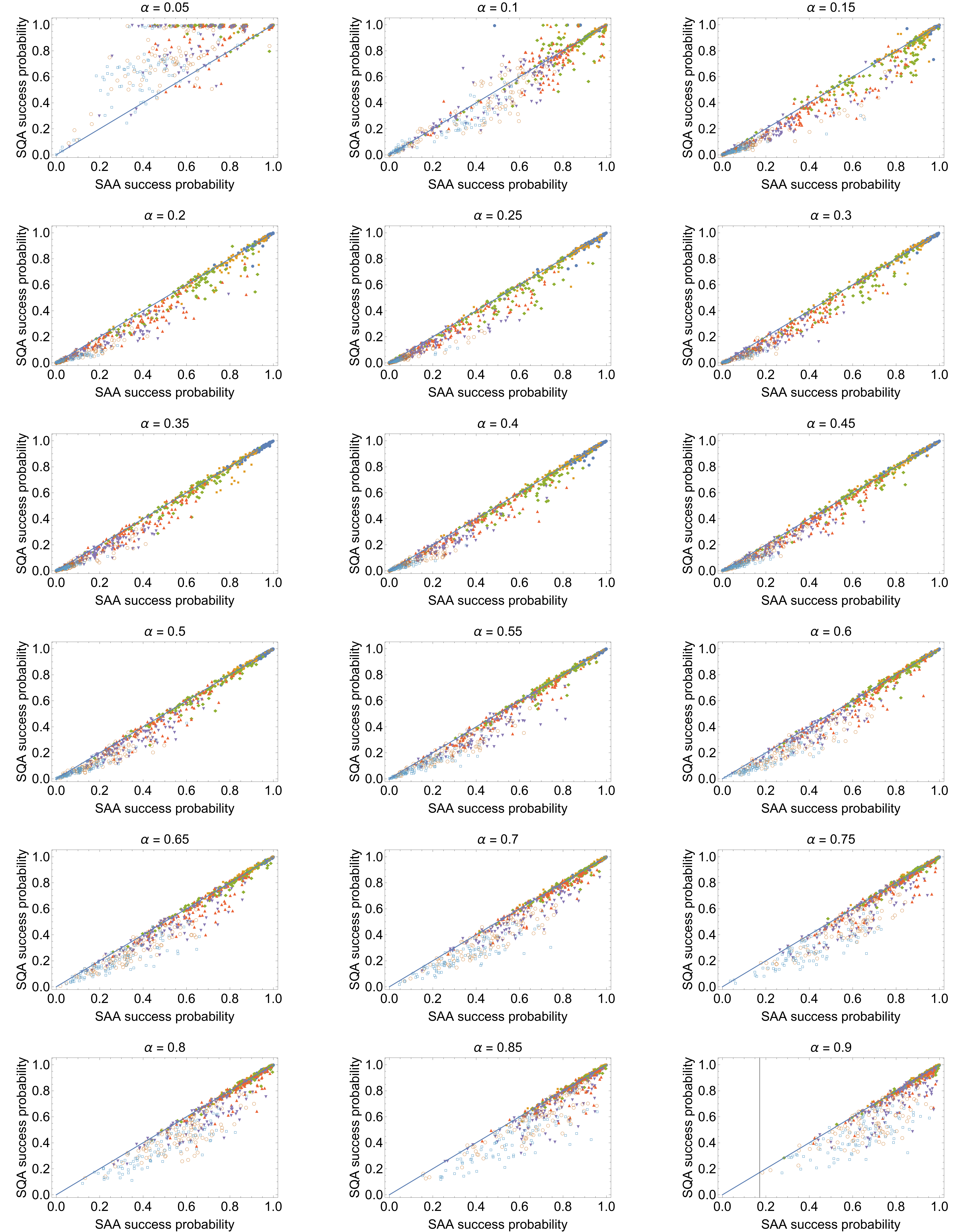}
\caption{\textbf{Success probability correlations.} The results for all instances at all $\alpha$ values are shown for the SQAA data at  $S=10,\!000$ and $\beta=5$ and SAA with $S=50,\!000$ and $\beta_f=5$. This complements Fig.~\ref{fig:DW-DW-SAA-corr}; the color scheme corresponds to different sizes $L$ as in that figure. The correlation gradually improves from poor for the lowest clause densities to strong at $\alpha=0.35$, then deteriorates again.  Note that we did not attempt to optimize the correlations between the two methods.}
\label{fig:corr-SAAvsSQA-all-alphas}
\end{center}
\end{figure*}

To complement Fig.~\ref{fig:DW-DW-SAA-corr}, we provide correlation plots for both the DW2 against itself at $t_a=20\mu$s and $t_a=40\mu$s (Fig.~\ref{fig:corr-DWvsDW-all-alphas}), the DW2 \vs SAA (Fig.~\ref{fig:corr-SAAvsDW-all-alphas}), the DW2 \vs SQAA (Fig.~\ref{fig:corr-SQAvsDW-all-alphas}), and SAA \vs SQAA (Fig.~\ref{fig:corr-SAAvsSQA-all-alphas}. The DW2 against itself displays an excellent correlation at all clause densities, while the DW2 \vs SAA and DW2 \vs SQAA continues to be skewed at low and high clause densities. Recall that Fig.~ref{fig:Euclid-dist} provides an objective Euclidean distance measure that is computed using all problem sizes and depends only on the clause density.

\subsection{Additional scaling analysis plots} \label{sec:scaling}
We provide a few additional plots in support of the scaling analysis presented in the main text. 

\begin{figure*}[t]
\begin{center}
\subfigure[]{\includegraphics[width=\columnwidth]{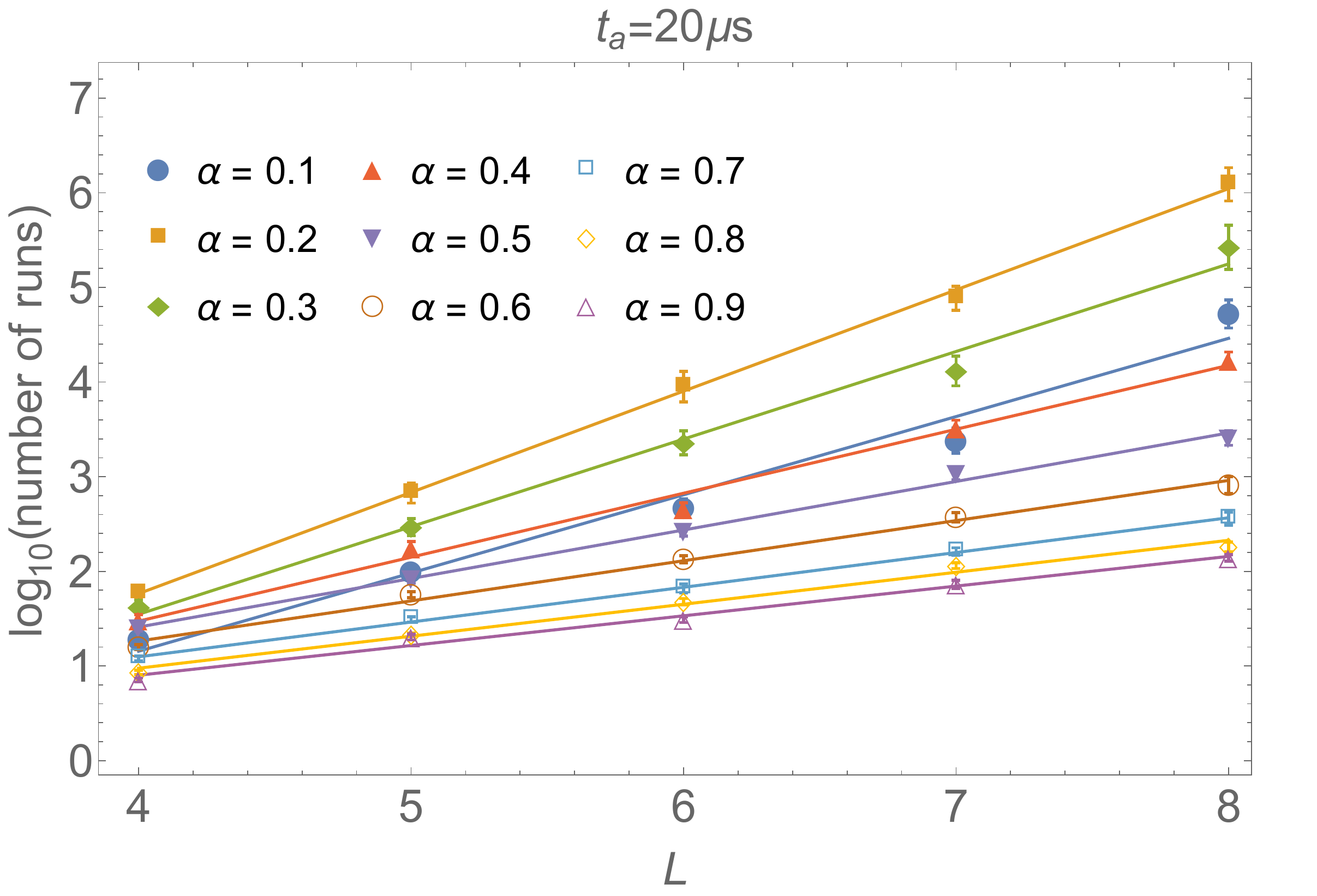}\label{fig:DW-fit}}
\subfigure[]{\includegraphics[width=\columnwidth]{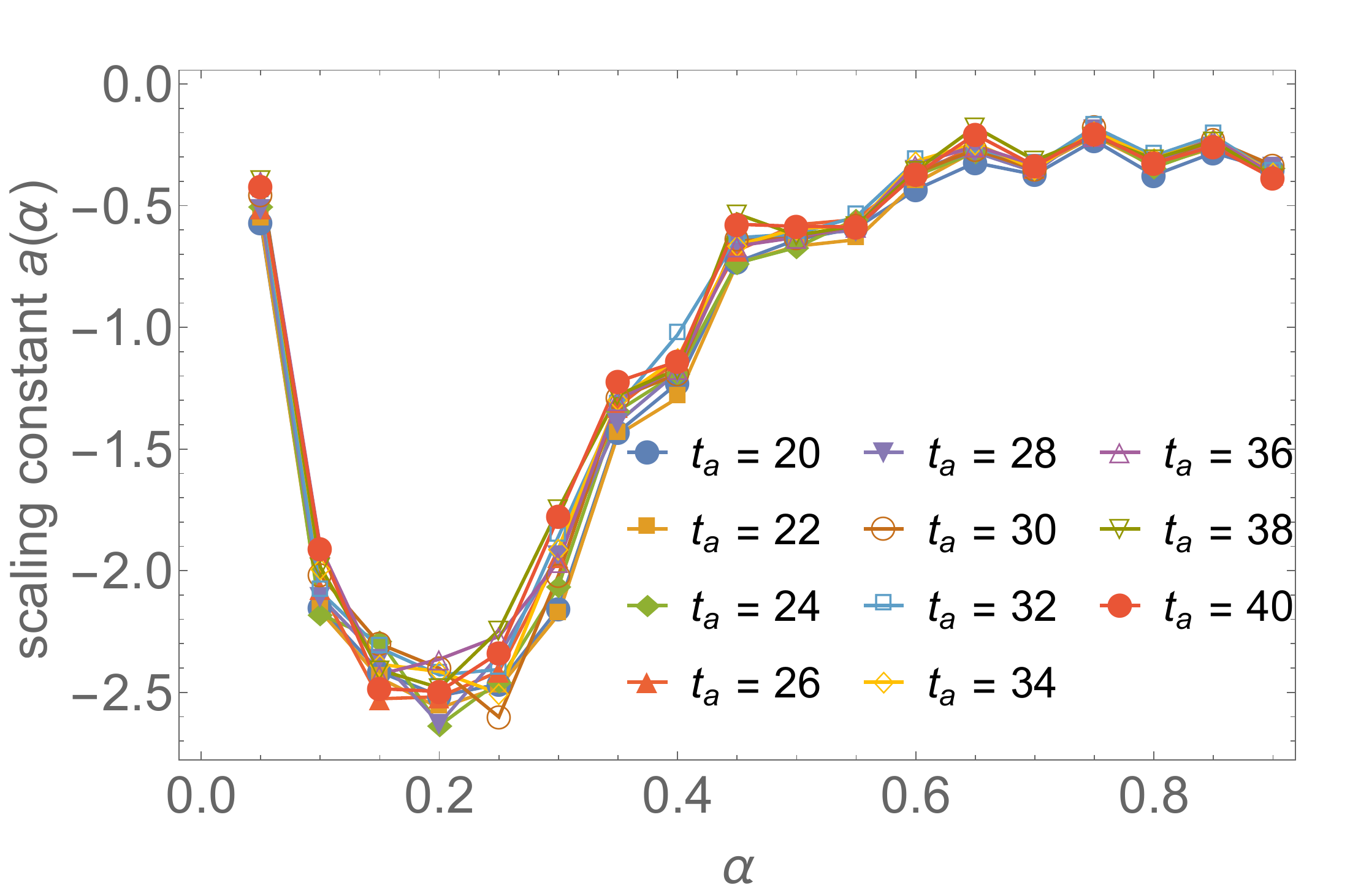}\label{fig:DW-scaling-constant}}
\caption{\textbf{Exponential fits to the DW2 number of runs.} In accordance with Eq.~\eqref{eq:fit-exp-r}, the least-squares linear fits to $\log[r(L,\alpha,0.5)]$ are plotted in (a) for $t_a=20\mu$s. To reduce finite-size scaling effects we exclude $L=2,3$ and perform the fit for $L\geq 4$. The intercept of the linear fits is the the DW2 scaling constant $a(\alpha)$ from Eq.~\eqref{eq:fit-exp-r}, and is shown in (b). Error bars represent $2\sigma$ confidence intervals.}
\end{center}
\end{figure*}

Figure~\ref{fig:DW-fit} shows the number of runs at different problem sizes and clause densities, and the corresponding least-squares fits. It can be seen that the straight lines fits are quite good. The slopes seen in this figure are the $b(\alpha)$ values for $t_a=20\mu$s plotted in Fig.~\ref{fig:DWslope}; the intercepts are plotted in Fig.~\ref{fig:DW-scaling-constant} for all annealing times, and collapse nicely, just like the $b(\alpha)$.

Finally, Fig.~\ref{fig:slope-DW_all_ta-SAA_many_sweeps} is a check of the convergence of SAA to its asymptotic scaling coefficient as the number of sweeps is increased from $5,000$ to $50,\!000$. Convergence is apparent within the $2\sigma$ error bars.

\subsection{SQAS \vs SAS} \label{sec:SQASSASscaling}
\begin{figure*}[t]
\begin{center}
\subfigure[]{\includegraphics[width=4in]{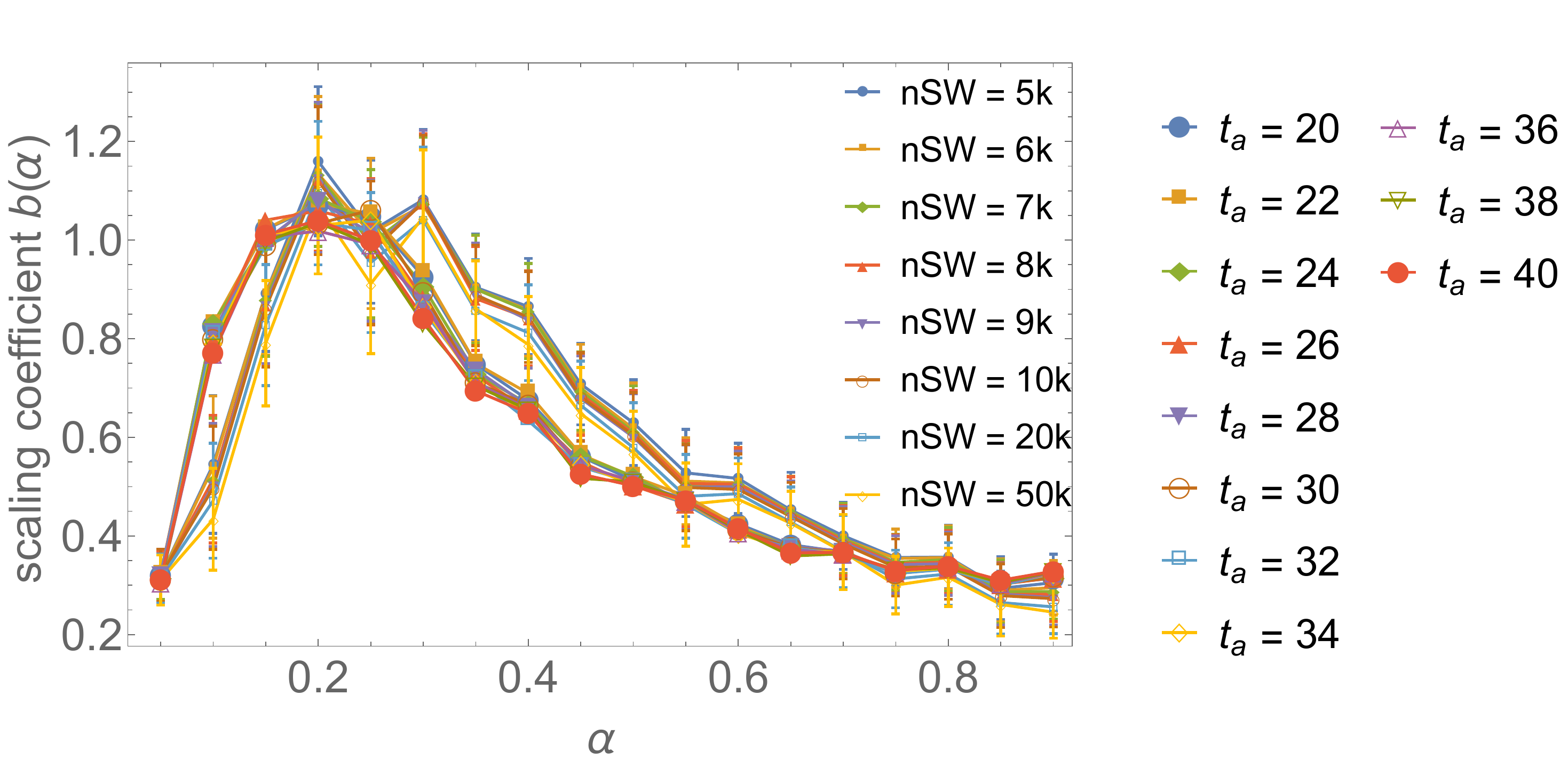}\label{fig:slope-DW_all_ta-SAA_many_sweeps}}
\subfigure[]{\includegraphics[width=2.75in]{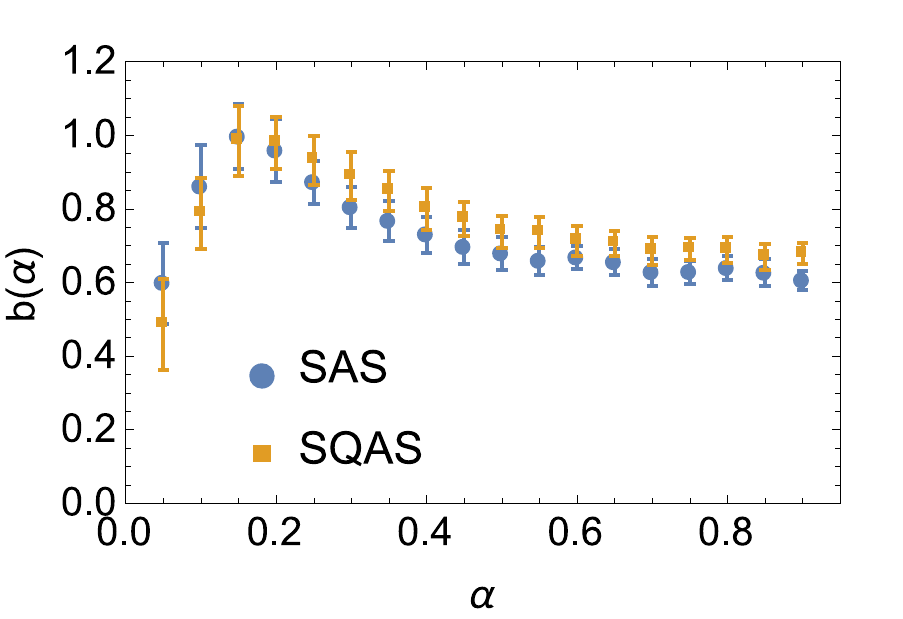}\label{fig:slope-SQA-SA}}
\caption{
(a) The DW2 scaling coefficients $b(\alpha)$ from Eq.~\eqref{eq:fit-exp-r} for SAA at various sweep numbers and $\beta_f=5$, along with the DW2 scaling coefficients for all $t_a$s we tried.
(b) The SQAS and SAS scaling coefficient $b(\alpha)$ at the optimal number of sweeps for each.  SAS had a final temperature of $\beta_f = 5$ and SQAS was operated at $\beta = 5$.}
\end{center}
\end{figure*}

%
In the main text we only considered SQA as an annealer since that is a more faithful representation of the DW2.  Here we present a comparison of SQA as a solver (SQAS), where we keep track of the lowest energy found during the entire anneal, with SAS.  We present the scaling coefficient $b(\alpha)$ from Eq.~\eqref{eq:fit-exp-r} of these two solvers in Fig.~\ref{fig:slope-SQA-SA}.   SAS has a smaller scaling coefficient than SQAS for the large $\alpha$ values, but at small $\alpha$ values we cannot make a conclusive determination because of the substantial overlap of the error bars.  We note that Ref.~\cite{Heim:2014jf} reported that discrete-time SQA (the version used here) can exhibit a scaling advantage over SA but that this advantage vanishes in the continuous-time limit.  We have not explored this possibility here. 

\subsection{Scale factor histograms} 
\label{sec:precision}

\begin{figure*}[t]
\begin{center}
\includegraphics[width=\textwidth]{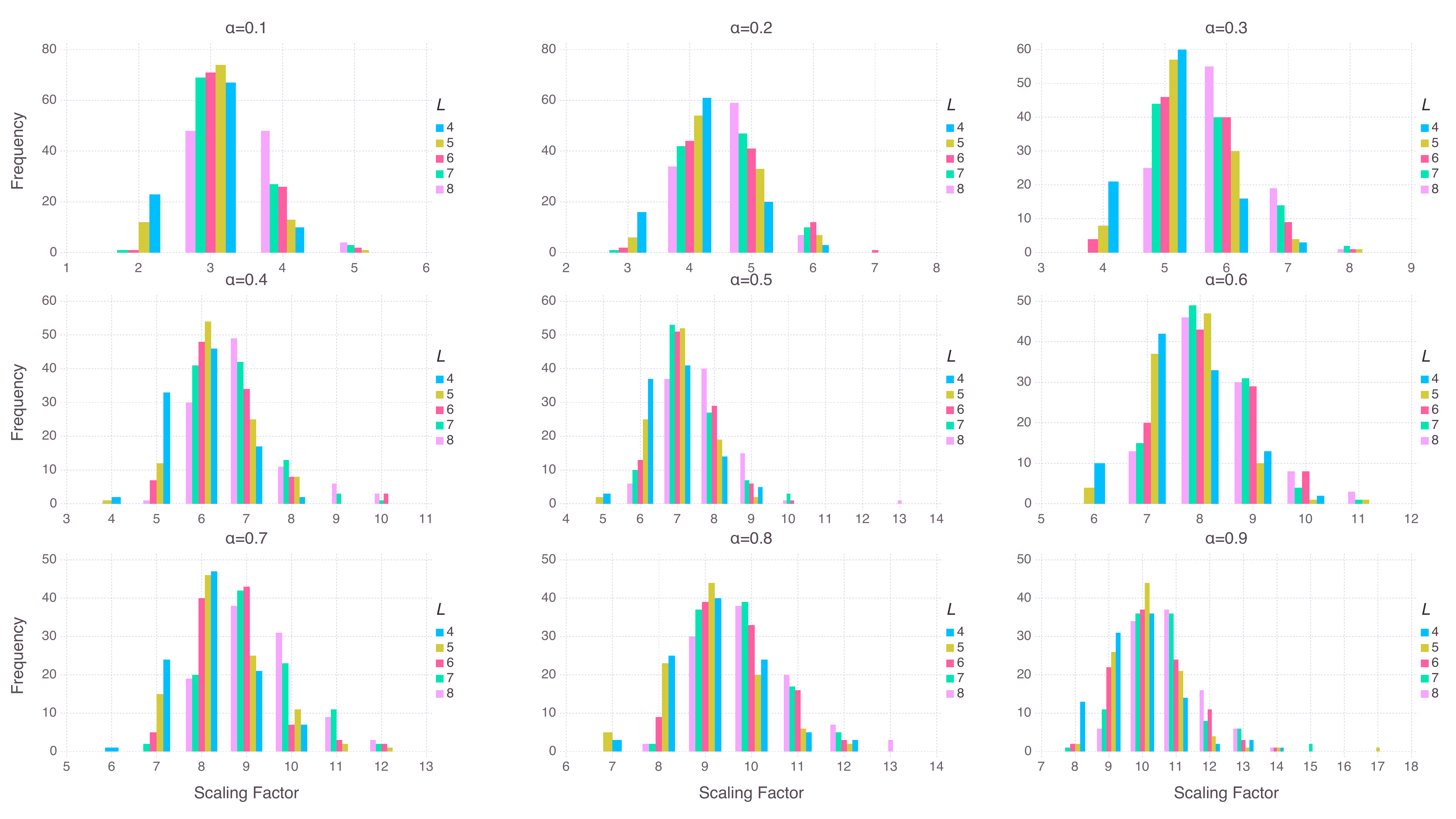}
\caption{
Histograms of the scale factor of the instances as a function of system size for various values of $\alpha$. All problems passed to DW2 must have all coupler in the range $[-1,1]$, so all couplers in a problem are rescaled down by a factor equal to the maximum absolute value of the couplers in the problem (and hence this quantity is called the scale factor). Since internal control error (ICE) is largely instance-independent, the larger the scaling factor of an instance, the worse the relative impact of ICE will be. We see a drift to larger scaling factors for with increasing size $L$ and increasing clause density $\alpha$. Larger values of $\alpha$ obviously will have larger scale factors as there are on average more loops per qubit (and thus, per edge) and thus a larger maximum potential coupler strength. Since the edges included in a loop are generated randomly, the more edges available at a fixed clause density, the more opportunities for a single edge to be included in many loops by chance, resulting the average scale factor to drift upward as function of problem size. 
}
\label{fig:precision}
\end{center}
\end{figure*}

We analyze the effect of increasing clause density and problem size on the required precision of the couplings. Fig.~\ref{fig:precision} shows a trend of scaling factor increasing as $\alpha$ increases for fixed $L$, and as $L$ increases for fixed $\alpha$. Increased scaling factor has the effect of relatively amplifying control error and thermal effects in DW2, and can therefore contribute to a decline in performance for the larger problems studied.
However, recall that the region where a speedup is possible according to our results is in fact that of high clause densities. Thus, whatever the effect of precision errors is, it does not appear to heavily impact the DW2's performance in the context of our problems. The same is true for SAS, since as can be seen in Fig.~\ref{fig:DiffDWslope}, the scaling coefficient is unaffected by the addition of noise when $\alpha \gtrsim 0.6$.

%

\end{document}